\documentclass[a4paper,11pt]{article}
\usepackage{jheppub}
\usepackage[utf8]{inputenc}

\usepackage{comment}

\usepackage{dsfont}

\usepackage{mathtools,amsmath,amssymb}
\usepackage{graphicx}

\usepackage{lmodern}
\usepackage[T1]{fontenc}
\usepackage{microtype}

\usepackage{hyperref}

\usepackage{subcaption}
\usepackage{xspace}

\usepackage{xspace}
\usepackage{tikz}
\usetikzlibrary{shapes}
\usetikzlibrary{decorations.markings}

\tikzset{->-/.style={decoration={
  markings,
  mark=at position .5 with {\arrow{>}}},postaction={decorate}}}

\usepackage{placeins}

\DeclareMathOperator{\phaneq}{\phantom{{}=}}

\newcommand{\refspina}{\xi_A}
\newcommand{\refspinb}{\xi_B}
\newcommand{\refspinal}{\xi_A}
\newcommand{\refspinbl}{\xi_B}

\newcommand{\e}{\operatorname{e}}

\newcommand{\de}{\operatorname{d}\!}

\newcommand{\dd}{\mathrm{d}}

\newcommand{\res}{\text{Res}}
\newcommand{\rest}{\widetilde{\res}}

\newcommand{\ba}{\mathbf{a}}
\newcommand{\bb}{\mathbf{b}}

\newcommand{\bd}{\mathbf{d}}

\newcommand{\aosc}{\ba}
\newcommand{\aoscdag}{{\aosc^\dagger}}
\newcommand{\bosc}{\bb}
\newcommand{\boscdag}{{\bosc^\dagger}}

\newcommand{\dosc}{\bd}
\newcommand{\doscdag}{{\dosc^\dagger}}

\newcommand{\nfsym}{$\mathcal{N}\!=\! 4$ SYM}

\newcommand{\abr}[1]{\langle #1 \rangle}
\newcommand{\sbr}[1]{\left[ #1 \right]}
\newcommand{\bra}[1]{\langle #1 |}

\newcommand{\ptwistor}{\mathcal{W}}
\newcommand{\mtwistor}{\mathcal{Z}}

\newcommand{\vll}{{\smash{\lambda}}}
\newcommand{\vlt}{{\smash{\tilde{\lambda}}}}
\newcommand{\vle}{{\smash{\tilde{\eta}}}}
\newcommand{\vlet}{{\smash{\eta}}}

\newcommand{\vllu}{\smash{\underline{\smash{\lambda}}}}
\newcommand{\vltu}{\smash{\underline{\smash{\tilde{\lambda}}}}}
\newcommand{\vleu}{\smash{\underline{\smash{\tilde{\eta}}}} }

\newcommand{\vlluu}{\smash{\underline{\underline{\smash{\lambda}}}}}
\newcommand{\vltuu}{\smash{\underline{\underline{\smash{\tilde{\lambda}}}}}}
\newcommand{\vleuu}{\smash{\underline{\underline{\smash{\tilde{\eta}}}}}}
\newcommand{\vletuu}{\smash{\underline{\underline{\smash{\eta}}}}}

\newcommand{\vlluuu}{\vlluu}
\newcommand{\vltuuu}{\vltuu}
\newcommand{\vleuuu}{\vleuu}
\newcommand{\vletuuu}{\vletuu}

\newcommand{\sigmapart}{\tilde{\sigma}}

\newcommand{\vmmuuu}{{\smash{\underline{\underline{\smash{\mu}}}}}}
\newcommand{\vmtuuu}{{\smash{\underline{\underline{\smash{\tilde{\mu}}}}}}}

\newcommand{\calA}{\mathcal{A}}
\newcommand{\calB}{\mathcal{B}}
\newcommand{\calC}{\mathcal{C}}
\newcommand{\calD}{\mathcal{D}}
\newcommand{\calO}{\mathcal{O}}

\newcommand{\calL}{\mathcal{L}}

\newcommand{\jj}{J}

\newcommand{\lax}{\mathcal{L}}
\newcommand{\mono}{\mathcal{M}}
\newcommand{\trans}{\mathcal{T}}
\newcommand{\transosc}{\mathbf{T}}
\newcommand{\laxosc}{\mathbf{L}}

\newcommand{\amp}{\mathcal{A}}
\newcommand{\op}{\calO}
\newcommand{\ff}{\mathcal{F}}
\newcommand{\ffgen}{\hat{\mathcal{F}}}

\newcommand{\rr}{\ensuremath{\mathrm{R}}\xspace}
\newcommand{\rrpdf}{R}

\newcommand{\MHV}{\ensuremath{{\mathrm{MHV}}}}
\newcommand{\MHVbar}{\ensuremath{\overline{\MHV}}}
\newcommand{\NmaxMHV}{\ensuremath{\text{N}^{\text{max}}\MHV}}

\DeclareMathOperator{\str}{str}

\newcommand{\eqndot}{\, . }
\newcommand{\eqncom}{\, , }
\newcommand{\eqnsem}{\, ; }
\newcommand{\splus}{\! + \!}
\newcommand{\sminus}{\! - \!}
\newcommand{\ssep}{\;}
\newcommand{\sssep}{\,}

\newcommand{\deltap}[1]{\delta^+_{#1}}
\newcommand{\deltam}[1]{\delta^-_{#1}}
\newcommand{\deltaf}[1]{\delta^{\ff}_{#1}}

\newcommand{\ketm}[1]{\mathopen{\mid}{}#1{}\mathclose{\rangle}}
\newcommand{\bram}[1]{\mathopen{\langle}{}#1{}\mathclose{\mid}}

\DeclareMathOperator{\tr}{tr}

\DeclareMathOperator{\idm}{\mathds{1}}

\definecolor{grayn}{gray}{0.7}
\definecolor{lightgrayn}{gray}{0.8}

\def\bridgedistance{0.75}
\def\vacuumheight{1}
\def\ddist{0.75}
\def\hdist{\ddist*0.70710678118}
\def\labelvdist{0.3}
\def\labelhdist{0.3}
\def\labelddist{\labelvdist*0.70710678118}

\newlength{\vacuumradius}
\newlength{\onshellradius}
\tikzstyle{db}=[circle, black, fill=black, minimum width=\onshellradius, draw, inner sep=0pt]
\tikzstyle{dw}=[circle, black, fill=white, minimum width=\onshellradius, draw, inner sep=0pt]
\tikzstyle{dvac}=[circle, black, fill=lightgrayn, minimum width=\vacuumradius, inner sep=0pt]
\tikzstyle{dl}=[circle, black, fill=white, inner sep=2pt]

\newcommand{\drawminimalff}[1]{
        \draw[thick,double] (#1-0.5,-0) -- (#1-0.5,-0.5); 
	\draw (#1-0.5,-0.5) -- (#1-1,-\vacuumheight);  
	\draw (#1-0.5,-0.5) -- (#1,-\vacuumheight);}

\newcommand{\drawvacp}[1]{       
        \draw (#1-1,-.25) -- (#1-1,-\vacuumheight); 
        \node[dvac] at (#1-1,-0.25) {$+$};}
\newcommand{\drawvacm}[1]{       
        \draw (#1-1,-.25) -- (#1-1,-\vacuumheight); 
        \node[dvac] at (#1-1,-0.25) {$-$};}
\newcommand{\drawbridge}[2]{
        \draw (#1-1,-\vacuumheight -#2*\bridgedistance+\bridgedistance) -- (#1,-\vacuumheight -#2*\bridgedistance+\bridgedistance); 
	\node[dw] at (#1-1,-\vacuumheight -#2*\bridgedistance+\bridgedistance) {};
	\node[db] at (#1,-\vacuumheight -#2*\bridgedistance+\bridgedistance) {};}
\newcommand{\drawvline}[2]{
        \draw (#1-1,-\vacuumheight) -- (#1-1,-\vacuumheight -#2*\bridgedistance);}

\newcommand{\athreetwofthreetwo}[4]{
 \begin{aligned}
 \begin{tikzpicture}[scale=0.8]
 			\draw[thick,double] (0,-0) -- (0,0.5); 
  			\draw (0,0) -- (-\hdist,-\hdist) -- (0,-2*\hdist) -- (+\hdist,-\hdist) -- (0,0); 
  			\draw (0,-2*\hdist) -- (0,-2*\hdist-\ddist) -- (+\ddist,-2*\hdist-\ddist) -- (+\hdist+\ddist,-\hdist-\ddist) -- (+\hdist,-\hdist); 
 			\draw (+\hdist+\ddist,-\hdist-\ddist) -- (+2*\hdist+\ddist,-2*\hdist-\ddist);   
 			\draw (-\hdist,-\hdist) -- (-2*\hdist,-2*\hdist);   
 			\draw (0,-2*\hdist-\ddist) -- (0,-2*\hdist-2*\ddist); 
 			\draw (\ddist,-2*\hdist-\ddist) -- (\hdist+\ddist,-3*\hdist-\ddist);
                         \node[dw] at (+\hdist,-\hdist) {}; 
                         \node[db] at (0,-2*\hdist) {};
                         \node[dw] at (-\hdist,-\hdist) {}; 
                         \node[db] at (0,-2*\hdist-\ddist) {}; 
                         \node[dw] at (+\ddist,-2*\hdist-\ddist) {};
                         \node[db] at (+\hdist+\ddist,-\hdist-\ddist) {}; 
                         \node at (+2*\hdist +\ddist +\labelddist,-2*\hdist-\ddist-\labelddist) {#1};
                         \node at (-2*\hdist-\labelddist,-2*\hdist-\labelddist) {#4};
                         \node at (0,-2*\hdist-2*\ddist-\labelvdist) {#3};
                         \node at (\hdist +\ddist +\labelddist,-3*\hdist-\ddist-\labelddist) {#2};
         \end{tikzpicture}
\end{aligned}
}
\newcommand{\afourtwoftwotwo}[4]{
 \begin{aligned}
 \begin{tikzpicture}[scale=0.8]
			\draw[thick,double] (0,-0+\hdist) -- (0,0.5+\hdist); 
 			\draw (0,\hdist) -- (+\hdist,0);
 			\draw (0,\hdist) -- (-\hdist,0);
 			\draw (+\hdist,0) -- (1*\hdist,0) -- (2*\hdist,\hdist);
 			\draw (-\hdist,0) -- (-3*\hdist,0) -- (-4*\hdist,\hdist);
 			\draw (2*\hdist,-3*\hdist) -- (1*\hdist,-2*\hdist) -- (-3*\hdist,-2*\hdist) -- (-4*\hdist,-3*\hdist);
 			\draw (-\hdist,0) -- (-\hdist,-2*\hdist);
			\draw (-3*\hdist,0) -- (-3*\hdist,-2*\hdist);
 			\draw (\hdist,0) -- (\hdist,-2*\hdist);
                         \node[db] at (+\hdist,0) {}; 
                         \node[dw] at (-\hdist,0) {}; 
                         \node[db] at (-3*\hdist,0) {}; 
                         \node[dw] at (+\hdist,-2*\hdist) {}; 
                         \node[db] at (-\hdist,-2*\hdist) {}; 
                         \node[dw] at (-3*\hdist,-2*\hdist) {}; 
                         \node at (-4*\hdist-\labelddist,\hdist +\labelddist) {#4};
                         \node at (-4*\hdist-\labelddist,-3*\hdist-\labelddist) {#3};
                         \node at (2*\hdist +\labelddist,-3*\hdist-\labelddist) {#2};
                         \node at (2*\hdist +\labelddist,\hdist +\labelddist) {#1};
         \end{tikzpicture}
\end{aligned}
}
\newcommand{\fthreethreeathreeone}[4]{
 \begin{aligned}
 \begin{tikzpicture}[scale=0.8]
 			\draw[thick,double] (0,-0) -- (0,0.5); 
  			\draw (0,0) -- (\hdist,-\hdist) -- (0,-2*\hdist) -- (-\hdist,-\hdist) -- (0,0); 
  			\draw (0,-2*\hdist) -- (0,-2*\hdist-\ddist) -- (-\ddist,-2*\hdist-\ddist) -- (-\hdist-\ddist,-\hdist-\ddist) -- (-\hdist,-\hdist); 
 			\draw (-\hdist-\ddist,-\hdist-\ddist) -- (-2*\hdist-\ddist,-2*\hdist-\ddist);   
 			\draw (+\hdist,-\hdist) -- (+2*\hdist,-2*\hdist);   
 			\draw (0,-2*\hdist-\ddist) -- (0,-2*\hdist-2*\ddist); 
 			\draw (-\ddist,-2*\hdist-\ddist) -- (-\hdist-\ddist,-3*\hdist-\ddist);
                         \node[db] at (-\hdist,-\hdist) {}; 
                         \node[dw] at (0,-2*\hdist) {};
                         \node[db] at (+\hdist,-\hdist) {}; 
                         \node[dw] at (0,-2*\hdist-\ddist) {}; 
                         \node[db] at (-\ddist,-2*\hdist-\ddist) {};
                         \node[dw] at (-\hdist-\ddist,-\hdist-\ddist) {};  
                         \node at (-2*\hdist -\ddist -\labelddist,-2*\hdist-\ddist-\labelddist) {#2};
                         \node at (+2*\hdist+\labelddist,-2*\hdist-\labelddist) {#3};
                         \node at (0,-2*\hdist-2*\ddist-\labelvdist) {#4};
                         \node at (-\hdist -\ddist -\labelddist,-3*\hdist-\ddist-\labelddist) {#1};
         \end{tikzpicture}
\end{aligned}
}
\newcommand{\ftwotwoafourtwo}[4]{
 \begin{aligned}
 \begin{tikzpicture}[scale=0.8]
 			\draw[thick,double] (0,-0+\hdist) -- (0,0.5+\hdist); 
 			\draw (0,\hdist) -- (+\hdist,0);
 			\draw (0,\hdist) -- (-\hdist,0);
 			\draw (+\hdist,0) -- (3*\hdist,0) -- (4*\hdist,\hdist);
 			\draw (-\hdist,0) -- (-1*\hdist,0) -- (-2*\hdist,\hdist);
 			\draw (4*\hdist,-3*\hdist) -- (3*\hdist,-2*\hdist) -- (-1*\hdist,-2*\hdist) -- (-2*\hdist,-3*\hdist);
 			\draw (-\hdist,0) -- (-\hdist,-2*\hdist);
 			\draw (\hdist,0) -- (\hdist,-2*\hdist);
			\draw (3*\hdist,0) -- (3*\hdist,-2*\hdist);
                         \node[db] at (+\hdist,0) {}; 
                         \node[dw] at (+3*\hdist,0) {};
                         \node[dw] at (-\hdist,0) {}; 
                         \node[dw] at (+\hdist,-2*\hdist) {}; 
                         \node[db] at (+3*\hdist,-2*\hdist) {};
                         \node[db] at (-\hdist,-2*\hdist) {}; 
                         \node at (4*\hdist +\labelddist,\hdist +\labelddist) {#3};
                         \node at (4*\hdist +\labelddist,-3*\hdist-\labelddist) {#4};
                         \node at (-2*\hdist -\labelddist,-3*\hdist-\labelddist) {#1};
                         \node at (-2*\hdist -\labelddist,\hdist +\labelddist) {#2};
         \end{tikzpicture}
\end{aligned}
}

\newcommand{\dualsupermomentum}{\vartheta}

\makeatletter
\DeclareRobustCommand*{\bfseries}{%
  \not@math@alphabet\bfseries\mathbf
  \fontseries\bfdefault\selectfont
  \boldmath
}

\makeatother

\usepackage{lipsum}
\makeatletter
\if@titlepage
  \renewenvironment{abstract}{%
      \titlepage
      \null\vfil
      \@beginparpenalty\@lowpenalty
      \begin{center}%
        \bfseries \abstractname
        \@endparpenalty\@M
      \end{center}}%
     {\par\vfil\null\endtitlepage}
\else
  \renewenvironment{abstract}{%
      \if@twocolumn
        \section*{\abstractname}%
      \else
        \small
        \begin{center}%
          {\bfseries \abstractname\vspace{-.5em}\vspace{\z@}}%
        \end{center}%
        \quotation
      \fi}
      {\if@twocolumn\else\endquotation\fi}
\fi
\makeatother

\title{On-shell Diagrams, Graßmannians and Integrability for Form Factors}

\begin{document}

\begingroup\parindent0pt
\begin{flushright}\footnotesize
\texttt{HU-MATH-2015-10}\\
\texttt{HU-EP-15/30}\\
\texttt{DCPT-15/37}
\end{flushright}
\vspace*{4em}
\centering
\begingroup\LARGE
\bf
On-shell Diagrams, Graßmannians and Integrability for Form Factors
\par\endgroup
\vspace{2.5em}
\begingroup\large{\bf
Rouven Frassek}$\,^a$, {\bf David Meidinger}$\,^b$, {\bf Dhritiman Nandan}$\,^b$,\\{\bf Matthias Wilhelm}$\,^b$
\par\endgroup
\vspace{1em}
\begingroup\itshape
$^a\,$Department of Mathematical Sciences, Durham University,\\
     South Road, Durham DH1 3LE, United Kingdom\\
\vspace{0.5em}
$^b\,$Institut für Mathematik und Institut für Physik,
Humboldt-Universität zu Berlin,\\
IRIS Gebäude,
Zum Großen Windkanal 6,
12489 Berlin\\

\par\endgroup
\vspace{1em}
\begingroup\ttfamily
rouven.frassek@durham.ac.uk, david.meidinger@physik.hu-berlin.de, dhritiman@physik.hu-berlin.de, mwilhelm@physik.hu-berlin.de \\
\par\endgroup
\vspace{2.5em}
\endgroup

\begin{abstract}
\noindent
We apply on-shell and integrability methods that have been developed in the context of scattering amplitudes in $\mathcal{N}=4$ SYM theory to tree-level form factors of this theory. 
Focussing on the colour-ordered super form factors of the chiral part of the stress-tensor multiplet as an example, we show how to systematically construct on-shell diagrams for these form factors with the minimal form factor as further building block in addition to the three-point amplitudes. 
Moreover, we obtain analytic representations in terms of Graßmannian integrals in spinor helicity, twistor and momentum twistor variables.
While Yangian invariance is broken by the operator insertion, we find that the form factors are eigenstates of the integrable spin-chain transfer matrix  built from the monodromy matrix that yields the Yangian generators.
Constructing them via the method of $\rr$ operators allows to introduce deformations that preserve the integrable structure.
We finally show that the integrable properties extend to minimal tree-level form factors of generic composite operators as well as certain leading singularities of their $n$-point loop-level form factors.
\end{abstract}


\thispagestyle{empty}

\newpage
\hrule
\tableofcontents
\afterTocSpace
\hrule
\afterTocRuleSpace

\section{Introduction}
In the last years, there has been tremendous progress in our understanding of $\mathcal{N}=4$ Super Yang-Mills (SYM) theory in the planar limit primarily based on two different approaches, namely the on-shell methods of modern quantum field theory and integrability techniques; see \cite{Elvang:2013cua,Henn:2014yza} and \cite{Beisert:2010jr} for respective reviews.
The former set of ideas and techniques has been successfully applied to the perturbative study of on-shell scattering amplitudes of elementary states. 
On the other hand, integrability-based methods, which rely on exploiting all the symmetries of the theory, have proven to be very powerful in particular in calculating the spectrum of anomalous dimensions of gauge-invariant local composite operators. 
Though some of the integrable structures have also appeared in the study of scattering amplitudes, the overlap between both approaches has been rather limited. 

Along with scattering amplitudes and correlation functions, another very interesting quantity in a quantum field theory is the form factor, which forms a bridge between the previously mentioned on-shell amplitudes and off-shell correlation functions.  
For a given gauge-invariant local composite operator $\calO(x)$ in a quantum field theory, the form factor $\ff_{\calO}$ is defined as the overlap of the off-shell state created by $\calO$ from the vacuum $\ketm{0}$ at the spacetime point $x$ with an on-shell $n$-particle state $\ketm{1,\dots,n}$.\footnote{As for amplitudes, the on-shell state is specified by the momenta, helicities and flavours of the $n$ elementary particles.} 
It can be Fourier transformed to momentum space, where the operator $\calO$ carries a momentum $q$ with $q^2 \neq 0$, yielding
\begin{equation}\label{eq: form factor intro}
 \ff_{\calO}(1,\dots,n;q)=\int \de^4x \e^{-ixq}\bram{1,\dots,n}\calO(x)\ketm{0}\eqndot
\end{equation}
This quantity will be our focus of attention for this paper as it is a perfect candidate to study the theory using both the on-shell and integrability techniques.

In \nfsym\ theory, form factors have been first studied more than thirty years back~\cite{vanNeerven:1985ja}, and have received increasing attention of late, both at weak coupling \cite{Brandhuber:2010ad,Bork:2010wf,Brandhuber:2011tv,Bork:2011cj,Henn:2011by,Gehrmann:2011xn,Brandhuber:2012vm,Bork:2012tt,Engelund:2012re,Johansson:2012zv,Boels:2012ew,Penante:2014sza,Brandhuber:2014ica,Bork:2014eqa,Wilhelm:2014qua,Nandan:2014oga,Loebbert:2015ova} and at strong coupling \cite{Alday:2007he,Maldacena:2010kp,Gao:2013dza} via the $\text{AdS}/\text{CFT}$ correspondence. 
They can be calculated using many of the successful on-shell techniques that were developed in the context of amplitudes.
In particular, BCFW~\cite{Britto:2004ap,Britto:2005fq} and MHV~\cite{Cachazo:2004kj} recursion relations can be applied to construct form factors at tree level~\cite{Brandhuber:2010ad,Brandhuber:2011tv} and the resulting expressions can also be interpreted in terms of the volume of polytopes~\cite{Bork:2014eqa}. Form factors have also been studied at loop level using generalised unitarity~\cite{Bern:1994zx,Bern:1994cg,Britto:2004nc} not just for the simplest BPS operator and its generalisations~\cite{Brandhuber:2010ad,Brandhuber:2012vm,Brandhuber:2014ica} but also for non-protected operators like the Konishi operator~\cite{Nandan:2014oga}, the operators in the $SU(2)$ sector~\cite{Loebbert:2015ova}, and even completely generic operators~\cite{Wilhelm:2014qua}. All these recent developments have shown that simplicity does exist also for form factors if one studies them using the language of modern on-shell techniques.  

However, not all of the interesting features of scattering amplitudes in \nfsym\ theory have a counterpart for form factors yet. 
A novel way of studying scattering amplitudes has been proposed in \cite{ArkaniHamed:2012nw} using so-called on-shell diagrams, which are bipartite  graphs built out of two kinds of trivalent on-shell vertices and encode the information of the scattering process using fully on-shell data. 
Moreover, each such scattering process is related to an integral over a Graßmannian manifold. In fact, it has been conjectured that all leading singularities of the scattering amplitudes of \nfsym\ theory as well as the tree-level scattering amplitudes can be obtained from an integral on a Graßmannian \cite{ArkaniHamed:2009dn,Mason:2009qx,ArkaniHamed:2009vw}. So far, there has been no direct analogue of this geometric picture of scattering for the case of form factors.
One of the goals of this paper is to provide such a formulation, starting with certain tree-level form factors.

Very little is known about the role of integrability in form factors of \nfsym\ theory.%
\footnote{That is, at least at weak coupling. For an application of integrability to form factors at strong coupling, see \cite{Maldacena:2010kp,Gao:2013dza}.}
For amplitudes as well as for correlation functions, integrability manifests itself in the appearance of an integrable spin chain at weak coupling.
In the spectral problem, single-trace operators are mapped to spin-chain eigenstates and the dilatation operator to the spin-chain Hamiltonian, see \cite{Beisert:2010jr}. 
This integrable Hamiltonian belongs to a whole family of commuting operators which also include the corresponding transfer matrices. This family can be diagonalised simultaneously using Bethe ansatz methods.
More recently, a somewhat different spin chain was discovered in the study of amplitudes in \nfsym\ theory \cite{Chicherin:2013ora,Frassek:2013xza,Ferro:2013dga,Beisert:2014qba,Broedel:2014pia,Kanning:2014maa,Bargheer:2014mxa,Ferro:2014gca}. 
It was pointed out in \cite{Drummond:2009fd} that the superconformal symmetry and the newly discovered dual superconformal symmetry \cite{Drummond:2008vq} combine into an infinite-dimensional Yangian symmetry, which yields the underlying integrable spin-chain picture \cite{Chicherin:2013ora,Frassek:2013xza}. 
Amplitudes at tree level are invariant under this Yangian symmetry.

Recently, it was proposed that the problem of computing the dilatation operator, i.e.\ the Hamiltonian of the integrable spin chain, in \nfsym\ theory  can be re-cast in a compact form using generalised unitarity methods with form factors being the main ingredients \cite{Wilhelm:2014qua,Nandan:2014oga,Loebbert:2015ova}.%
\footnote{For the calculation of the dilatation operator from on-shell methods via correlation functions, see \cite{Engelund:2012re,Koster:2014fva,Nandan:2014oga,Brandhuber:2014pta,Brandhuber:2015boa}.}
In fact, a special class of form factors called minimal form factors, with the number of external fields $n$ equal to the number of fields in the corresponding composite operator, realises this spin-chain picture of the spectral problem in the language of on-shell super fields used for amplitudes \cite{Wilhelm:2014qua}.

Naively, due to the nature of form factors and also motivated by the results mentioned above, we would expect to obtain a relation between the integrable spin chain of the spectral problem and the one that appeared in the study of tree-level scattering amplitudes.
Indeed, in this paper, we will show that form factors are special states of the latter integrable spin chain, namely eigenstates of the transfer matrix built from the monodromy that yields the Yangian generators studied in the context of amplitudes, provided that the corresponding composite operator is an eigenstate of the former integrable spin chain.
This implies enhanced symmetries for the form factors, analogous to the Yangian symmetry of scattering amplitudes.

This paper is structured as follows. 
In the remainder of this section, we discuss  the stress-tensor super multiplet and its super form factors, which we will be studying in most of the rest of the paper. In section \ref{sec: MHV}, we briefly review various ideas within the framework of on-shell techniques for scattering amplitudes, like on-shell diagrams, their construction via BCFW recursion relations and inverse soft limits as well as a Graßmannian integral representation, and present the corresponding extensions of these ideas for the case of form factors.
Our constructions rely on the use of the integrability inspired $\rr$-operator techniques and the association of a permutation to each on-shell graph as it was done for the scattering amplitudes. 
To allow for a pedagogical presentation, we restrict ourselves to the MHV level in this section.  
Next, in section \ref{sec: beyond MHV}, we further extend the techniques of the previous section in order to study similar form factors but at the NMHV and higher N$^k$MHV level. 
We also present some lower-point form factors as examples and conjecture a general Graßmannian integral formulation for tree-level form factors. 
In section \ref{sec: integrability}, we investigate  the role of integrability for tree-level form factors using the spin-chain monodromy matrix. 
Specifically, we show that all form factors of the chiral stress-tensor multiplet are annihilated by the transfer matrix given by its super trace. 
We also study the action of this transfer matrix on the minimal form factors of general operators as well as on on-shell diagrams involving them. 
Finally, in section \ref{sec: conclusion and outlook}, we conclude with a summary of our results and an outlook about future directions.

\paragraph{Note added}

On the day of submission, the paper \cite{Bork:2015fla} appeared, which has some overlap with this article.

\subsection*{Form factors of the stress-tensor super multiplet}

In most of this paper, we focus on the form factors of the chiral part of the stress-tensor super multiplet, which are the most widely studied ones. 
Using $\mathcal{N}=4$ harmonic superspace \cite{Hartwell:1994rp}, this part of the stress-tensor super multiplet can be written as
\begin{equation}
\label{eq: def stress-tensor multiplet}
  T(x,\theta^+)=\tr(\phi^{++}\phi^{++}) + \dots + \frac{1}{3}(\theta^+)^4 \calL \eqncom
\end{equation}
where $\theta^{+ a}_{\alpha}=\theta^{A}_{\alpha} u_A^{+ a}$, $\theta^{- a'}_{\alpha}=\theta^{A}_{\alpha} u_A^{- a'}$ with projectors $u_A^{+ a}$ and $u_A^{- a'}$.
The indices $a$, $a'$ and $\pm$ correspond to $SU(2)\times SU(2)' \times U(1) \subset SU(4)$, see \cite{Eden:2011yp,Brandhuber:2011tv,Bork:2014eqa} for details and further references.
The lowest component in $T(x,\theta^+)$ is the scalar operator $\tr(\phi^{++}\phi^{++})$ with $\phi^{++}=\frac{1}{2}\epsilon_{ab}u_A^{+a}u_B^{+b}\phi^{AB}$, whereas its highest component is the chiral part of the on-shell Lagrangian $\calL$. 

The super form factor of this super multiplet is defined as 
\begin{equation}
 \ff_{n,k}(1,\dots,n;q,\gamma^-)=\int \de^4x\de^4\theta^+ \e^{-iqx-i\theta^{+a}_\alpha \gamma_{a}^{-\alpha}}\bram{1,\dots,n}T(x,\theta^+)\ketm{0}\eqncom
\end{equation}
where $\gamma^{-\alpha a}$ is the supermomentum of the multiplet and $k$ denotes the (supersymmetric extension of the) MHV degree. 
For the minimal MHV degree $k=2$, the form factor of $T(x,\theta^+)$ reads \cite{Brandhuber:2011tv}:%
\begin{equation}
\label{eq: ff n,2 intro}
        \ff_{n,2}(1,\dots,n;q,\gamma^+)
        =\frac{
                \delta^4(P)\delta^4(Q^+)\delta^4(Q^{-})
        }{
          \abr{12}\abr{23}\cdots\abr{n \sminus 1 \ssep n}\abr{n1}
        } \eqncom
\end{equation}
where
\begin{equation}
\label{eq: deformed momenta and supermomenta intro}
        P=\sum_{i=1}^n \vll_i\vlt_i - q
        \eqncom \qquad
        Q^{+}=\sum_{i=1}^n \vll_i\vle_{i}^+
        \eqncom \qquad
        Q^{-}=\sum_{i=1}^n \vll_i\vle_{i}^-  - \gamma^-
\end{equation}
with $Q^{+ a \alpha}=\bar{u}_{A}^{+ a} Q^{A\alpha},Q^{- a' \alpha}=\bar{u}_{A}^{- a'} Q^{A\alpha}$ and $\vle^{+ a}=\bar{u}^{+ a}_A \vle^A,\vle^{- a'}=\bar{u}^{- a'}_A \vle^A$.\footnote{The projectors $\bar{u}$ are related to the $u$'s by conjugation.}

Note that throughout this paper we will be treating colour-ordered tree-level form factors and amplitudes. Hence, we will not indicate this at each expression individually.

\section{The MHV case}
\label{sec: MHV}

In this section, we demonstrate that many of the recent successful techniques that were developed for scattering amplitudes can also be applied to MHV form factors, namely on-shell diagrams, deformations, $\rr$ operators and a (deformed) Graßmannian integral representation.

\subsection{On-shell diagrams, inverse soft limits, BCFW bridges and permutations}
\label{sec: on shell diagrams}

\subsubsection*{On-shell diagrams}

\begin{figure}[htbp]
\begin{subfigure}[t]{0.44\textwidth}
 \begin{equation*}
  \begin{aligned}
   \begin{tikzpicture}[scale=0.8,rotate=0]
                         \draw (1,1) -- (1,2) -- (2,2) -- (2,1) -- (1,1);
                         \draw (0.5,0.5) -- (1,1);
                         \draw (2.5,0.5) -- (2,1);
                         \draw (2.5,2.5) -- (2,2);
                         \draw (0.5,2.5) -- (1,2);
                         \node[dw] at (2,1) {};
                         \node[dw] at (1,2) {};
                         \node[db] at (1,1) {};
                         \node[db] at (2,2) {};
                         \node at (2.5+\labelddist,0.5-\labelddist) {1};
                         \node at (0.5-\labelddist,2.5+\labelddist) {3};
                         \node at (0.5-\labelddist,0.5-\labelddist) {2};
                         \node at (2.5+\labelddist,2.5+\labelddist) {4};
   \end{tikzpicture}
   \end{aligned}
   =
     \begin{aligned}
   \begin{tikzpicture}[scale=0.8,rotate=0]
                         \draw (1,1) -- (1,2) -- (2,2) -- (2,1) -- (1,1);
                         \draw (0.5,0.5) -- (1,1);
                         \draw (2.5,0.5) -- (2,1);
                         \draw (2.5,2.5) -- (2,2);
                         \draw (0.5,2.5) -- (1,2);
                         \node[db] at (2,1) {};
                         \node[db] at (1,2) {};
                         \node[dw] at (1,1) {};
                         \node[dw] at (2,2) {};
                         \node at (2.5+\labelddist,0.5-\labelddist) {1};
                         \node at (0.5-\labelddist,2.5+\labelddist) {3};
                         \node at (0.5-\labelddist,0.5-\labelddist) {2};
                         \node at (2.5+\labelddist,2.5+\labelddist) {4};
   \end{tikzpicture}
   \end{aligned}
 \end{equation*} 
\caption{Square move.}
\label{fig: square move}
\end{subfigure}
\begin{subfigure}[t]{0.54\textwidth}
 \begin{equation*} 
 \begin{aligned}
   \begin{tikzpicture}[scale=0.8,rotate=0]
                         \draw (1,1) -- (1,2);
                         \draw (0.5,0.5) -- (1,1);
                         \draw (1.5,0.5) -- (1,1);
                         \draw (1.5,2.5) -- (1,2);
                         \draw (0.5,2.5) -- (1,2);
                         \node[db] at (1,1) {};
                         \node[db] at (1,2) {};
                         \node at (1.5+\labelddist,0.5-\labelddist) {1};
                         \node at (0.5-\labelddist,2.5+\labelddist) {3};
                         \node at (0.5-\labelddist,0.5-\labelddist) {2};
                         \node at (1.5+\labelddist,2.5+\labelddist) {4};
   \end{tikzpicture}
   \end{aligned}
   =
         \begin{aligned}
   \begin{tikzpicture}[scale=0.8,rotate=90]
                         \draw (0.5,0.5) -- (1,1);
                         \draw (1.5,0.5) -- (1,1);
                         \draw (1.5,1.5) -- (1,1);
                         \draw (0.5,1.5) -- (1,1);
                         \node[db] at (1,1) {};
                         \node at (1.5+\labelddist,0.5-\labelddist) {4};
                         \node at (0.5-\labelddist,1.5+\labelddist) {2};
                         \node at (0.5-\labelddist,0.5-\labelddist) {1};
                         \node at (1.5+\labelddist,1.5+\labelddist) {3};
                         \end{tikzpicture}
   \end{aligned}
   =
         \begin{aligned}
   \begin{tikzpicture}[scale=0.8,rotate=90]
                         \draw (1,1) -- (1,2);
                         \draw (0.5,0.5) -- (1,1);
                         \draw (1.5,0.5) -- (1,1);
                         \draw (1.5,2.5) -- (1,2);
                         \draw (0.5,2.5) -- (1,2);
                         \node[db] at (1,1) {};
                         \node[db] at (1,2) {};
                         \node at (1.5+\labelddist,0.5-\labelddist) {4};
                         \node at (0.5-\labelddist,2.5+\labelddist) {2};
                         \node at (0.5-\labelddist,0.5-\labelddist) {1};
                         \node at (1.5+\labelddist,2.5+\labelddist) {3};
   \end{tikzpicture}
   \end{aligned}
 \end{equation*} 
\caption{Merge/unmerge move for black vertices.}
\label{fig: merge move}
\end{subfigure}
\caption{Moves connecting equivalent on-shell diagrams. Similarly to the case for black vertices, the merge/unmerge move also exists for white vertices.}
\label{fig: amplitude moves}
\end{figure}
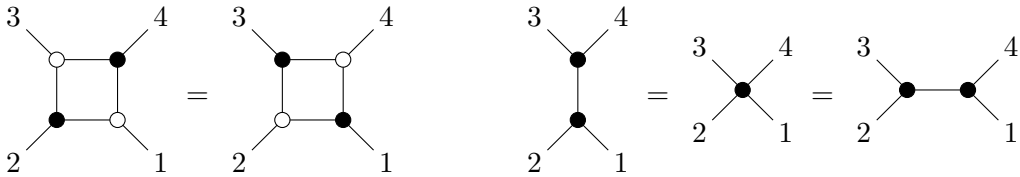

On-shell diagrams have proven to be a useful tool in the construction of scattering amplitudes.
They are built from two different elements, namely the three-point $\MHV$ amplitude $\amp_{3,2}$ and the three-point $\MHVbar$ amplitude $\amp_{3,1}$:
\begin{equation}
\label{eq: amplitude building blocks for on-shell diagrams}
 \begin{aligned}
  \begin{aligned}
        \begin{tikzpicture}[scale=0.8]
        \draw (1,1) -- (1,1+0.65);
        \draw (1,1) -- (1-0.5,1-0.5);
        \draw (1,1) -- (1+0.5,1-0.5);
        \node [] at (1,1+0.65+\labelvdist) {1};
        \node [] at (1-0.5-0.2,1-0.5-0.2) {3};
        \node [] at (1+0.5+0.2,1-0.5-0.2) {2};
        \node[db] at (1,1) {};
        \end{tikzpicture}
        \end{aligned}
        &=
        \amp_{3,2}(1,2,3)=\frac{
                \delta^4(\vll_1\vlt_1+\vll_2\vlt_2+\vll_3\vlt_3)
                \delta^8(\vll_1\vle_1+\vll_2\vle_2+\vll_3\vle_3)
        }{\abr{12}\abr{23}\abr{31}} \eqncom\\
   \begin{aligned}
        \begin{tikzpicture}[scale=0.8]
        \draw (1,1) -- (1,1+0.65);
        \draw (1,1) -- (1-0.5,1-0.5);
        \draw (1,1) -- (1+0.5,1-0.5);
        \node [] at (1,1+0.65+\labelvdist) {1};
        \node [] at (1-0.5-0.2,1-0.5-0.2) {3};
        \node [] at (1+0.5+0.2,1-0.5-0.2) {2};
        \node[dw] at (1,1) {};
        \end{tikzpicture}
        \end{aligned}
        &=
        \amp_{3,1}(1,2,3)
        =\frac{
                \delta^4(\vll_1\vlt_1+\vll_2\vlt_2+\vll_3\vlt_3)
                \delta^4(\sbr{12}\vle_3+\sbr{23}\vle_1+\sbr{31}\vle_2)
        }{\sbr{12}\sbr{23}\sbr{31}} \eqndot
 \end{aligned}
\end{equation}
All scattering amplitudes can be built from BCFW recursion relations \cite{Britto:2004ap,Britto:2005fq}, which can be depicted as \cite{ArkaniHamed:2012nw}
\begin{equation}
 \label{eq: BCFW for amplitudes}
 \amp_{n,k}= \sum_{\substack{n',n'',k',k''\\ n'+n''=n+2\\ k'+k''=k+1}}
 \begin{aligned}
 \begin{tikzpicture}[scale=0.8]
 \draw (0,-0) -- (2,-0); 
 \draw (0,-1.5) -- (2,-1.5); 
 \draw (0,-0) -- (0,-2.25); 
 \draw (2,-0) -- (2,-2.25); 
 \draw (0,-0) -- (-1,-0);
 \node[] at (-1-\labelhdist,-0) {3};
 \draw (0,-0) -- (-0,+1);
 \node[] at (-0,+1+\labelvdist) {$n'$};
 \draw (2,-0) -- (+3,-0);
 \node[] at (+3+\labelhdist,-0) {$n$};
 \draw (2,-0) -- (+2,+1);
 \node[] at (+2,+1+\labelvdist) {$n'+1$};
 \node[] at (-0.7,+0.7) {\rotatebox{45}{$\cdots$}};
 \node[] at (+2.7,+0.7) {\rotatebox{-45}{$\cdots$}};
 \node[dw] at (0,-1.5) {};
 \node[db] at (2,-1.5) {};
 \node[circle, black, fill=grayn, minimum width=4*\onshellradius, draw, inner sep=0pt] at (0,0) {$\scriptstyle \amp_{n',k'}$};
 \node[circle, black, fill=grayn, minimum width=4*\onshellradius, draw, inner sep=0pt] at (2,0) {$\scriptstyle \amp_{n'',k''}$};
 \node[] at (0,-2.25-\labelvdist) {2};
 \node[] at (2,-2.25-\labelvdist) {1};
\end{tikzpicture}
\end{aligned} \eqncom
\end{equation}
where the BCFW bridge attached at positions $1$ and $2$ implements the BCFW shift.%
\footnote{Note that in this work we use the parity flipped version of the BCFW bridge used in \cite{ArkaniHamed:2012nw}, i.e.\ the opposite assignment of the black and white vertices in the BCFW bridge.}
Hence, they can also be encoded in on-shell diagrams.
Similarly to the construction via BCFW recursion relations, the on-shell diagram encoding an amplitude is not unique.
Equivalent on-shell diagrams can be transformed into each other via the so-called square move and merge/unmerge move, which are depicted in figure~\ref{fig: amplitude moves} and can be applied to any subdiagram of a given on-shell diagram.

In order to construct form factors via BCFW recursion relations, the minimal form factor is required as an  additional building block.
Hence, it is also required to extend on-shell graphs to the construction of form factors.
We depict the minimal form factor of $T$ as 
\begin{equation}
\label{eq: form factor building block for on-shell diagrams}
 \begin{aligned}
          \begin{aligned}
        \begin{tikzpicture}[scale=0.8]
\drawminimalff{1} 
        \node [] at (-\labelddist,-\vacuumheight-\labelddist) {2};
        \node [] at (1+\labelddist,-\vacuumheight-\labelddist) {1};
        \end{tikzpicture}
        \end{aligned}
        &=
        \ff_{2,2}(1,2)=\frac{
                \delta^4(\vll_1\vlt_1 + \vll_2\vlt_2 - q)
                \delta^4(\vll_1\vle_1^+ + \vll_2\vle_2^+ )
                \delta^4(\vll_1\vle_1^- + \vll_2\vle_2^- - \gamma^-)
        }{\abr{12}\abr{21}} \eqndot
 \end{aligned}
\end{equation}

We can then use the construction of the form factors of $T$ via BCFW recursion relations \cite{Brandhuber:2010ad,Brandhuber:2011tv}, which we depict as
\begin{equation}
\label{eq: BCFW for form factors}
\ff_{n,k}= \sum_{\substack{n',n'',k',k''\\ n'+n''=n+2\\ k'+k''=k+1}}
 \begin{aligned}
 \begin{tikzpicture}[scale=0.8]
 \draw (0,-0) -- (2,-0); 
 \draw (0,-1.5) -- (2,-1.5); 
 \draw (0,-0) -- (0,-2.25); 
 \draw (2,-0) -- (2,-2.25); 
 \draw (0,-0) -- (-1,-0);
 \node[] at (-1-\labelhdist,-0) {3};
 \draw (0,-0) -- (-0,+1);
 \node[] at (-0,+1+\labelvdist) {$n'$};
 \draw (2,-0) -- (+3,-0);
 \node[] at (+3+\labelhdist,-0) {$n$};
 \draw (2,-0) -- (+2,+1);
 \node[] at (+2,+1+\labelvdist) {$n'+1$};
 \node[] at (-0.7,+0.7) {\rotatebox{45}{$\cdots$}};
 \node[] at (+2.7,+0.7) {\rotatebox{-45}{$\cdots$}};
 \node[dw] at (0,-1.5) {};
 \node[db] at (2,-1.5) {};
 \draw [thick,double] (0,0) -- (-1,-1);
 \node[circle, black, fill=grayn, minimum width=4*\onshellradius, draw, inner sep=0pt] at (0,0) {$\scriptstyle \ff_{n',k'}$};
 \node[circle, black, fill=grayn, minimum width=4*\onshellradius, draw, inner sep=0pt] at (2,0) {$\scriptstyle \amp_{n'',k''}$};
 \node[] at (0,-2.25-\labelvdist) {2};
 \node[] at (2,-2.25-\labelvdist) {1};
\end{tikzpicture}
\end{aligned}
+
\begin{aligned}
 \begin{tikzpicture}[scale=0.8]
 \draw (0,-0) -- (2,-0); 
 \draw (0,-1.5) -- (2,-1.5); 
 \draw (0,-0) -- (0,-2.25); 
 \draw (2,-0) -- (2,-2.25); 
 \draw (0,-0) -- (-1,-0);
 \node[] at (-1-\labelhdist,-0) {3};
 \draw (0,-0) -- (-0,+1);
 \node[] at (-0,+1+\labelvdist) {$n'$};
 \draw (2,-0) -- (+3,-0);
 \node[] at (+3+\labelhdist,-0) {$n$};
 \draw (2,-0) -- (+2,+1);
 \node[] at (+2,+1+\labelvdist) {$n'+1$};
 \node[] at (-0.7,+0.7) {\rotatebox{45}{$\cdots$}};
 \node[] at (+2.7,+0.7) {\rotatebox{-45}{$\cdots$}};
 \node[dw] at (0,-1.5) {};
 \node[db] at (2,-1.5) {};
 \draw [thick,double] (2,0) -- (+3,-1);
 \node[circle, black, fill=grayn, minimum width=4*\onshellradius, draw, inner sep=0pt] at (0,0) {$\scriptstyle \amp_{n',k'}$};
 \node[circle, black, fill=grayn, minimum width=4*\onshellradius, draw, inner sep=0pt] at (2,0) {$\scriptstyle \ff_{n'',k''}$};
 \node[] at (0,-2.25-\labelvdist) {2};
 \node[] at (2,-2.25-\labelvdist) {1};
\end{tikzpicture}
\end{aligned}
\eqndot
\end{equation}

\subsubsection*{Inverse soft limit}

The MHV form factors $\ff_{n,2}$ can also be constructed from the minimal form factors via the so-called inverse soft limit \cite{ArkaniHamed:2009si,ArkaniHamed:2010kv,Bullimore:2010pa} similarly to MHV amplitudes \cite{Nandan:2012rk}. 
In total, two types of inverse soft limits exist, which either preserve the MHV degree or increase it by one unit.
In terms of on-shell diagrams, the $k$-preserving inverse soft limit amounts to recursively adding the structure 
\begin{equation}
\label{eq: k-preserving inverse soft limit}
        \begin{aligned}
        \begin{tikzpicture}[scale=0.8,rotate=180]
        \draw (1,1) -- (0,0.7);
        \draw (1,1) -- (2,0.7);
        \draw (1,1) -- (1,1+0.65);
        \draw (0,0.7) -- (0-0.3,0.7+0.6);
        \draw (2,0.7) -- (2+0.3,0.7+0.6);
        \draw (0,0.7) -- (0+0.3,0.7-0.6);
        \draw (2,0.7) -- (2-0.3,0.7-0.6);
        \node[db] at (1,1) {};
        \node[dw] at (0,0.7) {};
        \node[dw] at (2,0.7) {};
        \end{tikzpicture}
        \end{aligned}
\end{equation}
to two adjacent legs of the diagram.

For the four-point amplitude $\amp_{4,2}$, this construction starts at the three-point amplitude $\amp_{3,2}$ and can be depicted as
\begin{equation}
\label{eq: on-shell diagram amp 4,2 from inverse soft limit}
   \begin{aligned}
        \begin{tikzpicture}[scale=0.8]
        \draw (1,1) -- (1,1+0.65);
        \draw (1,1) -- (1-0.5,1-0.5);
        \draw (1,1) -- (1+0.5,1-0.5);
        \node [] at (1,1+0.65+\labelvdist) {2};
        \node [] at (1-0.5-\labelddist,1-0.5-\labelddist) {1};
        \node [] at (1+0.5+\labelddist,1-0.5-\labelddist) {3};
        \node[db] at (1,1) {};
        \end{tikzpicture}
        \end{aligned}
        \quad
   \xrightarrow{%
   \resizebox{1cm}{!} {%
   \begin{tikzpicture}[scale=0.8,rotate=180]
        \draw (1,1) -- (0,0.7);
        \draw (1,1) -- (2,0.7);
        \draw (1,1) -- (1,1+0.65);
        \draw (0,0.7) -- (0-0.3,0.7+0.6);
        \draw (2,0.7) -- (2+0.3,0.7+0.6);
        \draw (0,0.7) -- (0+0.3,0.7-0.6);
        \draw (2,0.7) -- (2-0.3,0.7-0.6);
        \node[db] at (1,1) {};
        \node[dw] at (0,0.7) {};
        \node[dw] at (2,0.7) {};
        \end{tikzpicture}
        }%
   }
   \quad
   \begin{aligned}
   \begin{tikzpicture}[scale=0.8,rotate=45]
                         \draw (1,1) -- (1,2) -- (2,2) -- (2,1) -- (1,1);
                         \draw (0.5,0.5) -- (1,1);
                         \draw (2.5,0.5) -- (2,1);
                         \draw (2.5,2.5) -- (2,2);
                         \draw (0.5,2.5) -- (1,2);
        \node [] at (2.5+\labelddist,2.5+\labelddist) {2};
        \node [] at (2.5+\labelddist,0.5-\labelddist) {3};
        \node [] at (0.5-\labelddist,0.5-\labelddist) {4};
        \node [] at (0.5-\labelddist,2.5+\labelddist) {1};
                         \node[dw] at (2,1) {};
                         \node[dw] at (1,2) {};
                         \node[db] at (1,1) {};
                         \node[db] at (2,2) {};
   \end{tikzpicture}
   \end{aligned}
   \eqndot
\end{equation}
Similarly, the three-point form factor $\ff_{3,2}$ can be constructed from the minimal form factor $\ff_{2,2}$ as%
\footnote{The result of this construction can easily be seen to agree with the one obtained from the BCFW recursion relation \eqref{eq: BCFW for form factors} with a shift in the legs $3$ and $1$.}
\begin{equation}
\label{eq: on-shell diagram ff 3,2 from inverse soft limit}
        \begin{aligned}
        \begin{tikzpicture}[scale=0.8]
\drawminimalff{1} 
        \node [] at (-\labelddist,-\vacuumheight-\labelddist) {1};
        \node [] at (1+\labelddist,-\vacuumheight-\labelddist) {2};
        \end{tikzpicture}
        \end{aligned}
        \quad
   \xrightarrow{%
   \resizebox{1cm}{!} {%
   \begin{tikzpicture}[scale=0.8,rotate=180]
        \draw (1,1) -- (0,0.7);
        \draw (1,1) -- (2,0.7);
        \draw (1,1) -- (1,1+0.65);
        \draw (0,0.7) -- (0-0.3,0.7+0.6);
        \draw (2,0.7) -- (2+0.3,0.7+0.6);
        \draw (0,0.7) -- (0+0.3,0.7-0.6);
        \draw (2,0.7) -- (2-0.3,0.7-0.6);
        \node[db] at (1,1) {};
        \node[dw] at (0,0.7) {};
        \node[dw] at (2,0.7) {};
        \end{tikzpicture}
        }%
   }
   \quad
\begin{aligned}
 \begin{tikzpicture}[scale=0.8]
 			\draw[thick,double] (0,-0) -- (0,0.5); 
  			\draw (0,0) -- (-\hdist,-\hdist) -- (0,-2*\hdist) -- (+\hdist,-\hdist) -- (0,0); 
 			\draw (+\hdist,-\hdist) -- (+2*\hdist,-2*\hdist);   
 			\draw (-\hdist,-\hdist) -- (-2*\hdist,-2*\hdist);   
 			\draw (0,-2*\hdist) -- (0,-2*\hdist-\ddist);  
                         \node[dw] at (+\hdist,-\hdist) {}; 
                         \node[db] at (0,-2*\hdist) {};
                         \node[dw] at (-\hdist,-\hdist) {}; 
                         \node at (+2*\hdist  +\labelddist,-2*\hdist-\labelddist) {2};
                         \node at (0,-2*\hdist-\ddist-\labelvdist) {3};
                         \node at (-2*\hdist -\labelddist,-2*\hdist-\labelddist) {1};
         \end{tikzpicture}
\end{aligned}
   \eqndot
\end{equation}

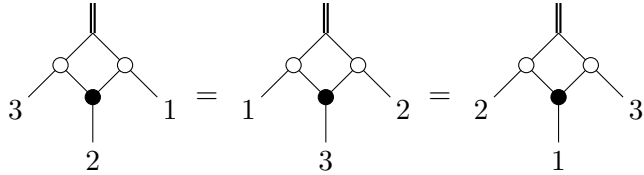
\begin{figure}[htbp]
\begin{equation*}
\begin{aligned}
 \begin{tikzpicture}[scale=0.8]
 			\draw[thick,double] (0,-0) -- (0,0.5); 
  			\draw (0,0) -- (-\hdist,-\hdist) -- (0,-2*\hdist) -- (+\hdist,-\hdist) -- (0,0); 
 			\draw (+\hdist,-\hdist) -- (+2*\hdist,-2*\hdist);   
 			\draw (-\hdist,-\hdist) -- (-2*\hdist,-2*\hdist);   
 			\draw (0,-2*\hdist) -- (0,-2*\hdist-\ddist);  
                         \node[dw] at (+\hdist,-\hdist) {}; 
                         \node[db] at (0,-2*\hdist) {};
                         \node[dw] at (-\hdist,-\hdist) {}; 
                         \node at (+2*\hdist  +\labelddist,-2*\hdist-\labelddist) {1};
                         \node at (0,-2*\hdist-\ddist-\labelvdist) {2};
                         \node at (-2*\hdist -\labelddist,-2*\hdist-\labelddist) {3};
         \end{tikzpicture}
\end{aligned}
=
\begin{aligned}
 \begin{tikzpicture}[scale=0.8]
 			\draw[thick,double] (0,-0) -- (0,0.5); 
  			\draw (0,0) -- (-\hdist,-\hdist) -- (0,-2*\hdist) -- (+\hdist,-\hdist) -- (0,0); 
 			\draw (+\hdist,-\hdist) -- (+2*\hdist,-2*\hdist);   
 			\draw (-\hdist,-\hdist) -- (-2*\hdist,-2*\hdist);   
 			\draw (0,-2*\hdist) -- (0,-2*\hdist-\ddist);  
                         \node[dw] at (+\hdist,-\hdist) {}; 
                         \node[db] at (0,-2*\hdist) {};
                         \node[dw] at (-\hdist,-\hdist) {}; 
                         \node at (+2*\hdist  +\labelddist,-2*\hdist-\labelddist) {2};
                         \node at (0,-2*\hdist-\ddist-\labelvdist) {3};
                         \node at (-2*\hdist -\labelddist,-2*\hdist-\labelddist) {1};
         \end{tikzpicture}
\end{aligned}
=
\begin{aligned}
 \begin{tikzpicture}[scale=0.8]
 			\draw[thick,double] (0,-0) -- (0,0.5); 
  			\draw (0,0) -- (-\hdist,-\hdist) -- (0,-2*\hdist) -- (+\hdist,-\hdist) -- (0,0); 
 			\draw (+\hdist,-\hdist) -- (+2*\hdist,-2*\hdist);   
 			\draw (-\hdist,-\hdist) -- (-2*\hdist,-2*\hdist);   
 			\draw (0,-2*\hdist) -- (0,-2*\hdist-\ddist);  
                         \node[dw] at (+\hdist,-\hdist) {}; 
                         \node[db] at (0,-2*\hdist) {};
                         \node[dw] at (-\hdist,-\hdist) {}; 
                         \node at (+2*\hdist  +\labelddist,-2*\hdist-\labelddist) {3};
                         \node at (0,-2*\hdist-\ddist-\labelvdist) {1};
                         \node at (-2*\hdist -\labelddist,-2*\hdist-\labelddist) {2};
         \end{tikzpicture}
\end{aligned}
\end{equation*}
 \caption{Rotation move for on-shell diagrams involving the minimal form factor. An analogous move exists for the inverse combination of black and white vertices.}
\label{fig: move for minimal ff}
\end{figure}

Note that the diagram in \eqref{eq: on-shell diagram ff 3,2 from inverse soft limit} is not invariant under a cyclical relabelling of the on-shell legs $1$, $2$ and $3$,
although the expression it is encoding is. Hence, we have to add an equivalence move for on-shell graphs that involve the minimal form factor, which can also be applied to any subgraph of a given on-shell graph.
We call it rotation move and depict it in figure~\ref{fig: move for minimal ff}.
Together with the equivalence moves for amplitudes shown in figure~\ref{fig: amplitude moves}, the rotation move guarantees the cyclic invariance of all on-shell diagrams of MHV form factors.
This is similar to the situation for scattering amplitudes. By itself, also the on-shell diagram \eqref{eq: on-shell diagram amp 4,2 from inverse soft limit} for the four-point MHV amplitude is not cyclically invariant. Its cyclic invariance has to be imposed in the form of the square move. However, this suffices to guarantee the cyclic invariance of all other MHV amplitudes when combined with the merge/unmerge move.

Finally, note that we can also construct $\NmaxMHV$ amplitudes and form factors via the inverse soft limit by adding the $k$-increasing structure
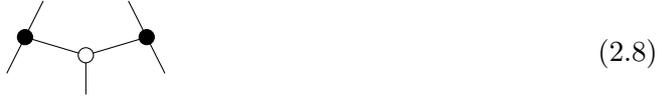
\begin{equation}
\label{eq: k-increasing inverse soft limit}
        \begin{aligned}
        \begin{tikzpicture}[scale=0.8,rotate=180]
        \draw (1,1) -- (0,0.7);
        \draw (1,1) -- (2,0.7);
        \draw (1,1) -- (1,1+0.65);
        \draw (0,0.7) -- (0-0.3,0.7+0.6);
        \draw (2,0.7) -- (2+0.3,0.7+0.6);
        \draw (0,0.7) -- (0+0.3,0.7-0.6);
        \draw (2,0.7) -- (2-0.3,0.7-0.6);
        \node[dw] at (1,1) {};
        \node[db] at (0,0.7) {};
        \node[db] at (2,0.7) {};
        \end{tikzpicture}
        \end{aligned}
\end{equation}
to two adjacent legs of the diagram. 
The resulting on-shell diagrams are related to those of MHV type by exchanging the black and white vertices.%
\footnote{In fact, all N$^k$MHV scattering amplitudes and form factors of $T$ can be constructed via the inverse soft limit using both \eqref{eq: k-preserving inverse soft limit} and \eqref{eq: k-increasing inverse soft limit} \cite{Nandan:2012rk}. In the case of non-extremal $k$, however, the position and order of adding these structures becomes important. }

\subsubsection*{Permutations}

For amplitudes, it is possible to associate a permutation 
\begin{equation}
 \sigma=\begin{pmatrix}
         1&2&3&\dots&n\\
         \downarrow&\downarrow&\downarrow&\dots &\downarrow\\
         \sigma(1)&\sigma(2)&\sigma(3)&\dots&\sigma(n)
         \end{pmatrix}
         \equiv (\sigma(1),\sigma(2),\sigma(3),\dots,\sigma(n))
\end{equation}
to every on-shell graph by starting at some external particle $i$ and turning right at every black vertex and left at every white vertex \cite{ArkaniHamed:2012nw}:
\begin{equation}
  \begin{aligned}
        \begin{tikzpicture}[scale=0.8]
        \draw (1,1) -- (1,1+0.65);
        \draw (1,1) -- (1-0.5,1-0.5);
        \draw (1,1) -- (1+0.5,1-0.5);
        \node[db] at (1,1) {};
        \node [] at (1,1+0.65+\labelvdist) {1};
        \node [] at (1-0.5-0.2,1-0.5-0.2) {3};
        \node [] at (1+0.5+0.2,1-0.5-0.2) {2};
        \draw[blue,->] (1-0.15,1+0.65) to[out=270,in=45] (1-0.5-0.1,1-0.5+0.1);
        \draw[blue,<-] (1+0.15,1+0.65) to[out=270,in=135] (1+0.5+0.1,1-0.5+0.1); 
        \draw[blue,<-] (1+0.5-0.1,1-0.5-0.1) to[in=45,out=135] (1-0.5+0.1,1-0.5-0.1);
        \end{tikzpicture}
  \end{aligned}
  \rightarrow \sigma=(3,1,2)\eqncom
  \qquad
  \begin{aligned}
        \begin{tikzpicture}[scale=0.8]
        \draw (1,1) -- (1,1+0.65);
        \draw (1,1) -- (1-0.5,1-0.5);
        \draw (1,1) -- (1+0.5,1-0.5);
        \node[dw] at (1,1) {};
        \node [] at (1,1+0.65+\labelvdist) {1};
        \node [] at (1-0.5-0.2,1-0.5-0.2) {3};
        \node [] at (1+0.5+0.2,1-0.5-0.2) {2};
        \draw[blue,<-] (1-0.15,1+0.65) to[out=270,in=45] (1-0.5-0.1,1-0.5+0.1);
        \draw[blue,->] (1+0.15,1+0.65) to[out=270,in=135] (1+0.5+0.1,1-0.5+0.1); 
        \draw[blue,->] (1+0.5-0.1,1-0.5-0.1) to[in=45,out=135] (1-0.5+0.1,1-0.5-0.1);
        \end{tikzpicture}
  \end{aligned}
  \rightarrow \sigma=(2,3,1)\eqndot
\end{equation}
If the path ends on particle $j$, set $\sigma(i)=j$.%
\footnote{In contrast to \cite{ArkaniHamed:2012nw}, we are using ordinary permutations instead of decorated permutations here; see the discussion below.}
We can extend this to form factor on-shell graphs by the prescription to turn back at every minimal form factor:
\begin{equation}
 \begin{aligned}
        \begin{tikzpicture}[scale=0.8]
\drawminimalff{1} 
        \node [] at (-\labelddist,-\vacuumheight-\labelddist) {2};
        \node [] at (1+\labelddist,-\vacuumheight-\labelddist) {1};
        \draw[blue] (0-0.1,-\vacuumheight+0.1) to[in=135,out=45] (0.25,-0.75\vacuumheight);
        \draw[blue,<-] (0+0.1,-\vacuumheight-0.1) to[in=-45,out=45] (0.25,-0.75\vacuumheight);
        \draw[blue] (1-0.1,-\vacuumheight-0.1) to[in=-135,out=135] (0.75,-0.75\vacuumheight);
        \draw[blue,<-] (1+0.1,-\vacuumheight+0.1) to[in=45,out=135] (0.75,-0.75\vacuumheight);
        \end{tikzpicture}
        \end{aligned}
        \rightarrow \sigma=(1,2)
        \eqndot
\end{equation}

As both MHV amplitudes and MHV form factors can be constructed by $k$-preserving inverse soft limits, we find that their on-shell graphs encode the same permutation, namely 
\begin{equation}
\label{eq: MHV permutation}
 \MHV: \qquad \sigma=(3,\dots,n,1,2) \eqndot
\end{equation}

\subsubsection*{Construction via BCFW bridges and permutations}

Using the permutation, a corresponding on-shell graph for general tree-level amplitudes can be constructed in a systematic way as follows \cite{ArkaniHamed:2012nw}.%
\footnote{We use the graphical notation of \cite{Kanning:2014maa}.}
\newcommand{\permprod}{\triangleleft}
First, the permutation is decomposed into a chain of transpositions of minimal length, where the multiplication of permutations corresponds to the right action, i.e.\ $\sigma_1 \permprod \sigma_2 = (\sigma_2(\sigma_1(1)),\dots,\sigma_2(\sigma_1(n)))$.
In this paper, we use only adjacent transpositions.
Second, each transposition $(i,j)$ is interpreted as a BCFW bridge 
\begin{equation}\label{eq: BCFW bridge}
 \begin{aligned}
 \begin{tikzpicture}[scale=0.8]
\drawvline{1}{2}
\drawvline{2}{2}
\drawbridge{1}{2}
\node[dl] at (0,-\vacuumheight-2*\bridgedistance-\labelvdist) {j};
\node[dl] at (1,-\vacuumheight-2*\bridgedistance-\labelvdist) {i};
\end{tikzpicture}
\end{aligned}
\end{equation}
connecting the legs $i$ and $j$.
Third, these BCFW bridges are applied to an empty diagram composed of $n$ lines that start in corresponding vacua,\footnote{We will give further meaning to these vacua below; they are the same as the ``lollipop'' diagrams of \cite{ArkaniHamed:2012nw}. 
Note that the kind of the different vacua, i.e.\ $+$ or $-$, is imposed by hand following the prescription of \cite{Kanning:2014maa} and not given by the permutations, which are not decorated in this work.
}
but in the inverse order compared to the multiplication in the chain of transpositions; this means
that the rightmost transposition corresponds to the BCFW bridge that is applied first to the vacua.%
\footnote{The inverse order follows from our choice of BCFW bridge.}
Fourth, the vacua, the edges starting at the vacua and every vertex that is connected to less then three edges is removed to obtain an on-shell diagram.
This construction is illustrated for $\amp_{3,2}$ in figure \ref{fig: A3,2}.

\begin{figure}[htbp]
 \begin{equation*}
 (3,1,2)=(2,3)\permprod(1,2)\longrightarrow
\begin{aligned}
 \begin{tikzpicture}[scale=0.8]
\drawvline{1}{2}
\drawvline{2}{2}
\drawvline{3}{2}
\drawvacm{1} 
\drawvacm{2} 
\drawvacp{3}
\drawbridge{2}{1}
\drawbridge{1}{2}
\node[dl] at (0,-\vacuumheight-2*\bridgedistance-\labelvdist) {3};
\node[dl] at (1,-\vacuumheight-2*\bridgedistance-\labelvdist) {2};
\node[dl] at (2,-\vacuumheight-2*\bridgedistance-\labelvdist) {1};
\end{tikzpicture}
\end{aligned}
\quad\longrightarrow\quad 
        \begin{aligned}
        \begin{tikzpicture}[scale=0.8]
        \draw (1,1) -- (1,1+0.65);
        \draw (1,1) -- (1-0.5,1-0.5);
        \draw (1,1) -- (1+0.5,1-0.5);
        \node[db] at (1,1) {};
        \node [] at (1,1+0.65+\labelvdist) {1};
        \node [] at (1-0.5-\labelddist,1-0.5-\labelddist) {3};
        \node [] at (1+0.5+\labelddist,1-0.5-\labelddist) {2};
        \end{tikzpicture}
        \end{aligned}
\end{equation*}
\caption{Constructing the on-shell diagram of $\amp_{3,2}$ via permutations and BCFW bridges.
}
\label{fig: A3,2}
\end{figure}

The on-shell diagrams of the MHV form factors can be systematically constructed via BCFW bridges using essentially the same construction as in the amplitude case.
The only difference to the amplitude case, where only one-site amplitude vacua appear, is that the minimal form factor occurs as a vacuum at positions $n$ and $n-1$.
This construction is illustrated in figures \ref{fig: F3,2}, \ref{fig: F4,2} and \ref{fig: F5,2}.

\begin{figure}[htbp]
\begin{equation*}
 (3,1,2)=(2,3)\permprod(1,2)  \longrightarrow 
\begin{aligned}
 \begin{tikzpicture}[scale=0.8]
\drawvline{1}{2}
\drawvline{2}{2}
\drawvline{3}{2}
\drawminimalff{1} 
\drawvacp{3}
\drawbridge{2}{1}
\drawbridge{1}{2}
\node[dl] at (0,-\vacuumheight-2*\bridgedistance-\labelvdist) {3};
\node[dl] at (1,-\vacuumheight-2*\bridgedistance-\labelvdist) {2};
\node[dl] at (2,-\vacuumheight-2*\bridgedistance-\labelvdist) {1};
\end{tikzpicture}
\end{aligned}
\quad\longrightarrow\quad 
\begin{aligned}
 \begin{tikzpicture}[scale=0.8]
 			\draw[thick,double] (0,-0) -- (0,0.5); 
  			\draw (0,0) -- (-\hdist,-\hdist) -- (0,-2*\hdist) -- (+\hdist,-\hdist) -- (0,0); 
 			\draw (+\hdist,-\hdist) -- (+2*\hdist,-2*\hdist);   
 			\draw (-\hdist,-\hdist) -- (-2*\hdist,-2*\hdist);   
 			\draw (0,-2*\hdist) -- (0,-2*\hdist-\ddist);  
                         \node[dw] at (+\hdist,-\hdist) {}; 
                         \node[db] at (0,-2*\hdist) {};
                         \node[dw] at (-\hdist,-\hdist) {}; 
                         \node at (+2*\hdist  +\labelddist,-2*\hdist-\labelddist) {1};
                         \node at (0,-2*\hdist-\ddist-\labelvdist) {2};
                         \node at (-2*\hdist -\labelddist,-2*\hdist-\labelddist) {3};
         \end{tikzpicture}
\end{aligned}
\end{equation*}
\caption{Constructing the on-shell diagram of $\ff_{3,2}$ via permutations and BCFW bridges.
}
\label{fig: F3,2}
\end{figure}

\begin{figure}[htbp]
\begin{equation*}
 (3,4,1,2)=(2,3)\permprod(3,4)\permprod(1,2)\permprod(2,3) \longrightarrow
\begin{aligned}
 \begin{tikzpicture}[scale=0.8]
\drawvline{1}{3}
\drawvline{2}{3}
\drawvline{3}{3}
\drawvline{4}{3}
\drawminimalff{1} 
\drawvacp{3}
\drawvacp{4}
\drawbridge{2}{1}
\drawbridge{1}{2}
\drawbridge{3}{2}
\drawbridge{2}{3}
\node[dl] at (0,-\vacuumheight-3*\bridgedistance-\labelvdist) {4};
\node[dl] at (1,-\vacuumheight-3*\bridgedistance-\labelvdist) {3};
\node[dl] at (2,-\vacuumheight-3*\bridgedistance-\labelvdist) {2};
\node[dl] at (3,-\vacuumheight-3*\bridgedistance-\labelvdist) {1};
\end{tikzpicture}
\end{aligned}
\quad\longrightarrow\quad 
\begin{aligned}
 \begin{tikzpicture}[scale=0.8]
 			\draw[thick,double] (0,-0) -- (0,0.5); 
  			\draw (0,0) -- (-\hdist,-\hdist) -- (0,-2*\hdist) -- (+\hdist,-\hdist) -- (0,0); 
  			\draw (0,-2*\hdist) -- (0,-2*\hdist-\ddist) -- (\ddist,-2*\hdist-\ddist) -- (\hdist+\ddist,-\hdist-\ddist) -- (+\hdist,-\hdist); 
 			\draw (+\hdist+\ddist,-\hdist-\ddist) -- (+2*\hdist+\ddist,-2*\hdist-\ddist);   
 			\draw (-\hdist,-\hdist) -- (-2*\hdist,-2*\hdist);   
 			\draw (0,-2*\hdist-\ddist) -- (0,-2*\hdist-2*\ddist); 
 			\draw (\ddist,-2*\hdist-\ddist) -- (+\hdist+\ddist,-3*\hdist-\ddist);
                         \node[dw] at (+\hdist,-\hdist) {}; 
                         \node[db] at (0,-2*\hdist) {};
                         \node[dw] at (-\hdist,-\hdist) {}; 
                         \node[dw] at (0,-2*\hdist-\ddist) {}; 
                         \node[db] at (\ddist,-2*\hdist-\ddist) {};
                         \node[dw] at (\hdist+\ddist,-\hdist-\ddist) {}; 
                         \node at (+2*\hdist +\ddist +\labelddist,-2*\hdist-\ddist-\labelddist) {1};
                         \node at (-2*\hdist-\labelddist,-2*\hdist-\labelddist) {4};
                         \node at (0,-2*\hdist-2*\ddist-\labelvdist) {3};
                         \node at (\hdist +\ddist +\labelddist,-3*\hdist-\ddist-\labelddist) {2};
         \end{tikzpicture}
\end{aligned}
\end{equation*}
\caption{Constructing the on-shell diagram of $\ff_{4,2}$ via permutations and BCFW bridges.
}
\label{fig: F4,2}
\end{figure}

\begin{figure}[htbp]
\begin{equation*}
\begin{aligned}
(3,4,5,1,2)=(2,3)\permprod(3,4)\permprod(4,5)\permprod(1,2)\permprod(2,3)\permprod(3,4)&\longrightarrow \\
 \begin{aligned}
 \begin{tikzpicture}[scale=0.8]
\drawvline{1}{4}
\drawvline{2}{4}
\drawvline{3}{4}
\drawvline{4}{4}
\drawvline{5}{4}
\drawminimalff{1} 
\drawvacp{3}
\drawvacp{4}
\drawvacp{5}
\drawbridge{2}{1}
\drawbridge{3}{2}
\drawbridge{4}{3}
\drawbridge{1}{2}
\drawbridge{2}{3}
\drawbridge{3}{4}
\node[dl] at (0,-\vacuumheight-4*\bridgedistance-\labelvdist) {5};
\node[dl] at (1,-\vacuumheight-4*\bridgedistance-\labelvdist) {4};
\node[dl] at (2,-\vacuumheight-4*\bridgedistance-\labelvdist) {3};
\node[dl] at (3,-\vacuumheight-4*\bridgedistance-\labelvdist) {2};
\node[dl] at (4,-\vacuumheight-4*\bridgedistance-\labelvdist) {1};
\end{tikzpicture}
\end{aligned}
\qquad&\longrightarrow\qquad 
\begin{aligned}
 \begin{tikzpicture}[scale=0.8]
 			\draw[thick,double] (0,-0) -- (0,0.5); 
  			\draw (0,0) -- (-\hdist,-\hdist) -- (0,-2*\hdist) -- (+\hdist,-\hdist) -- (0,0); 
  			\draw (0,-2*\hdist) -- (0,-2*\hdist-\ddist) -- (\ddist,-2*\hdist-\ddist) -- (\hdist+\ddist,-\hdist-\ddist) -- (+\hdist,-\hdist); 
  			\draw (0+\ddist,-2*\hdist-\ddist) -- (0+\ddist,-2*\hdist-\ddist-\ddist) -- (\ddist+\ddist,-2*\hdist-\ddist-\ddist) -- (\hdist+\ddist+\ddist,-\hdist-\ddist-\ddist) -- (+\hdist+\ddist,-\hdist-\ddist); 
 			\draw (+\hdist+2*\ddist,-\hdist-2*\ddist) -- (+2*\hdist+2*\ddist,-2*\hdist-2*\ddist);   
 			\draw (-\hdist,-\hdist) -- (-2*\hdist,-2*\hdist);   
 			\draw (0,-2*\hdist-\ddist) -- (0,-2*\hdist-2*\ddist); 
 			\draw (\ddist,-2*\hdist-2*\ddist) -- (\ddist,-2*\hdist-3*\ddist); 
 			\draw (2*\ddist,-2*\hdist-2*\ddist) -- (+\hdist+2*\ddist,-3*\hdist-2*\ddist);
                         \node[dw] at (+\hdist,-\hdist) {}; 
                         \node[db] at (0,-2*\hdist) {};
                         \node[dw] at (-\hdist,-\hdist) {}; 
                         \node[dw] at (0,-2*\hdist-\ddist) {}; 
                         \node[db] at (\ddist,-2*\hdist-\ddist) {};
                         \node[dw] at (\hdist+\ddist,-\hdist-\ddist) {}; 
                         \node[dw] at (\ddist,-2*\hdist-2*\ddist) {}; 
                         \node[db] at (2*\ddist,-2*\hdist-2*\ddist) {};
                         \node[dw] at (\hdist+2*\ddist,-\hdist-2*\ddist) {}; 
                         \node at (+2*\hdist +2*\ddist +\labelddist,-2*\hdist-2*\ddist-\labelddist) {1};
                         \node at (-2*\hdist-\labelddist,-2*\hdist-\labelddist) {5};
                         \node at (0,-2*\hdist-2*\ddist-\labelvdist) {4};
                         \node at (\ddist ,-2*\hdist-3*\ddist-\labelvdist) {3};
                         \node at (\hdist +2*\ddist +\labelddist,-3*\hdist-2*\ddist-\labelddist) {2};
         \end{tikzpicture}
\end{aligned}
\end{aligned}
\end{equation*}
\caption{Constructing the on-shell diagram of $\ff_{5,2}$ via permutations and BCFW bridges.
}
\label{fig: F5,2}
\end{figure}

\subsection{Deformed form factors and \texorpdfstring{$\rr$}{\rrpdf} operators}
\label{subsec: deformed form factors}

We can now introduce deformations of the form factors and construct these deformed form factors in analogy to the amplitude case \cite{Ferro:2012xw,Ferro:2013dga,Chicherin:2013ora,Kanning:2014maa,Broedel:2014pia,Bargheer:2014mxa,Ferro:2014gca}.
For amplitudes, a sequence of BCFW bridges can be translated into a chain of $\rr$ operators that acting on a suitable vacuum state produce 
a deformed version of the amplitude, or rather some BCFW term of it.
In this section, we will use the $\rr$ operators primarily as means to obtain analytic expressions
for the form factors, in particular representations in terms of Graßmannian integrals.
However, the $\rr$-operator formalism is based on the spin-chain picture of integrability
and we will use this fact in section \ref{sec: integrability} to show that form factors
are well defined states in the integrable model and posses enhanced symmetry properties. 
There, we will also give further details concerning the definition and the properties
of the $\rr$ operators and discuss the integrability-preserving deformations.

The $\rr$ operators \cite{Chicherin:2013ora} can be defined
by their action on general functions $f$ of the kinematic data,\footnote{Note that we extend the usual definition to harmonic superspace.}
\begin{equation}
        \rr_{ij}(u)
        f(\vll_i,\vlt_i,\vle_i,\vll_j,\vlt_j,\vle_j)
        =\int\frac{\dd\alpha}{\alpha^{1+u}}
        f(\vll_i-\alpha\vll_j,\vlt_i,\vle_i,\vll_j,\vlt_j+\alpha\vlt_i,\vle_j+\alpha\vle_j)
        \eqndot
        \label{eq: action r operator}
\end{equation}
Here, the parameter $u$ will eventually correspond to a (integrability-preserving) deformation of the 
physical form factor.
Moreover, the vacua that occurred in the previous discussion are given by 
\begin{equation}
  \begin{aligned}\begin{tikzpicture}[scale=0.8]
\drawvacp{1}
\node[dl] at (0,-\vacuumheight-0*\bridgedistance-\labelvdist) {$i$};
    \end{tikzpicture}\end{aligned} 
  =
      \deltap{i}=\delta^{2}(\vll_i)\eqncom\qquad
  \begin{aligned}\begin{tikzpicture}[scale=0.8]
\drawvacm{1}
\node[dl] at (0,-\vacuumheight-0*\bridgedistance-\labelvdist) {$i$};
    \end{tikzpicture}\end{aligned} 
  =
        \deltam{i}=\delta^{2}(\vlt_i)\delta^{4}(\vle_i)
  \eqndot
  \label{eq: vacua}
\end{equation}
There are two types of them, reflecting the different possible MHV degrees of the final expression. 

Let us consider the three-particle MHV amplitude $\amp_{3,2}$ in figure~\ref{fig: A3,2} as example. The sequence of transpositions $(2,3)\permprod(1,2)$ translates into
\begin{equation}
\label{eq: A 3,2 from r operators}
  \rr_{23}(u_{32})\rr_{12}(u_{31})\deltap{1}\deltam{2}\deltam{3}
  =
        \frac{
                \delta^4(\sum_{i=1}^3\vll_i\vlt_i)
                \delta^4(\sum_{i=1}^3\vll_i\vle_i^+)
                \delta^4(\sum_{i=1}^3\vll_i\vle_i^-)
              }{\abr{12}^{1-u_{23}}\abr{23}^{1-u_{31}}\abr{31}^{1-u_{12}}}
              \eqncom
\end{equation}
where $u_i$ are parameters associated to deformations of the local central charges (see section \ref{sec: integrability}) and 
\begin{equation}
 \label{eq: def uij}
 u_{ij}=u_i-u_j \eqndot
\end{equation}
The undeformed three-particle MHV amplitude $\amp_{3,2}$ 
is recovered in the limit $u_i\to 0$.

The previous discussion suggests that we can use essentially the same construction 
for the three-particle MHV form factor; the only diagrammatic
difference is the substitution of the minimal form factor
for the vacua at sites 2 and 3, cf.\ figure \ref{fig: F3,2}.
Using the minimal form factor \eqref{eq: form factor building block for on-shell diagrams},
which we label by the two sites it occupies, we find 
that the same chain of $\rr$ operators produces a deformed version of the three-point MHV form factor:\footnote{Here and in what follows, we will ignore phases that also appear in the amplitude case.}
\begin{equation}
\label{eq: F 3,2 from r operators}
  \rr_{23}(u_{32})\rr_{12}(u_{31})\deltap{1}\ff_{2,2}(2,3)
  =
        \frac{
                \delta^4(\sum_{i=1}^3\vll_i\vlt_i-q)
                \delta^4(\sum_{i=1}^3\vll_i\vle_i^+)
                \delta^4(\sum_{i=1}^3\vll_i\vle_i^- - \gamma^-)
              }{\abr{12}^{1-u_{23}}\abr{23}^{1-u_{31}}\abr{31}^{1-u_{12}}} \eqndot
\end{equation}
In the limit of vanishing deformation parameters, this reduces to \eqref{eq: ff n,2 intro} with $n=3$.

Since all $n$-point MHV form factors can be obtained by iterated inverse soft
limits, this construction generalises to all $n$, in particular to the further examples shown in figures \ref{fig: F4,2} and \ref{fig: F5,2}.
The result is 
\begin{equation} 
\label{eq: deformed ff n,2}
\ff_{n,2}(1,\dots,n)=\frac{
  \delta^4(\sum_{i=1}^n\vll_i\vlt_i-q)
  \delta^4(\sum_{i=1}^n\vll_i\vle_i^+)
  \delta^4(\sum_{i=1}^n\vll_i\vle_i^- - \gamma^-)
}{
\prod_{i=1}^n \abr{i \ssep i \splus 1}^{1-u_{i+1 \sssep i+2}}
}\eqncom
\end{equation}
which reduces to \eqref{eq: ff n,2 intro} in the limit of vanishing deformation parameters.

Instead of performing the construction of \eqref{eq: deformed ff n,2} via $\rr$ operators explicitly, we will
now use the $\rr$ operators to obtain a Graßmannian integral representation
that is valid for all $n$ and evaluates to \eqref{eq: deformed ff n,2}.

\subsection{A (deformed) Graßmannian integral representation for the MHV form factor}
\label{subsec: MHV Grassmannian}

The minimal form factor \eqref{eq: form factor building block for on-shell diagrams} can be rewritten in a  form that closely resembles the vacua \eqref{eq: vacua} used in the construction via $\rr$ operators:
\begin{equation}
 \label{eq: minimal form factor as vacuum}
        \ff_{2,2}(1,2)
        =
        \delta^2(\vltu_1)\delta^4(\vleu_1)
        \delta^2(\vltu_2)\delta^4(\vleu_2)
        \equiv \deltaf{12} \eqndot
\end{equation}
Here, we have absorbed the off-shell (super) momentum of the operator into
modified kinematic variables for the on-shell states,
\begin{equation}
  \begin{aligned}
        \vltu_1 &= \vlt_1-\frac{\bra{2}q}{\abr{21}} \eqncom \quad &
        \vleu_1^- &= \vle_1^--\frac{\bra{2}\gamma^-}{\abr{21}} \eqncom \quad &
        \vleu_1^+ & = \vle_1^+ \eqncom
        \\
        \vltu_2 &= \vlt_2-\frac{\bra{1}q}{\abr{12}} \eqncom \quad &
        \vleu_2^- &= \vle_2^--\frac{\bra{1}\gamma^-}{\abr{12}} \eqncom \quad &
        \vleu_2^+ &= \vle_2^+
        \eqndot
  \end{aligned}
  \label{eq: minimal ff as vacuum}
\end{equation}
Note that this expression looks exactly like $\deltam{1}\deltam{2}$, though with twisted kinematics $\vltu$ and $\vleu$ that contain the information about the off-shell (super) momentum insertion.

Using this form of the minimal form factor, we can apply the 
same sequence of $\rr$ operators as in \eqref{eq: F 3,2 from r operators} to obtain, before integration,
\begin{equation}
\label{eq: F 3,2 pre-grassmannian}
\begin{aligned}
  \ff_{3,2}(1,2,3)
  &=\rr_{23}(u_{32})\rr_{12}(u_{31})\deltap{1}\deltaf{23}\\
  &= 
  \int \frac{\dd \alpha_2}{\alpha_2^{1+u_{32}}}
  \int \frac{\dd \alpha_1}{\alpha_1^{1+u_{31}}}
  \;
  \delta^4(C(\alpha_1,\alpha_2)\cdot\vltu)\,\delta^8(C(\alpha_1,\alpha_2)\cdot\vleu)\,\delta^{2}(C^\perp(\alpha_1,\alpha_2)\cdot\vll)
  \eqncom
  \end{aligned}
\end{equation}
where the super spinor helicity variables 2 and 3 are twisted as in \eqref{eq: minimal form factor as vacuum} while 1 is untwisted.
The matrices $C$ and $C^\perp$ are orthogonal to each other, i.e.\ $C(C^\perp)^T=0$,
and 
given by
\begin{equation}
  C(\alpha_1,\alpha_2)=\begin{pmatrix}
    \alpha_1&1&0\\
    0&\alpha_2&1
  \end{pmatrix}
  \eqncom\qquad
  C^\perp(\alpha_1,\alpha_2)=\begin{pmatrix} 1&\,-\alpha_1\,&\,\alpha_1\alpha_2\end{pmatrix}
  \eqndot
\end{equation}
Their products with the external super spinor helicity variables are defined as
\begin{equation}
\label{eq: def products of C}
(C\cdot\vltu)_I^{\dot\alpha}=\sum_{i=1}^3 C_{Ii}\vltu_i^{\dot\alpha}\eqncom \qquad
(C\cdot\vleu)_I^{A}=\sum_{i=1}^3 C_{Ii}\vleu_i^A \eqncom \qquad
 (C^\perp\cdot\vllu)_J^\alpha=\sum_{i=1}^3 C^\perp_{Ji}\vllu_i^\alpha\eqncom
\end{equation}
where $I=1,\dots, k$ and $J=1,\dots, n-k$.
We can also write \eqref{eq: F 3,2 pre-grassmannian} in a $GL(2)$ invariant way, as an integral over the Graßmannian $G(2,3)$:
\begin{equation}
\label{eq: F 3,2 grassmannian}
  \ff_{3,2}(1,2,3)=\int \frac{\dd^{2\times 3 } C }{(12)^{1-u_{23}}(23)^{1-u_{31}}(31)^{1-u_{12}}}
  \;
  \delta^4(C\cdot\vltu)\,\delta^8(C\cdot\vleu)\,\delta^{2}(C^\perp\cdot\vll)
  \eqncom
\end{equation}
where $(i\,j)$ denotes the minor of $C$ that is built from the columns $i$ and $j$.
This is precisely the (deformed) Graßmannian integral for the 
three-point MHV amplitude \cite{Bargheer:2014mxa,Ferro:2014gca} with the twisted kinematics accounting 
for the operator insertion.

We can generalise the above derivation to an arbitrary number of external on-shell
fields:
\begin{equation}
\label{eq: F n,2 grassmannian}
 \ff_{n,2}(1,\dots,n)= \int \frac{\dd^{2\times n } C }{\prod_{i=1}^n (i \ssep i \splus 1)^{1-u_{i+1\ssep i+2}}}
  \;
  \delta^4(C\cdot\vltu)\,\delta^8(C\cdot\vleu)\,\delta^{2n-4}(C^\perp\cdot\vll) \eqndot
\end{equation}
Here, the shifted kinematic variables can actually be at any two positions.
One can easily check that the sequence of $\rr$ operators necessary to derive this expression does not contain BCFW shifts that would spoil this simple dependence on the modified kinematics $\vltu$, $\vleu$.
It is also trivial to check that this integral representation gives the correct result upon localising the integration on the support of the delta functions: we simply take the (deformed) Parke-Taylor formula and replace the kinematic variables, $\vlt\to\vltu$, $\vle\to\vleu$. 
Since the $\vll$'s are not modified, the Park-Taylor prefactor is unaffected by this replacement, and the only effects are shifts in the (super) momentum conserving delta functions, $P\to P-q$ and $Q^-\to Q^- - \gamma^-$. 
This follows from the identity
\begin{equation}
\label{eq: momentum conservation in new variables}
  \vll_i^\alpha \left( \vlt_i^{\dot\alpha}-\frac{\bra{j}q^{\dot\alpha}}{\abr{ji}} \right)
  +
\vll_j^\alpha \left( \vlt_j^{\dot\alpha}-\frac{\bra{i}q^{\dot\alpha}}{\abr{ij}} \right)
=\vll_i^\alpha \vlt_i^{\dot\alpha}+\vll_j^\alpha\vlt_j^{\dot\alpha}-\underbrace{\frac{\varepsilon_{\gamma\beta}(\vll_i^\alpha \vll_j^\gamma-\vll_j^\alpha\vll_i^\gamma)}{ \abr{ji} }}_{=\delta^\alpha_\beta}q^{\beta\dot\alpha}
\eqncom
\end{equation}
and a similar identity for the $\vle$'s. 
The above argument shows that \eqref{eq: F n,2 grassmannian} correctly reproduces \eqref{eq: deformed ff n,2} and in particular the undeformed result \eqref{eq: ff n,2 intro}.

\section{Beyond MHV}
\label{sec: beyond MHV}

In the previous section, we have considered the simplest form factors, namely the MHV form factors, to introduce many important concepts.
In this section, we will see that these concepts continue to apply beyond MHV, although with some modifications.
In particular, we will conjecture a Graßmannian integral representation for all form factors, both in spinor
helicity as well as in twistor and momentum twistor form, and provide several non-trivial checks.

\subsection{On-shell diagrams and \texorpdfstring{$\rr$}{\rrpdf} operators}
\label{subsec: on-shell diagrams and r operators}

Since all form factors can be constructed via BCFW recursion relations as shown in \eqref{eq: BCFW for form factors}, we can also directly associate on-shell diagrams to each BCFW term --- completely independent of the MHV degree $k$. One main difference between $k=2$ and $k>2$ is that all MHV form factors can be constructed via the inverse soft limit without regard to the order and insertion positions, 
which directly gives the on-shell diagram.
For $k>2$, the result of the BCFW construction, and hence the correct on-shell diagram, is less obvious.
A second main difference between $k=2$ and $k>2$ is that, both for amplitudes as well as for form factors, there are several BCFW terms and hence on-shell diagrams which have to be summed to obtain the complete expression. 
However, for amplitudes, they can be combined into a single top-cell diagram, which corresponds to a top-dimensional integral over the Graßmannian and yields all required BCFW terms when taking suitable residues.
We will find in section \ref{sec: grassmannian} that we can define such top-dimensional integrals also for form factors. However, we will see below that a sum of several top-cell diagrams will be required.

Likewise, it is always possible to construct a given on-shell graph by acting with a chain of BCFW bridges on suitable vacua. Translating these BCFW bridges to $\rr$ operators, we can build deformed BCFW terms and top-cell diagrams for form factors as in section \ref{subsec: deformed form factors}. 
Hence, in order to construct Graßmannian integrals and $\rr$ operator representations, the first important step is to identify the corresponding top-cell diagrams.
Let us look at several examples first.

\subsubsection*{$\NmaxMHV$}

A special class of form factors beyond $k=2$ is given by $\NmaxMHV$, which has $k=n$. For amplitudes, the corresponding case is $\MHVbar$, which has $k=n-2$.
Similarly to $\MHVbar$ amplitudes,  $\NmaxMHV$ form factors can be constructed via the inverse soft limit without regard to the order and insertion points. 
Hence, the on-shell graph is immediate.
The permutation associated to these on-shell diagrams is 
\begin{equation}
\NmaxMHV:\qquad \sigma=(n-1,n, 1,2,\dots,n-2)\eqndot
\end{equation}
In the construction via $\rr$ operators, we now have $n-2$ conjugate amplitude vacua $\deltam{i}$ on the right of the minimal form factor instead of $n-2$ amplitude vacua $\deltap{i}$ on its left.  
In the simplest case of $n=k=3$, this is depicted in figure \ref{fig: F3,3}. 

\begin{figure}[htbp]
\begin{equation*}
(2,3,1)=(1,2)\permprod(2,3)\longrightarrow
\begin{aligned}
 \begin{tikzpicture}[scale=0.8]
\drawvline{1}{2}
\drawvline{2}{2}
\drawvline{3}{2}
\drawminimalff{2} 
\drawvacm{1}
\drawbridge{1}{1}
\drawbridge{2}{2}
\node[dl] at (0,-\vacuumheight-2*\bridgedistance-\labelvdist) {3};
\node[dl] at (1,-\vacuumheight-2*\bridgedistance-\labelvdist) {2};
\node[dl] at (2,-\vacuumheight-2*\bridgedistance-\labelvdist) {1};
\end{tikzpicture}
\end{aligned}
\quad\longrightarrow\quad 
\begin{aligned}
 \begin{tikzpicture}[scale=0.8]
 			\draw[thick,double] (0,-0) -- (0,0.5); 
  			\draw (0,0) -- (-\hdist,-\hdist) -- (0,-2*\hdist) -- (+\hdist,-\hdist) -- (0,0); 
 			\draw (+\hdist,-\hdist) -- (+2*\hdist,-2*\hdist);   
 			\draw (-\hdist,-\hdist) -- (-2*\hdist,-2*\hdist);   
 			\draw (0,-2*\hdist) -- (0,-2*\hdist-\ddist);  
                         \node[db] at (+\hdist,-\hdist) {}; 
                         \node[dw] at (0,-2*\hdist) {};
                         \node[db] at (-\hdist,-\hdist) {}; 
                         \node at (+2*\hdist  +\labelddist,-2*\hdist-\labelddist) {1};
                         \node at (0,-2*\hdist-\ddist-\labelvdist) {2};
                         \node at (-2*\hdist -\labelddist,-2*\hdist-\labelddist) {3};
         \end{tikzpicture}
\end{aligned}
\end{equation*}
\caption{Constructing the on-shell diagram of $\ff_{3,3}$ via permutations and BCFW bridges.
}
\label{fig: F3,3}
\end{figure}

\subsubsection*{N\MHV}

The first case that is truly beyond $\MHV$ is $k=3$ for $n\geq4$. For the case of $n=4$, the BCFW sums for all adjacent shifts are shown in figure \ref{fig: BCFW ff 4,3}. 
\begin{figure}[htbp]
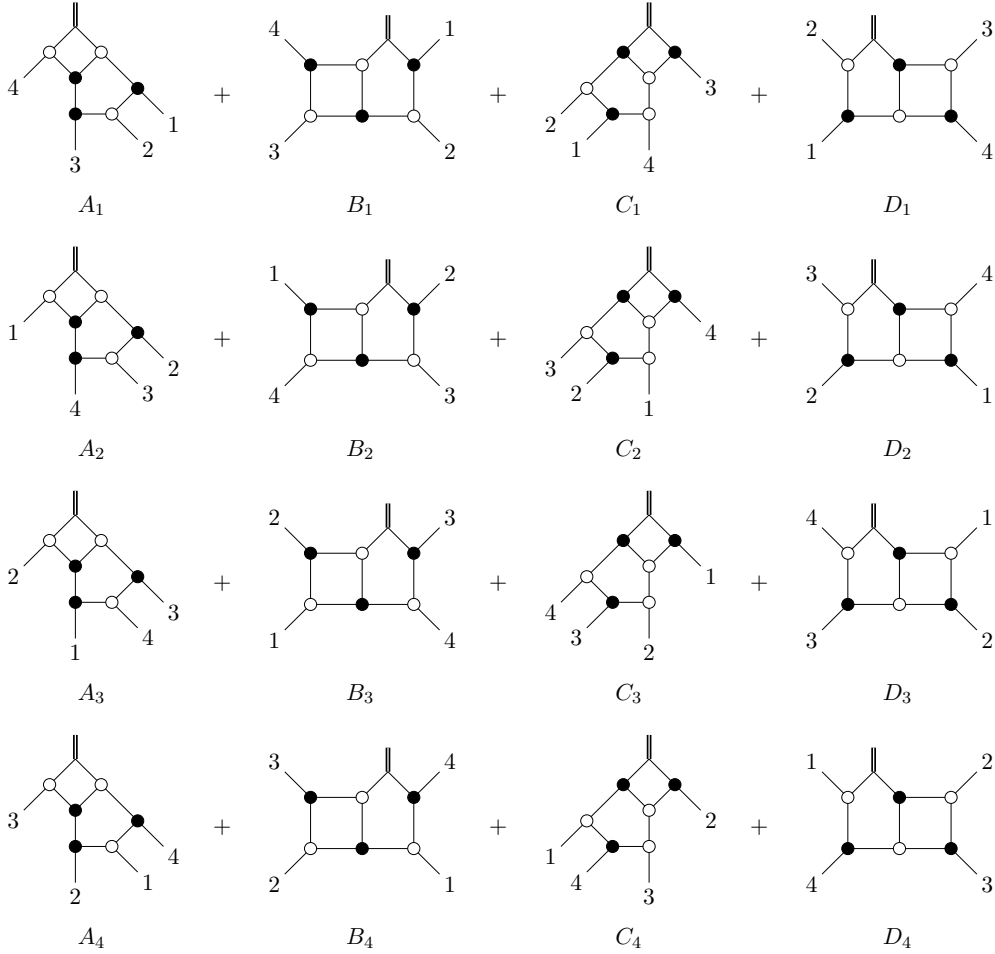

  \centering
  \scalebox{0.8}{
  \begin{tabular}{ccccccc}
 $\athreetwofthreetwo{1}{2}{3}{4}$   & $+$ &
 $\afourtwoftwotwo{1}{2}{3}{4}$      & $+$ &
 $\fthreethreeathreeone{1}{2}{3}{4}$ & $+$ &
 $\ftwotwoafourtwo{1}{2}{3}{4}$      \\
 $A_1$ && $B_1$ && $C_1$ && $D_1$    \\[11pt]
 $\athreetwofthreetwo{2}{3}{4}{1}$   & $+$ &
 $\afourtwoftwotwo{2}{3}{4}{1}$      & $+$ &
 $\fthreethreeathreeone{2}{3}{4}{1}$ & $+$ &
 $\ftwotwoafourtwo{2}{3}{4}{1}$      \\
 $A_2$ && $B_2$ && $C_2$ && $D_2$    \\[11pt]
 $\athreetwofthreetwo{3}{4}{1}{2}$   & $+$ &
 $\afourtwoftwotwo{3}{4}{1}{2}$      & $+$ &
 $\fthreethreeathreeone{3}{4}{1}{2}$ & $+$ &
 $\ftwotwoafourtwo{3}{4}{1}{2}$      \\
 $A_3$ && $B_3$ && $C_3$ && $D_3$    \\[11pt]
 $\athreetwofthreetwo{4}{1}{2}{3}$   & $+$ &
 $\afourtwoftwotwo{4}{1}{2}{3}$      & $+$ &
 $\fthreethreeathreeone{4}{1}{2}{3}$ & $+$ &
 $\ftwotwoafourtwo{4}{1}{2}{3}$      \\
 $A_4$ && $B_4$ && $C_4$ && $D_4$    
\end{tabular}
}
\caption{BCFW terms of $\ff_{4,3}$ for all adjacent shift. The $i^{\text{th}}$ line stems from a shift in $i$ and $i+1$.}
\label{fig: BCFW ff 4,3}
\end{figure}%
They have been generated using~\eqref{eq: BCFW for form factors}.
Applying the moves in figures \ref{fig: amplitude moves} and \ref{fig: move for minimal ff}, it is easy to see that 
\begin{equation}
  A_i=D_{(i+2) \text{ mod } 4}\eqncom \qquad B_i=C_{(i+2) \text{ mod } 4}\eqndot
\end{equation}
These BCFW terms can be obtained as residues from the sum of two different top-cell diagrams. The first of these is shown in figure \ref{fig: F4,3} together with its permutation and construction via $\rr$ operators; the second one can be obtained from it by a cyclic shift of the external on-shell legs by two.
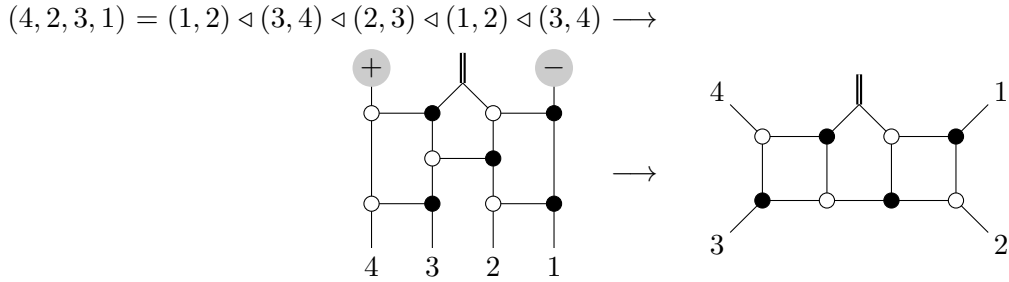
\begin{figure}[htbp]
\begin{equation*}
\begin{aligned}
(4,2,3,1)=(1,2)\permprod(3,4)\permprod(2,3)\permprod(1,2)\permprod(3,4) &\longrightarrow \\
\begin{aligned}
 \begin{tikzpicture}[scale=0.8]
\drawvline{1}{3}
\drawvline{2}{3}
\drawvline{3}{3}
\drawvline{4}{3}
\drawvacp{1}
\drawminimalff{2} 
\drawvacm{4}
\drawbridge{1}{1}
\drawbridge{3}{1}
\drawbridge{2}{2}
\drawbridge{1}{3}
\drawbridge{3}{3}
\node[dl] at (0,-\vacuumheight-3*\bridgedistance-\labelvdist) {4};
\node[dl] at (1,-\vacuumheight-3*\bridgedistance-\labelvdist) {3};
\node[dl] at (2,-\vacuumheight-3*\bridgedistance-\labelvdist) {2};
\node[dl] at (3,-\vacuumheight-3*\bridgedistance-\labelvdist) {1};
\end{tikzpicture}
\end{aligned}
\quad&\longrightarrow\quad
\begin{aligned}
 \begin{tikzpicture}[scale=0.8]
 			\draw[thick,double] (0,-0+\hdist) -- (0,0.5+\hdist); 
 			\draw (0,\hdist) -- (+\hdist,0);
 			\draw (0,\hdist) -- (-\hdist,0);
 			\draw (+\hdist,0) -- (3*\hdist,0) -- (4*\hdist,\hdist);
 			\draw (-\hdist,0) -- (-3*\hdist,0) -- (-4*\hdist,\hdist);
 			\draw (4*\hdist,-3*\hdist) -- (3*\hdist,-2*\hdist) -- (-3*\hdist,-2*\hdist) -- (-4*\hdist,-3*\hdist);
 			\draw (-\hdist,0) -- (-\hdist,-2*\hdist);
			\draw (-3*\hdist,0) -- (-3*\hdist,-2*\hdist);
 			\draw (\hdist,0) -- (\hdist,-2*\hdist);
			\draw (3*\hdist,0) -- (3*\hdist,-2*\hdist);
                         \node[dw] at (+\hdist,0) {}; 
                         \node[db] at (+3*\hdist,0) {};
                         \node[db] at (-\hdist,0) {}; 
                         \node[dw] at (-3*\hdist,0) {}; 
                         \node[db] at (+\hdist,-2*\hdist) {}; 
                         \node[dw] at (+3*\hdist,-2*\hdist) {};
                         \node[dw] at (-\hdist,-2*\hdist) {}; 
                         \node[db] at (-3*\hdist,-2*\hdist) {}; 
                         \node at (4*\hdist +\labelddist,1*\hdist +\labelddist) {1};
                         \node at (4*\hdist +\labelddist,-3*\hdist-\labelddist) {2};
                         \node at (-4*\hdist -\labelddist,-3*\hdist-\labelddist) {3};
                         \node at (-4*\hdist -\labelddist,1*\hdist +\labelddist) {4};
\end{tikzpicture}
\end{aligned}
\end{aligned}
\end{equation*}
\caption{Constructing the top-cell diagram of $\ff_{4,3}$ via permutations and BCFW bridges.
Note that the decomposition of the permutation that produces the correct on-shell diagram is not minimal in the amplitude sense.
}
\label{fig: F4,3}
\end{figure}
Concretely, all vertical edges in the top-cell diagram in figure \ref{fig: F4,3} are removable. Deleting them, we obtain from left to right $C_2$, $A_1$, $C_3$ and $A_2$. 

Several remarks are in order.
First, we do require more than one top-cell diagram to generate all BCFW terms.
Second, the top-cell diagram is not cyclically invariant, and neither is the corresponding permutation. Instead, we (in principle) have to explicitly consider all cyclic permutations of the top-cell diagram and the corresponding permutation.
Third, the permutation is not decomposed into a minimal number of transpositions. 

It is possible to construct other cases with higher $n,k$ in an analogous way.


\subsubsection*{N$^k$MHV and a relation to amplitude on-shell diagrams}

We conclude this subsection with a general observation relating the on-shell diagrams of form factors with those of amplitudes. 
In particular, this will lead to (a conjecture for) the form factor top-cell diagrams at general $n$, $k$.

We note that the $n$-point form factor shares interesting features with the $(n+2)$-point amplitude. 
To begin with, the MHV degree $k$ ranges from $2$ to $n$ in both cases.
Moreover, $n+2$ is the expected number of kinematic dependencies if we consider that the off-shell (super) momentum of the operator can be parametrised by two on-shell (super) momenta.

Finally, we have found that we can obtain the top-cell diagrams of $\ff_{n,k}$ from the top-cell diagram of $\amp_{n+2,k}$ by applying moves until a box appears and replacing this box with the minimal form factor. Graphically, this relation reads
\begin{equation}
 \label{eq: box eater}
    \scalebox{0.9}{\(
\begin{aligned} 
 \begin{tikzpicture}[scale=0.8, baseline=-0.7cm]
                        \draw (1,1) -- (1,2) -- (2,2) -- (2,1) -- (1,1);
                        \draw (1,0) -- (1,1);
                        \draw (2,0) -- (2,1);
                        \draw (2.5,2.5) -- (2,2);
                        \draw (0.5,2.5) -- (1,2);
                        \draw (-0.5,-0.75) -- (-0.5,-2.5);
                        \draw (1.5,-0.75) -- (1.5,-2.5);
                        \draw (2.5,-0.75) -- (2.5,-2.5);
                        \draw (3.5,-0.75) -- (3.5,-2.5);
                        \node at (-0.5,-2.5-\labelvdist) {$n$};
                        \node at (0.5,-2.5-\labelvdist) {$\cdots$};
                        \node at (1.5,-2.5-\labelvdist) {$3$};
                        \node at (2.5,-2.5-\labelvdist) {$2$};
                        \node at (3.5,-2.5-\labelvdist) {$1$};
                        \node at (2.5+\labelddist,2.5+\labelddist) {$n+2$};
                        \node at (0.5+\labelddist,2.5+\labelddist) {$n+1$};
                        \node[dw] at (2,1) {};
                        \node[dw] at (1,2) {};
                        \node[db] at (1,1) {};
                        \node[db] at (2,2) {};
                        \node[ellipse, black, fill=grayn, minimum width=4 cm, minimum height=2 cm, draw, inner sep=0pt] at (1.5,-0.75) {};
        \end{tikzpicture}
        \quad
        \longrightarrow
        \quad
        \begin{tikzpicture}[scale=0.8, baseline=-0.7cm]
        \draw[thick,double] (1.5,-0+1.5) -- (1.5,-0.5+1.5); 
	\draw (1.5,-0.5+1.5) -- (2,-\vacuumheight+1.4);  
	\draw (1.5,-0.5+1.5) -- (1,-\vacuumheight+1.4);
                        \draw (-0.5,-0.75) -- (-0.5,-2.5);
                        \draw (1.5,-0.75) -- (1.5,-2.5);
                        \draw (2.5,-0.75) -- (2.5,-2.5);
                        \draw (3.5,-0.75) -- (3.5,-2.5);
                        \node at (-0.5,-2.5-\labelvdist) {$n$};
                        \node at (0.5,-2.5-\labelvdist) {$\cdots$};
                        \node at (1.5,-2.5-\labelvdist) {$3$};
                        \node at (2.5,-2.5-\labelvdist) {$2$};
                        \node at (3.5,-2.5-\labelvdist) {$1$};
                        \node[ellipse, black, fill=grayn, minimum width=4 cm, minimum height=2 cm, draw, inner sep=0pt] at (1.5,-0.75) {};
        \end{tikzpicture}
        \quad
      \eqncom
      \end{aligned}
  \)}
\end{equation}
\nopagebreak
where we have replaced the box at the legs $n+1$ and $n+2$ for the sake of concreteness. 
This relation is valid for all form factors presented in this paper.

At the level of the BCFW terms, a respective relation can be seen to be true by considering the BCFW recursion relations \eqref{eq: BCFW for amplitudes} and \eqref{eq: BCFW for form factors}.
Note that for $n=2$ the amplitude on-shell diagram is nothing but a box and entirely replaced by the minimal form factor.
Constructing the amplitude $\amp_{n+2,k}$ recursively via \eqref{eq: BCFW for amplitudes}, we find that boxes can only occur at the boundary of the on-shell diagram.
Then constructing $\ff_{n,k}$ recursively via \eqref{eq: BCFW for form factors}, we find that each term in \eqref{eq: BCFW for form factors} can be obtained by replacing one of the boxes in a term of the construction of $\amp_{n+2,k}$ by the minimal form factor.
It would be interesting to prove this relation also at the level of the top-cell diagram(s).

Using the relation \eqref{eq: box eater} between $\ff_{n,k}$ and $\amp_{n+2,k}$, we can also relate the corresponding permutations.
By replacing the box by the minimal form factor, we can hide
the corresponding legs of the amplitude in the composite operator. At the level of permutations, this connects the preimage of the hidden leg to its image.
For $\amp_{n+2,k}$, the permutation corresponding to the top-cell diagram reads
\begin{equation}
 \amp_{n+2,k}:\qquad \sigma=(k+1,\dots, n,n+1, n+2, 1, 2, \dots, k)\eqndot
\end{equation}
If we hide the legs $n+1$ and $n+2$, we obtain
\begin{equation}
 \ff_{n,k}:\qquad \sigma=(k+1,\dots,n,k-1, k, 1, 2, \dots, k-2)\eqndot
\end{equation}
Moreover, the top-cell diagram with the minimal form factor replaced by the open legs $n+1$ and $n+2$ is characterised by 
\begin{equation}
 \ff_{n,k} \text{ without }\ff_{2,2}:\qquad \sigmapart=(k+1,\dots, n, n+2, n+1, 1, 2, \dots, k,k-1)\eqndot
 \label{eq: permutation diagram to be glued}
\end{equation}
The permutation $\sigmapart$ allows us to directly generate this on-shell diagram e.g.\ using the \texttt{Mathematica} package \texttt{positroid.m} \cite{Bourjaily:2012gy}.

Let us conclude with a comment about the role of the permutation for form factors.
As for scattering amplitudes, it is invariant under all equivalence moves.
In contrast to the case for tree-level scattering amplitudes, the permutation $\sigma$ for the complete form factor on-shell diagram requires a decomposition into more than the minimal number of transpositions to construct the corresponding on-shell diagram via BCFW bridges as shown above.%
\footnote{It would be interesting to see whether the decomposition is minimal when adding further conditions such as considering $1$ and $n$ to be non-adjacent for the purpose of the decomposition.}
However, the modified permutation $\sigmapart$ can be directly used to obtain the on-shell diagram with the minimal form factor replaced by the open legs $n+1$ and $n+2$.
This is similar to the situation for one-loop amplitudes, whose on-shell diagrams are also best not constructed from their permutations but from the permutations of the corresponding higher point amplitudes before taking the forward limit.
It would be interesting to fully classify the equivalence classes of form factor on-shell diagrams and find appropriate combinatorial labels for them.

\subsection{Graßmannian integrals for higher MHV degree}
\label{sec: grassmannian}
Having identified (a conjecture for) the general top-cell diagram, let us now turn to the Graßmannian integral representation for form factors.

\subsubsection*{General considerations on the Graßmannian}

The fundamental idea behind Graßmannian integral representations of scattering amplitudes \cite{ArkaniHamed:2009dn,ArkaniHamed:2012nw,Mason:2009qx} is to express momentum conservation in a geometric way.%
\footnote{Here, we focus on momentum for brevity. The same arguments apply to super momentum by replacing $\vlt$ with $\vle$.}
Regarding the external kinematic data as a pair of two-planes $\vll$ and $\vlt$ in $n$-dimensional space,
momentum conservation is expressed as the orthogonality of these planes:
\begin{equation}
\label{eq: momentum conservation as orthogonality of planes}
 \vll\cdot\vlt \equiv \sum_{i=1}^n\vll_i\vlt_i=0 \eqndot
\end{equation}
The Graßmannian representation linearises this constraint by introducing an auxiliary hyperplane $C\in G(k,n)$ such that
\begin{equation}
    (C\cdot\vlt)^{\dot\alpha}_I=\sum_{i=1}^nC_{Ii}\vlt_i^{\dot\alpha}=0 \quad\text{and}\quad (C^\perp\cdot\vll)^{\alpha}_J=\sum_{i=1}^nC^\perp_{Ji}\vll_i^\alpha=0 \quad\implies\quad
  \vll\cdot\vlt=0
  \eqncom
\end{equation}
where $C^\perp$ is the orthogonal complement of $C$ fulfilling $C (C^\perp)^T=0$ and $I=1,\dots, k$, $J=1,\dots, n-k$. 
The Graßmannian integral for scattering amplitudes integrates a holomorphic form on $G(k,n)$ on the support of these constraints.

As discussed in section \ref{subsec: MHV Grassmannian}, we can similarly geometrise momentum conservation for form factors. 
Setting
\begin{equation}
\label{eq: def arbitrary underscore variables}
  \vltu_k=\vlt_k \eqncom \quad k=1,\ldots,n\eqncom \quad k\neq i,j \eqncom\qquad
  \vltu_i=\vlt_i-\frac{\bra{j}q}{\abr{ji}}
  \eqncom\qquad
  \vltu_j=\vlt_j-\frac{\bra{i}q}{\abr{ij}}
\end{equation}
for arbitrary $i$ and $j$, we can express momentum conservation as $\vll\cdot\vltu=0$; cf.\ \eqref{eq: momentum conservation in new variables}.
We also saw that in the MHV case a naive way of introducing an auxiliary Graßmannian works. 
One can simply use the same Graßmannian and the same form as one would use for the MHV amplitude with the same number of legs.
It is clear, however, that this way of linearising the geometrical constraint cannot work beyond MHV. 
For instance, the MHV degree $k$ ranges up to $n$ for form factors, while for amplitudes it only ranges up to $n-2$. 
Since $G(n,n)$ is just a point, a larger Graßmannian is necessary for $\text{N}^{\text{max}}\MHV$, but in fact already starting from NMHV.

The different range of MHV degrees already suggests that the correct Graßmannian is $G(k,n+2)$; this also fits nicely with the observation \eqref{eq: box eater} as well as with the general fact that an off-shell momentum can be parametrised by two on-shell ones.
Indeed, instead of \eqref{eq: def arbitrary underscore variables}, we can define new kinematic variables as a pair of two-planes in an $(n+2)$-dimensional space as
\begin{equation}
 \label{eq: def underunderscore variables}
 \begin{aligned}
\vlluu_i&=\vll_i \eqncom \quad i=1,\ldots,n \eqncom &
\vlluu_{n+1}&=\refspina\eqncom&
\vlluu_{n+2}&=\refspinb\eqncom 
\\
\vltuu_i&=\vlt_i \eqncom \quad i=1,\ldots,n \eqncom &
\vltuu_{n+1}&=-\frac{\bra{\refspinbl}q}{\abr{\refspinbl\refspinal}}\eqncom&
\vltuu_{n+2}&=-\frac{\bra{\refspinal}q}{\abr{\refspinal\refspinbl}}\eqncom
\end{aligned}
\end{equation}
where $\refspina$ and $\refspinb$ are arbitrary non-collinear reference spinors.
Momentum conservation is then expressed as $\vlluu\cdot\vltuu=0$. As shown in \eqref{eq: momentum conservation in new variables}, the two additional on-shell momenta indeed encode the off-shell momentum: $\vlluu_{n+1}\vltuu_{n+1}+\vlluu_{n+2}\vltuu_{n+2}=-q$. 
Analogously, we can define fermionic variables 
\begin{equation}
 \label{eq: def underunderscore variables etas}
 \begin{aligned}
\vleuu_i^+&=\vle_i^+ \eqncom \quad i=1,\ldots,n \eqncom &
\vleuu_{n+1}^+&=0\eqncom&
\vleuu_{n+2}^+&=0\eqncom 
\\
\vleuu_i^-&=\vle_i^- \eqncom \quad i=1,\ldots,n \eqncom &
\vleuu_{n+1}^-&=-\frac{\bra{\refspinbl}\gamma^-}{\abr{\refspinbl\refspinal}}\eqncom&
\vleuu_{n+2}^-&=-\frac{\bra{\refspinal}\gamma^-}{\abr{\refspinal\refspinbl}}\eqncom
\end{aligned}
\end{equation}
which encode the off-shell super momentum as $\vlluu_{n+1}\vleuu_{n+1}^-+\vlluu_{n+2}\vleuu^{-}_{n+2}=-\gamma^-$.
Super-momentum conservation can then be written as $\vlluu\cdot\vleuu=0$.

We can now linearise the constraint imposed by (super) momentum conservation by requiring $C'\cdot\vltuu=0$, $C'\cdot\vleuu=0$ and $C'^\perp\cdot\vlluu=0$ with $C'\in G(k,n+2)$. 

\subsubsection*{From on-shell graphs to Graßmannian integrals}

To show that form factors can be written as integrals over the Graßmannian $G(k,n+2)$, we break the corresponding diagram into two pieces: the minimal form factor \eqref{eq: minimal form factor as vacuum} and a purely on-shell piece with $n+2$ legs for which a Graßmannian integral representation is known. 
We then glue these two pieces together, i.e.\ we perform the on-shell phase space integration. 
We start by discussing this procedure in a general form that can be applied to any diagram; then, we will look at some low-point examples to see how the explicit form for top-cell diagrams looks like.

The on-shell piece that will be glued with the minimal form factor can be written as~\cite{ArkaniHamed:2012nw}
\begin{equation}
  I=
        \int \frac{\dd\alpha_1}{\alpha_1}\cdots\frac{\dd\alpha_m}{\alpha_m} \;
        \delta^{k\times 2}(C\cdot\vlt) \,
        \delta^{k\times 4}(C\cdot\vle) \,
        \delta^{(n+2-k)\times 2}(C^\perp\cdot\vll)
        \eqncom
\end{equation}
where the matrix $C$ depends on the $\alpha_i$'s, $C=C(\alpha_i)\in G(k,n+2)$ and $m$ is the dimension of the corresponding cell in the Graßmannian. 
Gluing the minimal form factor to the legs $n+1$ and $n+2$ corresponds to calculating
\begin{equation}
\label{eq: gluing minimal ff to on-shell piece}
        \int 
        \prod_{i=n+1}^{n+2}\left( \frac{\dd^2\vll_{i}\,\dd^2\vlt_{i}}{\mathrm{Vol}[GL(1)]}
        \,
        \dd^4\vle_{i}\right)
\;\;
       \deltaf{n+1\; n+2} \Big\vert_{\vll\to -\vll}
        \;\;
        I(1,\ldots,n+2) \eqncom
\end{equation}
where $\deltaf{}$ was defined in \eqref{eq: minimal form factor as vacuum} and the signs of the corresponding  $\lambda$'s are inverted since the two particles are ingoing with respect to $\deltaf{}$.
We can perform the $\vlt$ and $\vle$ integration using the delta functions of the minimal form factor \eqref{eq: minimal form factor as vacuum}; this replaces 
\begin{equation}
\label{eq: replacement n+1 n+2}
 \begin{aligned}
  \vlt_{n+1} &\to -  \frac{\bra{n+2}q}{\abr{n \splus 2 \ssep n \splus 1}}\eqncom&
  \vle_{n+1}^- &\to - \frac{\bra{n+2}\gamma^-}{\abr{n \splus 2 \ssep n \splus 1}}\eqncom&
  \vle_{n+1}^+ &\to 0\eqncom\\
  \vlt_{n+2} &\to - \frac{\bra{n+1}q}{\abr{n \splus 1 \ssep n \splus 2}}\eqncom&
  \vle_{n+2}^- &\to - \frac{\bra{n+1}\gamma^-}{\abr{ n \splus 1 \ssep n \splus 2}}\eqncom&
  \vle_{n+2}^+ &\to 0\eqndot
\end{aligned}
\end{equation}
To remove the $GL(1)^2$ redundancy in the remaining $\vll$ integrations, we parametrise 
\begin{equation}
  \vll_{n+1}=\refspina-\beta_1 \refspinb
        \eqncom\qquad
        \vll_{n+2}=\refspinb-\beta_2 \refspina
        \eqncom
\end{equation}
where $\refspina$ and $\refspinb$ are two arbitrary but linearly independent reference spinors, which will be identified with the ones in \eqref{eq: def underunderscore variables}.
With this, $\abr{n \splus 1 \ssep n \splus 2 }=(\beta_1\beta_2-1)\abr{\refspinbl\refspinal}$ and the replacement \eqref{eq: replacement n+1 n+2} becomes
\begin{equation}
\label{eq: last to spinors in terms of reference spinors}
\begin{aligned}
  \vlt_{n+1} &\to  
                \frac{1}{\beta_1\beta_2-1}\,\frac{\bra{\refspinbl}q}{\abr{\refspinbl\refspinal}}
                +\frac{\beta_2}{\beta_1\beta_2-1}\,\frac{\bra{\refspinal}q}{\abr{\refspinal\refspinbl}}\eqncom
                \\
                \vle_{n+1}^{-} &\to  
                \frac{1}{\beta_1\beta_2-1}\,\frac{\bra{\refspinbl}\gamma^{-}}{\abr{\refspinbl\refspinal}}
                +\frac{\beta_2}{\beta_1\beta_2-1}\,\frac{\bra{\refspinal}\gamma^{-}}{\abr{\refspinal\refspinbl}}\eqncom
                \\
                \vlt_{n+2} &\to  
                \frac{1}{\beta_1\beta_2-1}\,\frac{\bra{\refspinal}q}{\abr{\refspinal\refspinbl}}
                +\frac{\beta_1}{\beta_1\beta_2-1}\,\frac{\bra{\refspinbl}q}{\abr{\refspinbl\refspinal}}\eqncom
                \\
                \vle_{n+2}^{-} &\to  
                \frac{1}{\beta_1\beta_2-1}\,\frac{\bra{\refspinal}\gamma^{-}}{\abr{\refspinal\refspinbl}}
                +\frac{\beta_1}{\beta_1\beta_2-1}\,\frac{\bra{\refspinbl}\gamma^{-}}{\abr{\refspinbl\refspinal}}
                \eqndot
\end{aligned}
\end{equation}
At the same time, the measure transforms to
\begin{equation}
         \int 
        \frac{\dd^2\vll_{n+1}}{\mathrm{Vol}[GL(1)]}
        \frac{\dd^2\vll_{n+2}}{\mathrm{Vol}[GL(1)]}
        =
        \abr{\refspinal\refspinbl}\abr{\refspinbl\refspinal}
        \int\dd\beta_1\dd\beta_2
        \eqndot
\end{equation}
Applying the substitutions in \eqref{eq: last to spinors in terms of reference spinors}, we can write \eqref{eq: gluing minimal ff to on-shell piece} as
\begin{multline}
\label{eq: general Grassmannian gluing formula}
  I_{\ff}=
  \abr{\refspinal\refspinbl}\abr{\refspinbl\refspinal}
  \int \frac{\dd\alpha_1}{\alpha_1}\cdots\frac{\dd\alpha_m}{\alpha_m}\;\frac{\dd\beta_1\,\dd\beta_2}{(1-\beta_1\beta_2)^2} 
    \\ \times\;
        \delta^{k\times 2}(C'(\alpha_i,\beta_i)\cdot\vltuu) \,
        \delta^{k\times 4}(C'(\alpha_i,\beta_i)\cdot\vleuu)\,
        \delta^{(n+2-k)\times 2}(C'^\perp(\alpha_i,\beta_i)\cdot\vlluu)
        \eqncom
\end{multline}
where we recombined the columns of the matrix $C$ such that they form the coefficients of the kinematic data
$\vltuu$, $\vleuu$ defined in \eqref{eq: def underunderscore variables} and \eqref{eq: def underunderscore variables etas}. 
This new matrix $C'=(C'_1 \cdots C'_{n+2})$ depends both on the $\alpha_i$'s as well as the $\beta_i$'s.
Its first $n$ columns coincide with those of $C$ and the last two columns are given respectively by 
\begin{equation}
\begin{aligned}
  C'_{n+1} & = \frac{1}{1-\beta_1\beta_2}C_{n+1}+\frac{\beta_1}{1-\beta_1\beta_2} C_{n+2} \eqncom \qquad
  C'_{n+2} & = \frac{1}{1-\beta_1\beta_2}C_{n+2}+\frac{\beta_2}{1-\beta_1\beta_2} C_{n+1} \eqndot
\end{aligned}
\end{equation}
Hence, also the first $n$ columns of $C'^\perp$ coincide with those of $C^\perp$ and the last two columns are 
\begin{equation}
 \begin{aligned}
C'^\perp_{n+1} &= C^\perp_{n+1}-\beta_2 C^\perp_{n+2} \eqncom 
\qquad
C'^\perp_{n+2} &= C^\perp_{n+2}-\beta_1 C^{\perp}_{n+1} \eqndot
 \end{aligned}
\end{equation}
The factor of $(1-\beta_1\beta_2)^2$ in \eqref{eq: general Grassmannian gluing formula} is a Jacobian from reorganising the $C^\perp\cdot\vll$ 
delta functions, which we can write as 
\begin{equation}
\label{eq: delta function C perp alternative for Jacobian}
        \delta^{(n+2-k)\times 2}(C^\perp\cdot\vll) =
        \prod_{K=1}^k
        \int \dd^{2}\rho_K  \;\,
        \delta^{(n+2)\times 2}\left(\vll_i-\rho_{L}C_{Li}\right)
        \eqncom
\end{equation}
where $\rho_K^\alpha$ with $K=1,\dots,k$ is a set of auxiliary variables.
For the delta functions corresponding to the columns $n+1$ and $n+2$, we have
\begin{equation}
 \begin{aligned}
    &
    \!\!\!
    \!\!\!
    \!\!\!
    \!\!\!
    \!\!\!
    \!\!\!
    \!\!\!
    \!\!\!
    \delta^{2}(\vll_{n+1}-\rho_{L}C_{L\ssep n+1})
    \delta^{2}(\vll_{n+2}-\rho_{L}C_{L\ssep n+2}) \\[10pt]
     \rightarrow& \phaneq
     \delta^{2}(
        \vlluu_{n+1}-\rho_{L}C'_{L\ssep n+1}
        -\beta_1(
            \vlluu_{n+2}-\rho_{L}C'_{L\ssep n+2}            
            )
        )\\
        &\phaneq
    \delta^{2}(
        \vlluu_{n+2}-\rho_{L}C'_{L\ssep n+2}
        -\beta_2(
            \vlluu_{n+1}-\rho_{L}C'_{L\ssep n+1}            
            )
        )
        \\[5pt]
    &= 
    \frac{1}{(1-\beta_1\beta_2)^2}
    \delta^{2}(\vlluu_{n+1}-\rho_{L}C'_{L\ssep n+1})
    \delta^{2}(\vlluu_{n+2}-\rho_{L}C'_{L\ssep n+2})
    \eqndot
 \end{aligned}
\end{equation}

The above shows that diagrams contributing to the form factor can be expressed naturally as some Graßmannian integrals. 
Of course, it remains to find some general expression for the form that is to be integrated over the support of the delta functions. 
For this we will look at some concrete examples first; although the gluing procedure works for any on-shell
diagram, we will from now on focus on top-cell diagrams based on the conjectured relation with the amplitude diagrams 
in section \ref{subsec: on-shell diagrams and r operators}.

\subsubsection*{MHV revisited}

In this section, we will show that the general procedure outlined in the last section reduces to the results from section \ref{sec: MHV} for MHV degree $k=2$. 
This will also give us a first idea of how the form of the $G(k,n+2)$ Graßmannian integral looks like in the general case.

Consider the $C'$ matrix in the standard gauge fixing,
\begin{equation}
  C'=\begin{pmatrix} 1&0&c'_{13}&\cdots&c'_{1 \sssep n+2 }\\0&1&c'_{23}&\cdots&c'_{2 \sssep n+2}\end{pmatrix}
  \eqncom
\end{equation}
and kinematic data $\vlluu$, $\vltuu$, $\vleuu$ with the off-shell information encoded at positions $n+1$ and $n+2$. 
Compared to the $G(2,n)$ integral given in \eqref{eq: F n,2 grassmannian}, we have four additional integrations as well as four additional bosonic delta functions $\delta(C'^\perp\cdot\vlluu)$ involving the $(n+1)^{\text{th}}$ and $(n+2)^{\text{th}}$ rows of $C'$. 
If we choose the reference spinors $\refspina$, $\refspinb$ such that $\refspina\equiv \vlluu_{n+1}=\vll_2$ and $\refspinb\equiv \vlluu_{n+2}=\vll_1$, these four additional delta functions impose
\begin{equation}
  -c'_{1 \sssep n+1 }\,\vll_1^\alpha-c'_{2 \sssep n+1}\,\vll_2^\alpha+\vll_2^\alpha=0
  \eqncom\qquad
  -c'_{1 \sssep n+2}\,\vll_1^\alpha-c'_{2 \sssep n+2}\,\vll_2^\alpha+\vll_1^\alpha=0
  \eqndot
  \label{eq: reduction MHV}
\end{equation}
Upon integrating out these delta functions, we obtain a Jacobian $\abr{12}^{-2}$ which cancels the prefactor in the general expression \eqref{eq: general Grassmannian gluing formula}, and the $C'$ matrix is set to
\begin{equation}
  C'=\begin{pmatrix} 1&0&c'_{13}&\cdots&c'_{1n}&0&1\\0&1&c'_{23}&\cdots&c'_{2n}&1&0\end{pmatrix}
\eqndot
\end{equation}
By defining $C$ as $C'$ without the last two columns, the delta functions are now identical to the ones in \eqref{eq: F n,2 grassmannian},
\begin{equation}
  \delta^{4}(C\cdot\vltu) \, \delta^{8}(C\cdot\vleu) \, \delta^{2n-4}(C^\perp\cdot\vll)
\end{equation}
with the twisted kinematics at position $1$ and $2$, as in \eqref{eq: minimal ff as vacuum}.

We now have to check whether the form obtained by the gluing procedure yields the same form as in section \ref{sec: MHV} after this integration.
\begin{figure}[htbp]
\(
I=
\begin{aligned}
        \begin{tikzpicture}[scale=0.8]
        \draw (1,-1) -- (0,-0.7);
        \draw (1,-1) -- (2,-0.7);
        \draw (1,-1) -- (1,-1-0.65);
        \draw (0,-0.7) -- (0-0.3,-0.7-0.6);
        \draw (2,-0.7) -- (2+0.3,-0.7-0.6);
        \draw (0,-0.7) -- (0+0.3,-0.7+0.6);
        \draw (2,-0.7) -- (2-0.3,-0.7+0.6);
        \node[db] at (1,-1) {};
        \node[dw] at (0,-0.7) {};
        \node[dw] at (2,-0.7) {};
                         \node[] at (-0.45,-1.6) {3};
                         \node[] at (2.45,-1.6) {1};
                         \node[] at (0.4,0.2) {4};
                         \node[] at (1.6,0.2) {5};
                         \node[] at (1,-2) {2};
        \end{tikzpicture}
      \end{aligned}
\eqncom      
\qquad
\sigmapart=(3,5,4,2,1)
\eqncom
\qquad
C=\begin{pmatrix}
          1&0&-\alpha_2&-\alpha_2\alpha_4&-\alpha_1\\
          0&1&\alpha_3&\alpha_3\alpha_4&0
        \end{pmatrix}
  \eqndot
\)
  \centering  
\caption{On-shell sub-diagram $I$ obtained by removing the minimal form factor from the on-shell diagram of $\ff_{3,2}$ shown in figure~\ref{fig: F3,2}, corresponding permutation $\sigmapart$ and $C$ matrix.}
\label{fig:gluing32}
\end{figure}
\begin{figure}[htbp]
  \centering
\begin{align*}
  I&=
  \begin{aligned}
        \begin{tikzpicture}[scale=0.8]
        \draw (1,-1) -- (0,-0.7);
        \draw (1,-1) -- (2,-0.7);
        \draw (1,-1) -- (1,-1-0.65);
        \draw (0,-0.7) -- (0-0.3,-0.7-0.6);
        \draw (2,-0.7) -- (2+0.3,-0.7-0.6);
        \draw (0,-0.7) -- (0+0.3,-0.7+0.6);
        \draw (2,-0.7) -- (2-0.3,-0.7+0.6);
        \draw (2.4,-1.5) -- (1.8,-1.8);
        \draw (1,-1.8) -- (1.8,-1.8);
        \draw (2.3,-1.3)--(2.7,-2.1);
        \draw (1,-1.8) -- (1,-2.5);
        \draw (1.8,-1.8) -- (2.0,-2.5) ;
        \node[] at (2.87,-2.4) {1};
        \node[] at (1,-2.8) {3};
        \node[] at (2.1,-2.8) {2};
        \node[db] at (1,-1) {};
        \node[dw] at (0,-0.7) {};
        \node[dw] at (2,-0.7) {};
                         \node[] at (-0.45,-1.6) {4};
                         \node[dw] at (2.4,-1.5) {};
                         \node[] at (0.4,0.2) {5};
                         \node[] at (1.6,0.2) {6};
                         \node[dw] at (1,-1.8) {};
        \node[db] at (1.8,-1.8) {};
        \end{tikzpicture}
      \end{aligned}
  \eqncom      
\qquad
\sigmapart=(3, 4, 6, 5, 2, 1)
\eqncom\\
  C&= \begin{pmatrix}
          1&0&-\alpha_3&-\alpha_2-\alpha_3\alpha_5&-\alpha_6(\alpha_2+\alpha_3\alpha_5)&-\alpha_1\\
          0&1&\alpha_4&\alpha_4\alpha_5&\alpha_4\alpha_5\alpha_6&0
  \end{pmatrix}
  \eqndot
  \end{align*}
\caption{On-shell sub-diagram $I$ obtained by removing the minimal form factor from the on-shell diagram of $\ff_{4,2}$ shown in figure~\ref{fig: F4,2}, corresponding permutation $\sigmapart$ and $C$ matrix.}
\label{fig:gluing42}
\end{figure}
We have performed the gluing explicitly for the MHV form factors with up to six external particles. 
The on-shell subdiagrams obtained by removing the minimal form factor from the on-shell diagrams, the corresponding permutations $\sigmapart$ as well as the $C$ matrices obtained from the \texttt{Mathematica} package \texttt{positroid.m} \cite{Bourjaily:2012gy} are shown in figures \ref{fig:gluing32} and \ref{fig:gluing42} for 3 and 4 points, respectively.
We invariably found that after changing from edge variables $\alpha_i$, $\beta_1$, $\beta_2$ to canonically gauge-fixed $c_{ij}$ variables, the integral could be written in the following form:%
\footnote{Note that we have ignored overall signs in the gluing procedure since the sign of residues expressed in edge variables is not readily determined. See \cite{Olson:2014pfa} for an elaborate algorithm that determines these signs, though.
}
\begin{equation}
\begin{aligned}
 \abr{\refspinal\refspinbl}^2 \int\frac{\dd^{2(n+2)}C'}{\text{Vol}[GL(2)]}\;
  \frac{Y(1-Y)^{-1}}{(12)(23)\cdots(n \splus 1 \ssep n \splus 2)(n \splus 2 \ssep 1)} 
  \delta^{4}(C'\!\cdot\vltuu) \, \delta^{8}(C'\!\cdot\vleuu) \, \delta^{2n}(C'^\perp\!\cdot\vlluu)\eqncom
\end{aligned}
\end{equation}
where
\begin{equation}
  Y=\frac{(n \ssep n \splus 1)(n \splus 2 \ssep 1)}{(n \ssep n \splus 2)(n \splus 1 \ssep 1)}  \eqndot
\end{equation}
We checked up to $n=6$ that, after integrating out the four additional delta functions as in \eqref{eq: reduction MHV}, the form reduces to one from section \ref{subsec: MHV Grassmannian}: 
\begin{equation}
  \frac{Y(1-Y)^{-1}}{(12)(23)\cdots(n \ssep n \splus 1)(n \splus 1 \ssep n \splus 2)(n \splus 2 \ssep 1)} \Bigg\vert_{C'}
  \quad\!\!\longrightarrow\quad\frac{1}{(12)(23)\cdots(n \sminus 1 \ssep n)(n 1)} \Bigg\vert_{C}
  \eqndot
\end{equation}

\subsubsection*{Three-point NMHV}

\begin{figure}[htbp]
\centering
\begin{align*}
I=
\begin{aligned}
        \begin{tikzpicture}[scale=0.8]
        \draw (1,-1) -- (0,-0.7);
        \draw (1,-1) -- (2,-0.7);
        \draw (1,-1) -- (1,-1-0.65);
        \draw (0,-0.7) -- (0-0.3,-0.7-0.6);
        \draw (2,-0.7) -- (2+0.3,-0.7-0.6);
        \draw (0,-0.7) -- (0+0.3,-0.7+0.6);
        \draw (2,-0.7) -- (2-0.3,-0.7+0.6);
        \node[dw] at (1,-1) {};
        \node[db] at (0,-0.7) {};
        \node[db] at (2,-0.7) {};
                         \node[] at (-0.45,-1.6) {3};
                         \node[] at (2.45,-1.6) {1};
                         \node[] at (0.4,0.2) {4};
                         \node[] at (1.6,0.2) {5};
                         \node[] at (1,-2) {2};
        \end{tikzpicture}
        \end{aligned}
\eqncom        
\qquad
    \sigmapart=(5,4,1,3,2)
\eqncom    
\qquad    
    C=\begin{pmatrix}
                1&\alpha_3&\alpha_2&0&0\\
                0&0&1&\alpha_1&0\\
                \alpha_4&0&0&0&1
        \end{pmatrix}
  \eqndot
\end{align*}
\caption{On-shell sub-diagram $I$ obtained by removing the minimal form factor from the on-shell diagram of $\ff_{3,3}$ shown in figure~\ref{fig: F3,3}, corresponding permutation $\sigmapart$ and $C$ matrix.}
\label{fig:gluing33}
\end{figure}

The simplest NMHV form factor is $\ff_{3,3}$. 
Diagrammatically, it can be obtained from a $k$-increasing inverse soft limit of the minimal form factor.
Using the general gluing procedure outlined above, we find after a change of variables the following Graßmannian integral representation:
\begin{equation}
\label{eq: ff 3,3 Grassmannian}
  \abr{\refspinal\refspinbl}^2
  \int\frac{\dd^{3\times 5}C'}{\text{Vol}[GL(3)]}\;
  \frac{Y(1-Y)^{-1}}{(123)(234)(345)(451)(512)}\;
  \delta^{6}(C'\cdot\vltuu) \, \delta^{12}(C'\cdot\vleuu) \, \delta^{4}(C'^\perp\cdot\vlluu) \eqncom
\end{equation}
where
\begin{equation}
  Y=\frac{(234)(512)}{(235)(412)}
\end{equation}
and the off-shell (super) momentum is encoded in the on-shell variables at position 4 and 5; cf.\ \eqref{eq: def underunderscore variables}.

Let us now evaluate the Graßmann integral \eqref{eq: ff 3,3 Grassmannian}.
After gauge fixing, the matrix $C'$ reads
\begin{equation}
 C'=
 \begin{pmatrix}
  1 & 0 & 0 & c'_{14} & c'_{15} \\
  0 & 1 & 0 & c'_{24} & c'_{25} \\
  0 & 0 & 1 & c'_{34} & c'_{35} 
 \end{pmatrix}
\eqndot
\end{equation}
We can solve for $c'_{i4}$ and $c'_{i5}$ with $i=1,2,3$ by contracting the terms inside $\delta^{6}(C'\cdot\vltuu)$ with $\vltuu_4$ and $\vltuu_5$. This yields
\begin{equation}
\label{eq: c solution ff 3,3}
 c'_{i4}=-\frac{\sbr{i5}}{\sbr{45}}=-\frac{\bra{\refspinal}q|i]}{q^2}\eqncom \qquad 
 c'_{i5}=-\frac{\sbr{i4}}{\sbr{54}}=-\frac{\bra{\refspinbl}q|i]}{q^2}\eqncom
\end{equation}
where we have used \eqref{eq: def underunderscore variables} in the second step.
Inserting \eqref{eq: c solution ff 3,3} into $\delta^{4}(C'^\perp\cdot\vlluu)$, we obtain the momentum-conserving delta function contracted with $\vltuu_4$ and $\vltuu_5$.
Undoing this contraction yields a Jacobian of $\sbr{45}^{2}$, which, together with the Jacobian $\sbr{45}^{-3}$ from the previous contraction with $\vltuu_4$ and $\vltuu_5$, gives $\sbr{45}^{-1}$.%
\footnote{Note that we have dropped the double underscore in the notation for the spinor brackets.}

Inserting the solutions \eqref{eq: c solution ff 3,3} into \eqref{eq: ff 3,3 Grassmannian} and applying the Schouten identity, we find
\begin{equation}
\label{eq: final expression ff3,3 Grassmann evaluation}
 \begin{aligned}
  &\abr{\refspinal\refspinbl}^2
  \int\frac{\dd^{3\times 5}C'}{\text{Vol}[GL(3)]}\;
  \frac{Y(1-Y)^{-1}}{(123)(234)(345)(451)(512)}\;
  \delta^{6}(C'\cdot\vltuu) \, \delta^{12}(C'\cdot\vleuu) \, \delta^{4}(C'^\perp\cdot\vlluu) \\
  &=\frac{(q^2)^2}{\sbr{12}\sbr{23}\sbr{31}}\delta^{12}(C'\cdot\vleuu) \, \delta^{4}(\sum_{i=1}^3p_i-q)\eqncom
 \end{aligned}
\end{equation}
which agrees with the result of \cite{Brandhuber:2011tv}.

Note that the cyclic invariance of the form factor is not manifest in \eqref{eq: ff 3,3 Grassmannian}. The final expression \eqref{eq: final expression ff3,3 Grassmann evaluation} obtained from its evaluation, however, is manifestly invariant under cyclic relabelling of the legs $1$, $2$ and $3$, as can be seen from \eqref{eq: c solution ff 3,3}.

\subsubsection*{Four-point NMHV}

\begin{figure}[htbp]
  \centering
\begin{align*}
I&=\begin{aligned}
 \begin{tikzpicture}[scale=0.8]
 			\draw (+\hdist,1.4*\hdist) -- (+\hdist,0);
 			\draw (-\hdist,1.4*\hdist) -- (-\hdist,0);
 			\draw (+\hdist,1.4*\hdist) -- (+\hdist,0);
 			\draw (-\hdist,1.4*\hdist) -- (-\hdist,0);
 			\draw (+\hdist,0) -- (3*\hdist,0) -- (4*\hdist,\hdist);
 			\draw (-\hdist,0) -- (-3*\hdist,0) -- (-4*\hdist,\hdist);
 			\draw (4*\hdist,-3*\hdist) -- (3*\hdist,-2*\hdist) -- (-3*\hdist,-2*\hdist) -- (-4*\hdist,-3*\hdist);
 			\draw (-\hdist,0) -- (-\hdist,-2*\hdist);
			\draw (-3*\hdist,0) -- (-3*\hdist,-2*\hdist);
 			\draw (\hdist,0) -- (\hdist,-2*\hdist);
			\draw (3*\hdist,0) -- (3*\hdist,-2*\hdist);
                         \node[db] at (+\hdist,0) {}; 
                         \node[dw] at (+3*\hdist,0) {};
                         \node[dw] at (-\hdist,0) {}; 
                         \node[db] at (-3*\hdist,0) {}; 
                         \node[dw] at (+\hdist,-2*\hdist) {}; 
                         \node[db] at (+3*\hdist,-2*\hdist) {};
                         \node[db] at (-\hdist,-2*\hdist) {}; 
                         \node[dw] at (-3*\hdist,-2*\hdist) {}; 
                         \node[] at (-4*\hdist-0.2,\hdist+0.2) {4};
                         \node[] at (4*\hdist+0.2,\hdist+0.2) {1};
                         \node[] at (-4*\hdist-0.2,-3*\hdist-0.2) {3};
                         \node[] at (4*\hdist+0.2,-3*\hdist-0.2) {2};
                         \node[] at (+\hdist,1.4*\hdist+\labelvdist) {6};
                         \node[] at (-\hdist,1.4*\hdist+\labelvdist) {5};
\end{tikzpicture}
\end{aligned}
\eqncom
\qquad
\sigmapart=(4,6,5,1,3,2)\eqncom \\
        C&=\begin{pmatrix}
                1&\alpha_2+\alpha_4&0&-\alpha_2\alpha_3&-\alpha_2\alpha_3\alpha_6&0\\
                0&1&0&-\alpha_3&-\alpha_3\alpha_6&-\alpha_1\\
                0&0&1&\alpha_5+\alpha_7&\alpha_5\alpha_6&0
        \end{pmatrix}
        \eqndot
\end{align*}
\caption{On-shell sub-diagram $I$ obtained by removing the minimal form factor from the top-cell diagram of $\ff_{4,3}$ shown in figure \ref{fig: F4,3}, corresponding permutation $\sigmapart$ and $C$ matrix.}
\label{fig:gluing43}
\end{figure}

As discussed in subsection \ref{subsec: on-shell diagrams and r operators}, the four-point NMHV form factor is the first example of a form factor for which it appears natural to combine different BCFW terms diagrammatically into a top-cell diagram with additional edges.
Note that the gluing procedure outlined in the beginning of this subsection together with the connection \eqref{eq: box eater} will indeed generally lead to a top-dimensional integral over the Graßmannian $G(k,n+2)$.

From the general expression \eqref{eq: general Grassmannian gluing formula} and the on-shell diagram and $C$ matrix shown in figure~\ref{fig:gluing43}, 
we obtain a result that can be written in the following form:
\begin{equation}
  \abr{\refspinal\refspinbl}^2
  \int\frac{\dd^{3\times 6}C'}{\text{Vol}[GL(3)]}\;\,
  \Omega_{4,3} \;\,
  \delta^{6}(C'\cdot\vltuu) \, \delta^{12}(C'\cdot\vleuu)\, \delta^{6}(C'^\perp\cdot\vlluu)\eqncom
\end{equation}
where
\begin{equation}
  \Omega_{4,3} = \frac{Y(1-Y)^{-1}}{(123)(234)(345)(456)(561)(612)} \eqncom
  \qquad
  Y=\frac{(345)(612)}{(346)(512)}
  \eqndot
\end{equation}

Gluing the same diagram at legs 2 and 3 and relabelling to obtain the other top-cell diagram, we find with the same $\vlluu$, $\vltuu$, $\vleuu$ as for the first diagram:
\begin{equation}
  \abr{\refspinal\refspinbl}^2
  \int\frac{\dd^{3\times 6}C'}{\text{Vol}[GL(3)]}\;\,
\Bigg(  \Omega_{4,3}\Big\vert_{\tiny\begin{pmatrix}
      1&2&3&4&5&6\\ 
      \downarrow&\downarrow&\downarrow&\downarrow&\downarrow&\downarrow\\
      3&4&1&2&5&6
    \end{pmatrix}    
    }\Bigg)
  \;
  \delta^{6}(C'\cdot\vltuu) \, \delta^{12}(C'\cdot\vleuu) \, \delta^{6}(C'^\perp\cdot\vlluu)
  \eqndot
\end{equation}

\subsubsection*{Conjecture for all $n,k$}

Based on the (conjectured) relation \eqref{eq: box eater} between the top-cell diagrams of amplitudes and form factors for generic $n,k$,%
\footnote{Above, we have explicitly checked that this conjecture leads to the correct result for $k=2$, $k=n=3$ and $k=n-1=3$. Further checks will be given below.}
we have computed the Graßmannian integrals for all form factor top-cell diagrams up to six points, with the minimal form factor glued at positions $n+1$ and $n+2$. 
The on-shell diagrams that need to be glued in this case are labelled by the permutation given in \eqref{eq: permutation diagram to be glued}.
We have invariably found the following representation:
\begin{equation}
\label{eq: general grassmannian}
  \abr{\refspinal\refspinbl}^2
  \int\frac{\dd^{k\times(n+2)}C'}{\text{Vol}[GL(k)]}\;\,
  \Omega_{n,k}\,\;
  \delta^{2\times k}(C'\cdot\vltuu) \, \delta^{4\times k}(C'\cdot\vleuu) \, \delta^{2\times(n+2-k)}(C'^\perp\cdot\vlluu)\eqncom
\end{equation}
where
\begin{equation}
\label{eq: general grassmannian form}
  \begin{aligned}
    \Omega_{n,k} &=
  \frac{Y(1-Y)^{-1}}{(1\cdots k)(2\cdots k \splus 1)\cdots(n \cdots k \sminus 3)(n \splus 1 \cdots k \sminus 2)(n \splus 2\cdots k \sminus 1)} 
  \eqncom
  \\[0.5em]
  Y&=\frac{
  (n \sminus k \splus 2 \cdots n \ssep n \splus 1)(n \splus 2 \ssep 1 \cdots k \sminus 1)
}{
  (n \sminus k \splus 2 \cdots n \ssep n \splus 2)(n \splus 1 \ssep 1 \cdots k \sminus 1)
  } 
\end{aligned}
\end{equation}
and the off-shell data is encoded in the kinematical variables at the position $n+1$ and $n+2$ as in \eqref{eq: def underunderscore variables}.%
\footnote{The quotient $Y$ always corresponds to the product  $\beta_1\beta_2$ from
the gluing procedure \eqref{eq: general Grassmannian gluing formula} and thus the factor $Y(1-Y)^{-1}$ always cancels a factor of $[\beta_1\beta_2(1-\beta_1\beta_2)]^{-1}$ that arise when the 
consecutive minors are translated into edge variables $\alpha_i$, $\beta_1$ and $\beta_2$ and that is not present in \eqref{eq: general Grassmannian gluing formula}.}
In general, up to $n$ copies of this form have to be considered, which arise from cyclic shifts in the labels $1$ to $n$.

We will present further checks for this conjecture in the next subsection.

\paragraph{Note on deformations}
In contrast to the MHV form factors considered in section \ref{sec: MHV}, 
we did not employ the method of $\rr$ operators for more general form factors and hence we have so far not included deformations in the Graßmannian for N$^k$MHV as counterpart to what was done for amplitudes in \cite{Bargheer:2014mxa,Ferro:2014gca}. 
The reason we preferred a direct gluing procedure is that it immediately leads to a top-dimensional integral over the Graßmannian. 
The method of $\rr$ operators, while still applicable, will in general result in a Graßmannian integral with some of the delta functions already integrated out. 
Nevertheless, a general expression for the deformed top-dimensional form could be obtained in this way; we leave this for future work.

\subsection{Twistor and momentum twistor Graßmannians}

Next, let us transform the previously obtained Graßmannian integral representation from spinor helicity to twistor and momentum twistor variables. 
Throughout this subsection, all kinematic variables are defined as in \eqref{eq: def underunderscore variables} and \eqref{eq: def underunderscore variables etas}. In order to facilitate notation, we will hence drop the double underscore from all spinor (and twistor) brackets.

\subsubsection*{Twistor space}

Given the Graßmannian integral \eqref{eq: general grassmannian} in momentum space, we can transform it to twistor space in analogy to what was done in the amplitude case in \cite{ArkaniHamed:2009dn}.

The super twistor space we use here corresponds to our special choice of spinor helicity variables \eqref{eq: def underunderscore variables} and \eqref{eq: def underunderscore variables etas} and is given by the set of all super twistors
\begin{equation}
 \ptwistor_i=(\vmtuuu_i,\vltuuu_i,\vleuuu_i)\eqncom
\end{equation}
where $\vmtuuu_i$ is related to $\vlluuu_i$ via Witten's half Fourier transformation \cite{Witten:2003nn} 
\begin{equation}
\label{eq: half Fourier transformation}
 \bullet \to \int \de^2 \vlluuu_j \exp(-i\vmtuuu^\alpha_j \vlluuu_{j\alpha}) \bullet \eqndot
\end{equation}
Via \eqref{eq: half Fourier transformation}, the prefactor in \eqref{eq: general grassmannian} can be written as
\begin{equation}
 \abr{\refspinal\refspinbl}^2
 =\abr{\frac{\partial}{\partial \vmtuuu_{n+1}}\frac{\partial}{\partial \vmtuuu_{n+2}}}^2 \eqndot
\end{equation}
The delta function $\delta^{2\times(n+2-k)}(C'^\perp\cdot\vlluuu)$ can be written as in \eqref{eq: delta function C perp alternative for Jacobian}.
Applying \eqref{eq: half Fourier transformation} to this representation and performing the integrals over $\vlluuu_i$ via the delta functions, we find 
\begin{equation}
 \prod_{K=1}^k \int \de^2\rho_K \exp(-i\sum_{j=1}^{n+2}\sum_{L=1}^k\rho^\alpha_{L}C'_{Lj}\vmtuuu_{\alpha j})=\delta^{2k}(C'\cdot\vmtuuu)\eqndot
\end{equation}
Hence, we can write \eqref{eq: general grassmannian} as 
\begin{equation}
\label{eq: general grassmannian position twistor}
  \abr{\frac{\partial}{\partial \vmtuuu_{n+1}}\frac{\partial}{\partial \vmtuuu_{n+2}}}^2
  \int\frac{\dd^{k\times(n+2)}C'}{\text{Vol}[GL(k)]}\;\,
  \Omega_{n,k}\,\;
  \delta^{4 k| 4 k }(C'\cdot\ptwistor)\eqncom
  \end{equation}
where $ \Omega_{n,k}$ is given in \eqref{eq: general grassmannian form}.

It would be interesting to further investigate the structure of this expression; we leave this
for future work. Instead, we will now transform the Graßmannian integral to momentum
twistor space, which will in particular facilitate the explicit calculation of some
example form factors.

\subsubsection*{Momentum twistor space}

Next, we transform our result to momentum twistor space following the strategy of \cite{ArkaniHamed:2009vw,Elvang:2014fja}.
\begin{figure}[htbp]
\begin{equation*}
        \begin{tikzpicture}[scale=0.8]
        \draw[draw=none,fill=black!20, fill opacity=0.15] (-0.4,-1.7) rectangle (6.4,3.7);
        \draw[draw=none,fill=black!20, fill opacity=0.15] (2.2,-0.4) rectangle (8.9,4.4);
        \draw[dotted] (-0.8,1.2) -- (0,0);
        \draw[<-] (0,0) -- (1.7,3);
        \draw[<-] (1.7,3) -- (2.6,1.9);
        \draw[<-] (2.6,1.9) -- (3.9,3.3);
        \draw[<-] (3.9,3.3) -- (6,0);
        \draw[<-] (6,0) -- (7.7,3);
        \draw[<-] (7.7,3) -- (8.6,1.9);
        \draw[<-] (8.6,1.9) -- (9.9,3.3);
        \draw[<-] (9.9,3.3) -- (12,0);
        \draw[dotted,<-] (12,0) -- (12.8,1.2);
        \draw[dashed,<-] (2.2,-1.3) -- (0,0);
        \draw[dashed,<-] (6,0) -- (2.2,-1.3);
        \draw[dashed,<-] (4.8,0.6) -- (2.6,1.9);
        \draw[dashed,<-] (8.6,1.9) -- (4.8,0.6);
        \node[label={[label distance=-7pt]135:$p_1$}] at (0.85,1.5) {};
        \node[label={[label distance=-8pt]45:$p_2$}] at (2.15,2.45) {};
        \node[label={[label distance=-7pt]340:$p_3$}] at (3.25,2.6) {};
        \node[label={[label distance=-2pt]360:$p_4$}] at (4.95,1.65) {};
        \node[label={[label distance=-2pt]354:$p_5$}] at (4.1,-0.65) {};
        \node[label={[label distance=-2pt]270:$p_6$}] at (1.1,-0.65) {};
        \node[label={[label distance=-6pt]220:$y_1$}] at (0,0) {};
        \node[label={[label distance=-2pt]90:$y_2$}] at (1.7,3) {};
        \node[label={[label distance=-2pt]270:$y_3$}] at (2.6,1.9) {};
        \node[label={[label distance=-2pt]90:$y_4$}] at (3.9,3.3) {};
        \node[label={[label distance=-4pt]330:$y_5$}] at (6,0) {};
        \node[label={[label distance=-2pt]270:$y_6$}] at (2.2,-1.3) {};
    \end{tikzpicture}
\end{equation*}
\caption{Momenta and dual momenta in the case of form factors, shown for $n=4$. In contrast to the case of amplitudes, the $n$ on-shell momenta do not add up to zero but to the off-shell momentum $q$ of the operator. Hence, the contour is not closed but periodic. In order to obtain a closed contour, two on-shell momenta $p_{n+1}$ and $p_{n+2}$ can be inserted between any $y_i$ and $y_{i+1}$ of the periodic contour. Two different choices are shown in shaded frames.}
\label{fig: dual contour}
\end{figure}
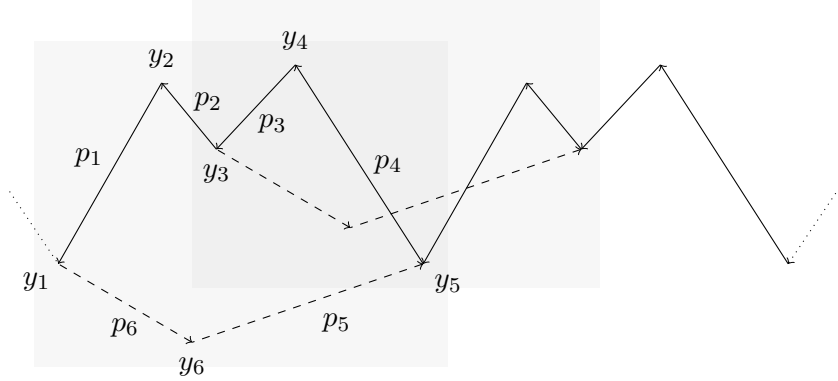

In order to introduce the momentum twistor variables $\mtwistor_i=(\vlluuu_i,\vmmuuu_i,\vletuuu_i)$ \cite{Hodges:2009hk} corresponding to our choice of variables \eqref{eq: def underunderscore variables} and \eqref{eq: def underunderscore variables etas}, we define the dual (super) momenta $y_i$ ($\dualsupermomentum_i$) via%
\footnote{We use the definitions of \cite{Elvang:2014fja}, which coincide with the ones of \cite{Brandhuber:2011tv} but differ from the ones of \cite{Bork:2014eqa} by a global sign and a cyclic relabelling.}
\begin{equation}
 \vlluuu_i\vltuuu_i=y_i-y_{i+1}\eqncom \qquad \vlluuu_i\vleuuu_i=\dualsupermomentum_i-\dualsupermomentum_{i+1} \eqndot
\end{equation}
Note that we base the dual (super) momenta on the closed contour obtained by adding $p_{n+1}$ and $p_{n+2}$ instead of the periodic contour as done in \cite{Brandhuber:2011tv,Bork:2014eqa}; cf.\ figure \ref{fig: dual contour}. 
Then, we define $\vmmuuu_i$ and $\vletuuu_i$ via the incidence relations 
\begin{equation}
 \vmmuuu_i=\vlluuu_iy_i=\vlluuu_iy_{i+1}\eqncom \qquad \vletuuu_i=\vlluuu_i\dualsupermomentum_i=\vlluuu_i\dualsupermomentum_{i+1}\eqndot 
\end{equation}
Inverting these relations, we have
\begin{equation}
\label{eq: mtwistor inverted}
 \begin{aligned}
  \vltuuu_i&=\frac{\abr{i\splus1\,i}\vmmuuu_{i-1}+\abr{i\,i\sminus1}\vmmuuu_{i+1}+\abr{i\sminus1\, i\splus1}\vmmuuu_{i}}{\abr{i\sminus1\,i}\abr{i\,i\splus1}}\eqncom \\
  \vleuuu_i&=\frac{\abr{i\splus1\,i}\vletuuu_{i-1}+\abr{i\,i\sminus1}\vletuuu_{i+1}+\abr{i\sminus1\, i\splus1}\vletuuu_{i}}{\abr{i\sminus1\,i}\abr{i\,i\splus1}}\eqndot
 \end{aligned}
\end{equation}

We start the transformation of the Graßmannian integral from momentum space using the representation of $\delta^{2\times(n+2-k)}(C'^\perp\cdot\vlluuu)$ as \eqref{eq: delta function C perp alternative for Jacobian}. We can use part of the $GL(k)$ redundancy to fix 
\begin{equation}
 \rho=\begin{pmatrix}
       0 & \cdots & 0 & 1 & 0 \\
       0 & \cdots & 0 & 0 & 1 \\ 
      \end{pmatrix}
      \eqndot
\end{equation}
As a consequence, the delta functions in \eqref{eq: delta function C perp alternative for Jacobian} fix the last two rows of $C'$ as
\begin{equation}
 C'_{k\sminus1 \, i}=\vlluuu^1_i\eqncom \qquad C'_{k \, i}=\vlluuu^2_i\eqndot 
\end{equation}
Then, \eqref{eq: general grassmannian} becomes
\begin{equation}
  \abr{\refspinal\refspinbl}^2
  \delta^{4}(\vlluuu\cdot\vltuuu) \, \delta^{8}(\vlluuu\cdot\vleuuu) 
  \int\frac{\dd^{(k-2)\times(n+2)}C'}{\text{Vol}[GL(k-2)\ltimes T_{k-2}]}\;\,
  \Omega_{n,k}\,\;
  \delta^{2\times (k-2)}(C'\cdot\vltuuu) \, \delta^{4\times (k-2)}(C'\cdot\vleuuu)\eqncom
\end{equation}
where the integral and the delta functions contain only the first $k-2$ rows of $C'$. 
The shift symmetry $T_{k-2}$ acts on these $k-2$ rows as
\begin{equation}
 C'_{Ii}\longrightarrow C'_{Ii}+r_{1I}\vlluuu_i^1+r_{2I}\vlluuu_i^2\eqncom \quad I=1,\dots, k-2\eqncom
\end{equation}
with $r_{1I}$, $r_{2I}$ arbitrary.
According to \cite{Elvang:2014fja}, \eqref{eq: mtwistor inverted} leads to 
\begin{equation}
 \sum_{i=1}^{n+2}C'_{Ii}\vltuuu_i=-\sum_{i=1}^{n+2}D_{Ii}\vmmuuu_i \eqncom \qquad
 \sum_{i=1}^{n+2}C'_{Ii}\vleuuu_i=-\sum_{i=1}^{n+2}D_{Ii}\vletuuu_i \eqncom \qquad I=1,\dots, k-2\eqncom
\end{equation}
where the matrix $D$ is defined via
\begin{equation}
 D_{Ii}=\frac{\abr{i\, i\splus1}C'_{I\,i\sminus1}+\abr{i\sminus1\,i}C'_{I\,i\splus1}+\abr{i\splus1\,i\sminus1}C'_{I\,i}}{\abr{i\sminus1\,i}\abr{i\,i\splus1}}\eqndot
\end{equation}
Next, we rewrite the minors of $C'$ in terms of minors of $D$. In \cite{Elvang:2014fja}, it was found that the consecutive minors are related as
\begin{equation}
\label{eq: old relation}
 (C'_1\dots C'_{k})=-\abr{1\,2}\cdots\abr{k\sminus1\,k}(D_2\dots D_{k\sminus1})
\end{equation}
and its natural extension via cyclic shifts.
However, we do also need non-consecutive minors, as can be seen from \eqref{eq: general grassmannian form}. For these, we find
\begin{equation}
\label{eq: new relation}
\begin{aligned}
 (C'_1\dots C'_{k-1}C'_{k+1})&=
 -\abr{1\,2}\cdots\abr{k\sminus2\,k\sminus1}\abr{k\sminus1\,k\splus1}(D_2\dots D_{k\sminus1})\\
 &\phaneq-\abr{1\,2}\cdots\abr{k\sminus2\,k\sminus1}\abr{k\,k\splus1}(D_2\dots D_{k\sminus2}D_{k})\eqncom\\
  (C'_1C'_3 \dots C'_{k+1})&=
 -\abr{1\,3}\abr{3\,4}\cdots\abr{k\,k\splus1}(D_3\dots D_{k})\\
 &\phaneq-\abr{1\,2}\abr{3\,4}\cdots\abr{k\,k\splus1}(D_2D_4\dots D_{k})\eqndot
\end{aligned}
\end{equation}
Using \eqref{eq: old relation}, the product of consecutive minors in \eqref{eq: general grassmannian form} becomes
\begin{equation}
 (1\cdots k)_{C'}\cdots(n \splus 2\cdots k \sminus 1)_{C'}=(-1)^{n+2}(\abr{1\,2}\cdots\abr{n\splus2\,1})^{k-1}(1\cdots k)_{D}\cdots(n \splus 2\cdots k \sminus 1)_{D}\eqndot
\end{equation}
For $Y$, we find using \eqref{eq: new relation}
\begin{equation}
\label{eq: Y mtwistor}
\begin{aligned}
 Y&=\frac{
  (n \sminus k \splus 2 \cdots n \ssep n \splus 1)_{C'}(n \splus 2 \ssep 1 \cdots k \sminus 1)_{C'}
}{
  (n \sminus k \splus 2 \cdots n \ssep n \splus 2)_{C'}(n \splus 1 \ssep 1 \cdots k \sminus 1)_{C'}
  } \\
&=\frac{
  \abr{n\,n\splus1}(n \sminus k \splus 3 \cdots n )_{D}
}{
  \abr{n\,n\splus2}(n \sminus k \splus 3 \cdots n )_{D}+\abr{n\splus1\,n\splus2}(n \sminus k \splus 3 \cdots n\sminus1 \, n\splus1)_{D}
  } \\
  &\phaneq
  \frac{
  \abr{n\splus2\,1}(1 \cdots k \sminus 2)_{D}
  }{
    \abr{n\splus1\,1}(1 \cdots k \sminus 2)_{D}
    +\abr{n\splus1\,n\splus2}(n\splus2\, 2 \cdots k \sminus 2)_{D}
  }
  \eqndot
\end{aligned}
\end{equation}

The remaining steps in the derivation of \cite{Elvang:2014fja} go through unchanged. 
First, we use the $T_{k-2}$ shift symmetry to set $C'_{I1}=C'_{I2}=0$, which changes the measure as
\begin{equation}
 \frac{\dd^{(k-2)\times(n+2)}C'}{\text{Vol}[GL(k-2)\ltimes T_{k-2}]} = \abr{12}^{k-2}\frac{\dd^{(k-2)\times(n)}C'}{\text{Vol}[GL(k-2)]} \eqndot
\end{equation}
Then, we perform the change of integration variables from $C'$ to $D$, which yields
\begin{equation}
 \frac{\dd^{(k-2)\times(n)}C'}{\text{Vol}[GL(k-2)]}=\left(\frac{\abr{12} \cdots \abr{n\splus2\,1}}{\abr{12}^2}\right)^{k-2}\frac{\dd^{(k-2)\times(n)}D}{\text{Vol}[GL(k-2)]}\eqndot
\end{equation}
Finally, we undo the gauge fixing of the first two columns of the $C'$ matrix, which yields factors of
\begin{equation}
 \abr{12}\delta^2(D_{Ii}\vlluuu_i)
\end{equation}
 for $I=1,\dots,k-2$.
See \cite{Elvang:2014fja} for details of these steps.

The final expression we find is
\begin{equation}
 \ff_{n,2}
  \int\frac{\dd^{(k-2)\times(n+2)}D}{\text{Vol}[GL(k-2)]}\;\,
  \Omega_{n,k}\,\;
  \delta^{4(k-2)|4(k-2)}(D\cdot\mtwistor) \eqncom
\end{equation}
where
\begin{equation}
\label{eq: general grassmannian form mtwistor}
  \begin{aligned}
    \Omega_{n,k} &=
  \frac{\abr{n\,1}\abr{n\splus1\,n\splus2}}{ \abr{n\,n\splus1}  \abr{n\splus2\,1}}
  \frac{Y(1-Y)^{-1}}{(1\cdots k\sminus2)(2\cdots k \sminus 1)\cdots(n \cdots k \sminus 5)(n \splus 1 \cdots k \sminus 4)(n \splus 2\cdots k \sminus 3)} 
\end{aligned}
\end{equation}
and $Y$ is given in \eqref{eq: Y mtwistor}.

\subsubsection*{A convenient choice of reference spinors}

It turns out that one choice of reference spinors $\refspinal$, $\refspinbl$ is particularly convenient.
If we set
\begin{equation}
   \refspinal \equiv \vlluuu_{n+1} = \vll_1
    \eqncom\qquad
   \refspinbl \equiv \vlluuu_{n+2} = \vll_n
    \eqncom
    \label{eq: convenient ref spinors}
\end{equation}
the above momentum twistor Graßmannian integral for N$^k$MHV becomes
\begin{equation}
\label{eq: Grassmannian integral momentum twistors gauged}
 \begin{aligned}
     {\ff_{n,2}}
  \int\frac{\dd^{(k-2)\times(n+2)}D}{\text{Vol}[GL(k-2)]}\;\,
     \frac{-\tilde{Y}(1-\tilde{Y})^{-1}\;\delta^{4(k-2)|4(k-2)}(D\cdot\mtwistor)}{(1\cdots k\sminus 2)\cdots(n \cdots k \sminus 5)(n \splus 1 \cdots k \sminus 4)(n \splus 2\cdots k \sminus 3)} 
 \eqncom
   \end{aligned}
\end{equation}
with
\begin{equation}
    \tilde{Y}=\frac{(n\sminus k \splus 3 \cdots n)(1\cdots k \sminus 2)}{(n \sminus k \splus 3 \cdots n \sminus 1 \ssep n \splus 1)(n\splus 2\ssep 2 \cdots k\sminus 2)}
    \eqndot
\end{equation}

\subsubsection*{Examples at MHV, NMHV and NNMHV}

Let us look at some special cases of the above Graßmannian integral representation.
For $k=2$, the matrix $D$ is zero-dimensional and all consecutive minors of $D$ are $1$ whereas all non-consecutive minors are $0$. Hence, the integral in \eqref{eq: Grassmannian integral momentum twistors gauged} is zero-dimensional while the integrand is $1$.%
\footnote{Note that, although $\tilde{Y}$ is singular when inserting the above values for consecutive and non-consecutive minors, the ratio $\frac{-\tilde{Y}}{1-\tilde{Y}}=\frac{-(n\sminus k \splus 3 \cdots n)(1\cdots k \sminus 2)}{(n \sminus k \splus 3 \cdots n \sminus 1 \ssep n \splus 1)(n\splus 2\ssep 2 \cdots k\sminus 2)-(n\sminus k \splus 3 \cdots n)(1\cdots k \sminus 2)}$ is $1$ in this case.}
Considering the prefactor $\ff_{n,2}$, this is precisely the correct result.

For $k=3$, 
\begin{equation}
D=
\begin{pmatrix}
 d_1 & d_2 & \cdots & d_{n+2}
\end{pmatrix}
\eqndot
\end{equation}
The consecutive minors of $D$ are equal to the single $d_i$ included in them and the non-consecutive minors are equal to the $d_i$ that is alone on its side of the gap, cf.\ \eqref{eq: new relation}.
Hence, the Graßmannian integral \eqref{eq: Grassmannian integral momentum twistors gauged} becomes
 \begin{equation}
 \label{eq: NMHV Grassmann integral gauged}
 \begin{aligned}
  \ff_{n,2}
   \int\frac{\dd^{1\times(n+2)}D}{\text{Vol}[GL(1)]}\;\,
    \frac{1}{1-\frac{d_{n+1}d_{n+2}}{d_1 d_n}}
  \frac{1}{d_1\cdots d_{n}}\frac{1}{d_{n+1}d_{n+2}}
   \delta^{4|4}(D\cdot\mtwistor) \eqndot
   \end{aligned}
 \end{equation}
Note that in all of the examples considered in this subsection we will use the convenient choice of reference spinors
\eqref{eq: convenient ref spinors} to obtain compact expressions. We have explicitly 
checked that our results are indeed independent of this choice.

 The simplest example for $k=3$ is $n=3$: 
 \begin{equation}
 \begin{aligned}
  \ff_{3,2}
   \int\frac{\dd^{1\times5}D}{\text{Vol}[GL(1)]}\;\,
    \frac{1}{
    1-\frac{d_{4}d_{5}}{d_1 d_3}
 }
  \frac{1}{d_1 d_2 d_3 d_{4} d_{5}}
   \delta^{4|4}(d_1 \mtwistor_1+d_2 \mtwistor_2+d_3 \mtwistor_3+d_4 \mtwistor_4+d_5 \mtwistor_5) \eqndot
   \end{aligned}
 \end{equation}
We can use the $GL(1)$ redundancy to fix $d_5=\abr{1\,2\,3\,4}$,
where the four-bracket is defined as 
\begin{equation}
 \abr{i\,j\,k\,l}=\det(Z_iZ_jZ_kZ_l)=\epsilon_{ABCD}Z_i^AZ_j^BZ_k^CZ_l^D 
\end{equation}
with $Z_i=(\vlluuu_i,\vmmuuu_i)$.
The remaining four integration variables are then completely determined by the delta function:
\begin{equation}
 d_1=\abr{2\,3\,4\,5}\eqncom \quad d_2=\abr{3\,4\,5\,1}\eqncom \quad d_3=\abr{4\,5\,1\,2}\eqncom \quad d_4=\abr{5\,1\,2\,3}\eqndot
\end{equation}
Thus,
\begin{equation}
 \begin{aligned}
  \ff_{3,3}=\ff_{3,2}
    \frac{\sbr{1\,2\,3\,4\,5}}{
    1-  \frac{\abr{5\,1\,2\,3}\abr{1\,2\,3\,4}}{\abr{2\,3\,4\,5}\abr{4\,5\,1\,2}}
  } \eqncom
   \end{aligned}
 \end{equation}
where the five-bracket is defined as 
\begin{equation}
 \sbr{i\,j\,k\,l\,m}=\frac{\delta^4(\abr{i\,j\,k\,l}\vlet_m+\text{cyclic})}{\abr{i\,j\,k\,l}\abr{j\,k\,l\,m}\abr{k\,l\,m\,i}\abr{l\,m\,i\,j}\abr{m\,i\,j\,k}}\eqndot
\end{equation}
This result numerically agrees with the one found in \cite{Bork:2014eqa}.

For general $n$, the denominator of \eqref{eq: NMHV Grassmann integral gauged} has poles for
\begin{equation}
\begin{aligned}
 d_i&=0\eqncom\quad i=2,\dots, n-1, n+1, n+2\eqncom \\
 d_1&=\frac{d_{n+1}d_{n+2}}{d_n}
\eqncom \quad
 d_n= \frac{d_{n+1}d_{n+2}}{d_1}
 \eqncom \quad
 d_{n+1}= \frac{d_{1}d_{n}}{d_{n+2}}
\eqncom \quad
 d_{n+2}= \frac{d_{1}d_{n}}{d_{n+1}}
\eqndot
\end{aligned}
\end{equation}
In principle, one can consider (composite) residues of \eqref{eq: NMHV Grassmann integral gauged} for zero and non-zero values of the $d_i$. 
However, we find that it is sufficient to consider residues which are composed of individual residues taken at zero.%
\footnote{This can also be understood from the corresponding on-shell diagrams.}
As in the amplitude case discussed in \cite{Elvang:2014fja}, these can be characterised by the five $d_i$'s with respect to which no residues are taken.
In contrast to the amplitude case, these have to include $d_1$ and $d_n$. 
We have to consider two cases. In the first case, no residues are taken with respect to $d_{n+1}$ and $d_{n+2}$. The resulting expressions are 
\begin{equation}
\res_i= \frac{1}{1-\frac{\abr{n\splus2\,1\,n\,i}\abr{1\,n\,i\,n\splus1}}{\abr{n\,i\,n\splus1\,n\splus2} \abr{i\,n\splus1\,n\splus2\,1}}}
  \sbr{i\,n\splus1\,n\splus2\,1\,n}\eqncom
\end{equation}
where $i\in\{2,\dots,n-1\}$.
In the second case, at least one residue is taken with respect to either $d_{n+1}$ or $d_{n+2}$. The resulting expressions are 
\begin{equation}
\rest_{i,j,k}=\sbr{i\,j\,k\,1\,n}\eqncom
\end{equation}
where $i,j,k\in\{2,\dots,n-1,n+1,n+2\}$.

An additional property of form factor top-cell diagrams arising here is that we have to take the sum of more than one form. 
This can also be achieved by shifting the legs between which the minimal form factor is glued-in from $(n,1)$ to $(n+s \mod n, 1+s\mod n)$, cf.\ figure \ref{fig: dual contour}.

Numerically comparing with the results of \cite{Bork:2014eqa}, we find
\begin{equation}
\begin{aligned}
 \ff_{4,3}&=\ff_{4,2} (
    +\res_3
    +\rest_{2,3,5}
    +\res_3^{s=2}
    +\rest_{2,3,5}^{s=2})\eqncom\\
 \ff_{5,3}&=\ff_{5,2} (
    +\res_4
    +\rest_{3,4,6}
    +\rest_{2,3,6}^{s=3}
    +\res_3^{s=3}
    -\rest_{2,3,4}
    \\ & \qquad\qquad\qquad\qquad\qquad
    +\rest_{2,3,6}
    +\rest_{3,4,7}^{s=3}
    -\rest_{2,3,4}^{s=3}
    +\res_5^{s=1})\eqncom
 \end{aligned}
\end{equation}
where the superscript $s$ specifies the shift.
This also gives further support for the relation \eqref{eq: box eater}.

Finally, we look at the simplest example of $k=4$, namely $n=4$.
In this case, the matrix $D$ can be gauge-fixed to be 
\begin{equation}
 D=\begin{pmatrix}
    1&0&d_{13}&d_{14}&d_{15}&d_{16}\\
    0&1&d_{23}&d_{24}&d_{25}&d_{26}
   \end{pmatrix}
\eqndot
\end{equation}
The delta functions completely fix their entries to  
\begin{equation}
 \begin{aligned}
  d_{i3}=-\frac{\abr{i\,4\,5\,6}}{\abr{3\,4\,5\,6}}\eqncom \quad
  d_{i4}=+\frac{\abr{i\,3\,5\,6}}{\abr{3\,4\,5\,6}}\eqncom \quad
  d_{i5}=-\frac{\abr{i\,3\,4\,6}}{\abr{3\,4\,5\,6}}\eqncom \quad
  d_{i6}=+\frac{\abr{i\,3\,4\,5}}{\abr{3\,4\,5\,6}}\eqncom
 \end{aligned}
\end{equation}
where $i=1,2$.
Hence,%
\footnote{Note that there is a subtle sign occurring in the evaluation of the momentum twistor Graßmannian integral for $\ff_{4,4}$, which is also present in the case of the corresponding amplitude $\amp_{6,4}$.}
\begin{equation}
\begin{aligned}
 \ff_{4,4}
 =\ff_{4,2} \frac{\abr{1\,3\,4\,5} \abr{1\,3\,4\,6} \abr{1\,3\,5\,6} \abr{2\,3\,4\,6} \abr{2\,3\,5\,6} \abr{2\,4\,5\,6}\sbr{1\,3\,4\,5\,6}\sbr{2\,3\,4\,5\,6}}{\abr{1\,2\,3\,4} \abr{1\,2\,3\,6} \abr{3\,4\,5\,6}^2 (\abr{1\,2\,4\,6} \abr{1\,3\,4\,5}+\abr{1\,2\,5\,6} \abr{3\,4\,5\,6})}
 \eqndot
 \end{aligned}
\end{equation}
We have successfully checked (components of this expression) against \cite{Brandhuber:2011tv}.

\section{Integrability and form factors}
\label{sec: integrability}

After having already used the integrability-related $\rr$ operators
in section \ref{sec: MHV}, we now study the integrable structure of form factors in \nfsym\ theory more carefully.
Apart from linking integrability approaches from the spectral
problem and the study of amplitudes, this is also motivated by the search for symmetries.

We approach this problem by introducing the spin-chain monodromy matrix as it appeared in the context of tree-level amplitudes \cite{Frassek:2013xza,Chicherin:2013ora}. 
While the on-shell part of the form factors, studied in the previous sections and built from the $\rr$ operators \eqref{eq: action r operator}, is Yangian invariant, we find that this symmetry is broken by the insertion of the minimal form factor, i.e.\ the off-shell part.
However, the off-shell part can be interpreted as an eigenvector of the corresponding transfer matrix. 
From this we show that all form factors of the chiral stress-tensor super multiplet are annihilated by the corresponding transfer matrix.%
\footnote{For this particular super multiplet, the corresponding eigenvalue is zero.}
This is the analogue of Yangian invariance for form factors. 
Finally, we show that the transfer matrix acts diagonally on a given minimal form factor of a generic operator if the corresponding operator renormalises multiplicatively, i.e.\ is an eigenstate of the integrable Hamiltonian studied at one-loop order in the spectral problem. 
As a consequence, also certain planar leading singularities of loop-level form factors of generic operators are eigenstates of the transfer matrix.

\subsection{Spin chains and Yangian invariance}
In the following, we introduce the spin chain that appeared in the context of tree-level amplitudes \cite{Frassek:2013xza,Chicherin:2013ora,Broedel:2014pia,Kanning:2014maa}.
In the spin-chain language, the integrability construction is naturally formulated using the complex Lie algebra $\mathfrak{gl}(4|4)$ instead of $\mathfrak{psu}(2,2|4)$.

The $\mathfrak{gl}(4|4)$-invariant Lax operator relevant for the construction of the integrable spin chain naturally acts on the tensor product of two spaces and depends on the spectral parameter~$u$:  
\begin{equation}
    \lax_i(u) =
  \begin{tikzpicture}[scale=0.8,baseline=-34pt]
  \drawvline{1}{1}   
  \draw[dashed] (-0.5, -1*\vacuumheight-0.5*\bridgedistance) -- (0.5, -1*\vacuumheight-0.5*\bridgedistance);
\node[dl] at (0,-\vacuumheight-\bridgedistance-\labelvdist) {$i$};
  \end{tikzpicture}
    = u + (-1)^{|\calB|}e^{\calA\calB} \;  \hat{x}_i^{\calB}\hat{p}_i^\calA
    \eqncom
        \label{eq: def lax}
\end{equation}
where $\calA=(\alpha,\dot\alpha,A)$.
We have introduced a graphical notation, usually used in the context of vertex models, to depict the tensor structure of the Lax operator, see e.g.\ \cite{Frassek:2013xza}, and $|\cdot |$ denotes the grading.
While the auxiliary space is finite-dimensional with the generators $(e^{\calA\calB})_{\calC\calD}=\delta_\calC^\calA\delta_\calD^\calB$ 
and illustrated by the dashed horizontal line, the quantum space at site $i$ is infinite-dimensional and is denoted by the vertical line. The corresponding generators are realised in the Jordan-Schwinger form $\jj^{\calA\calB}=\hat{x}^\calA\, \hat{p}^\calB$ using the Heisenberg pairs
\begin{equation}
    \hat{x}^{\calA}= \left(\vll^\alpha,-\frac{\partial}{\partial\vlt^{\dot\alpha}} ,\frac{\partial}{\partial\vle^A}\right)
        \, ,\qquad
        \hat{p}^{\calA}= \left(\frac{\partial}{\partial\vll^\alpha},\vlt^{\dot\alpha} ,\vle^A \right)
        \, ,\qquad \text{with}\,\, \qquad
        [\hat{x}^\calA,\hat{p}^\calB] = (-1)^{|\calA|}\delta^{\calA\calB}
       \eqncom
        \label{eq: def xp}
\end{equation}
where $[\cdot,\cdot]$ denotes the graded commutator; see e.g.\ \cite{Ferro:2013dga}.

The spin-chain monodromy matrix is built from the $n$-fold tensor product of the Lax operators \eqref{eq: def lax} in the infinite-dimensional quantum space and matrix multiplication in the auxiliary space. 
Graphically, multiplication from the left or the right in the auxiliary space (quantum space) correspond to attaching vertices from left (bottom) or right (top). 
We define 
\begin{equation}
\begin{aligned}
    \mono_n(u,\{v_i\})
    =
  \begin{tikzpicture}[scale=0.8,baseline=-33.5pt]
  \drawvline{1}{1}   
  \drawvline{3}{1}   
  \drawvline{4}{1}   
  \draw[dashed] (-0.5, -1*\vacuumheight-0.5*\bridgedistance) -- (3.5, -1*\vacuumheight-0.5*\bridgedistance);
  \node at (1, -1*\vacuumheight-0.25*\bridgedistance) {$\cdots$};
  \node at (1, -1*\vacuumheight-0.85*\bridgedistance) {$\cdots$};
\node[dl] at (0,-\vacuumheight-\bridgedistance-\labelvdist) {$n$};
\node[dl] at (2,-\vacuumheight-\bridgedistance-\labelvdist) {$2$};
\node[dl] at (3,-\vacuumheight-\bridgedistance-\labelvdist) {$1$};
  \end{tikzpicture}
    = \lax_n(u-v_n) \cdots \lax_2(u-v_2)\lax_1(u-v_1)\eqncom
    \label{eq: def monodromy}
\end{aligned}    
\end{equation}
with inhomogeneities $v_i$ that are local shifts of the spectral parameter $u$.  

For later purposes, we also introduce the corresponding transfer matrix, which is constructed from the monodromy matrix as the super trace over the auxiliary space:
 \begin{equation}
   \trans_n(u,\{v_i\})
   =
  \begin{tikzpicture}[scale=0.8,baseline=-33.5pt]
  \drawvline{1}{1}   
  \drawvline{3}{1}   
  \drawvline{4}{1}   
  \draw[rounded corners=2mm, dashed] (-0.5, -1*\vacuumheight-0.5*\bridgedistance) rectangle (3.5, -1*\vacuumheight+0.25*\bridgedistance);
  \node at (1, -1*\vacuumheight-0.25*\bridgedistance) {$\cdots$};
  \node at (1, -1*\vacuumheight-0.85*\bridgedistance) {$\cdots$};
\node[dl] at (0,-\vacuumheight-\bridgedistance-\labelvdist) {$n$};
\node[dl] at (2,-\vacuumheight-\bridgedistance-\labelvdist) {$2$};
\node[dl] at (3,-\vacuumheight-\bridgedistance-\labelvdist) {$1$};
  \end{tikzpicture}
    =\str\mono_n(u,\{v_i\})  \eqndot  
    \label{eq: def transfer}
\end{equation} 
As a consequence of the Yang-Baxter equation, this transfer matrix is $\mathfrak{gl}(4|4)$ invariant:
\begin{equation}
    [\trans(u,\{ v_i\}), \sum_{i=1}^n \jj_i^{\calA \calB}]=0
    \eqndot
    \label{eq: transfer invariance}
\end{equation}

Tree-level scattering amplitudes are Yangian invariant \cite{Drummond:2009fd}.
Instead of using Drinfeld's first realisation, this can be expressed as a set of eigenvalue equations involving the monodromy matrix in \eqref{eq: def monodromy}, cf.\ \cite{Frassek:2013xza,Chicherin:2013ora}:
\begin{equation}
\mono_n(u,\{v_i\})   \mathcal{A}  \propto \idm\mathcal{A}  \eqndot
    \label{eq: def monodom}
\end{equation}
While for the physical amplitude the inhomogeneities $v_i$ are zero, they can be set to non-zero
values to obtain deformations of the amplitude with non-vanishing central charges \cite{Ferro:2012xw,Ferro:2013dga}.

From \eqref{eq: def monodom}, it follows that the transfer matrix \eqref{eq: def transfer} acts diagonally on tree-level amplitudes yielding a vanishing eigenvalue. However, this condition is weaker than the set of eigenvalue equations in \eqref{eq: def monodom}.

\subsection{Form factors of the chiral stress-tensor multiplet}
\label{subsec: ff integrability T}

Next, we study the action of the monodromy matrix \eqref{eq: def monodromy}
on more general form factor expressions $\ffgen$ of the chiral stress-tensor multiplet, which we define as
\begin{align}
\label{eq: ff function}
&\ffgen= \rr_{i_1j_1}(z_1)\cdots \rr_{i_mj_m}(z_m) \;\; \ff_{2,2}^\delta(k-1,k)\qquad\intertext{with}
& \ff_{2,2}^\delta(k-1,k)=\delta^+_1 \cdots \delta^+_{k-2} \;
    \ff_{2,2}(k-1,k)\;
    \delta^-_{k+1}\cdots\delta^-_{n}
    \eqndot
    \label{eq: ff delta}
\end{align}
They are constructed from a chain of $\rr$ operators
acting on a vacuum state that is composed of the amplitude vacua
$\delta^+_i$, $\delta^-_i$ as well as the minimal form factor \eqref{eq: minimal form factor as vacuum}. 
The $\rr$ operators and amplitude vacua correspond to the on-shell part of the diagram, with the minimal 
form factor cut out. 
The number $m$ of $\rr$ operators depends on the diagram under consideration.
Moreover, the $\rr$ operators have to be chosen in such a way that the corresponding diagram is planar.%
\footnote{In particular, we assume that for each operator $\rr_{ij}$ the indices satisfy $i<j$; 
this corresponds to the chosen convention for the BCFW shifts.}
This generalises the construction we used in section \ref{sec: MHV} (see e.g.\ \eqref{eq: F 3,2 from r operators} and figures \ref{fig: F3,2}, \ref{fig: F4,2}, \ref{fig: F4,3}), and is another way of writing the gluing of diagrams that was performed in section \ref{sec: beyond MHV}.
Note that these objects correspond to any planar on-shell diagram containing an insertion of the minimal form factor, including top-cell diagrams, individual BCFW terms, factorisation channels etc. 

\subsubsection*{On-shell part}

It was discussed in \cite{Chicherin:2013ora,Broedel:2014pia,Kanning:2014maa} that tree-level scattering amplitudes can be constructed via the method of $\rr$ operators which naturally include the inhomogeneities $v_i$ as deformations of the local central charges \cite{Ferro:2013dga}.%
\footnote{See also \cite{Frassek:2013xza}, where Yangian invariants were constructed using Bethe ansatz methods.} 
These operators, defined by their action on functions of the kinematic data in \eqref{eq: action r operator}, can be formally written as 
\begin{equation}\label{eq: rop}
    \rr_{ij}(u)
    = 
 \begin{tikzpicture}[scale=0.8,baseline=-43pt]
\drawvline{1}{2}
\drawvline{2}{2}
\drawbridge{1}{2}
\node[dl] at (0,-\vacuumheight-2*\bridgedistance-\labelvdist) {$j$};
\node[dl] at (1,-\vacuumheight-2*\bridgedistance-\labelvdist) {$i$};
\end{tikzpicture}
= \int \frac{\dd\alpha}{\alpha^{1+u}}\e^{-\alpha (\hat{x}_j \cdot \hat{p}_i)}
   \eqndot
\end{equation}
The operator $\rr$ can be seen as one of two basic building blocks for Yangian invariants. It satisfies the Yang-Baxter equation
\begin{equation}
        \rr_{ij}(u_j-u_i)
        \lax_j(u_j)\lax_i(u_i)
        =
        \lax_j(u_i)\lax_i(u_j)
        \rr_{ij}(u_j-u_i)\eqncom
        \label{eq: rll}
\end{equation}
with the Lax operators defined in \eqref{eq: def lax}. This equation can be depicted as
\begin{equation} \label{eq: property of r operators in picture} 
  \begin{tikzpicture}[scale=0.8, baseline=-43pt]
	\drawvline{1}{2}   
 	\drawvline{2}{2}   
 	\drawbridge{1}{2}
	\draw[dashed] (-0.5, -1*\vacuumheight-0.5*\bridgedistance) -- (1.5, -1*\vacuumheight-0.5*\bridgedistance);
\node[dl] at (1,-\vacuumheight-2*\bridgedistance-\labelvdist) {$i$};
\node[dl] at (0,-\vacuumheight-2*\bridgedistance-\labelvdist) {$j$};
         \end{tikzpicture}
         =
  \begin{tikzpicture}[scale=0.8, baseline=-43pt]
 	\drawvline{1}{2}   
 	\drawvline{2}{2}   
 	\drawbridge{1}{2}
	\draw[dashed] (-0.5, -1*\vacuumheight-1.5*\bridgedistance) -- (1.5, -1*\vacuumheight-1.5*\bridgedistance);
\node[dl] at (1,-\vacuumheight-2*\bridgedistance-\labelvdist) {$i$};
\node[dl] at (0,-\vacuumheight-2*\bridgedistance-\labelvdist) {$j$};
         \end{tikzpicture}
         \eqndot
\end{equation}
The other basic building blocks are the vacuum states 
$\delta_{i}^+=\delta^2(\vll_i)$ and $\delta_{i}^-=\delta^2(\vlt_i)\delta^4(\vle_i)$ introduced in \eqref{eq: vacua}.
They satisfy 
\begin{equation}
        \lax_i(u)\,\delta_{i}^+
        = (u-1)\; \idm \; \delta_{i}^+\eqncom
        \qquad
        \lax_i(u)\, \delta_{i}^-
        = u \; \idm \; \delta_{i}^-
        \eqncom
        \label{eq: lax on vacua}
\end{equation}
which can be depicted as 
\begin{equation}
  \begin{tikzpicture}[scale=0.8,baseline=-20pt]
\drawvacp{1}
\draw[dashed] (-0.5, -0.75*\vacuumheight) -- (0.5, -0.75*\vacuumheight);
\node[dl] at (0,-\vacuumheight-0*\bridgedistance-\labelvdist) {$i$};
 \end{tikzpicture}
=(u-1)
  \begin{tikzpicture}[scale=0.8,baseline=-20pt]
\drawvacp{1}
\draw[dashed] (-0.5, 0.3*\vacuumheight) -- (0.5, 0.3*\vacuumheight);
\node[dl] at (0,-\vacuumheight-0*\bridgedistance-\labelvdist) {$i$};
 \end{tikzpicture}
\eqncom \qquad
  \begin{tikzpicture}[scale=0.8,baseline=-20pt]
\drawvacm{1}
\draw[dashed] (-0.5, -0.75*\vacuumheight) -- (0.5, -0.75*\vacuumheight);
\node[dl] at (0,-\vacuumheight-0*\bridgedistance-\labelvdist) {$i$};
 \end{tikzpicture}
=u
  \begin{tikzpicture}[scale=0.8,baseline=-20pt]
\drawvacm{1}
\draw[dashed] (-0.5, 0.3*\vacuumheight) -- (0.5, 0.3*\vacuumheight);
\node[dl] at (0,-\vacuumheight-0*\bridgedistance-\labelvdist) {$i$};
 \end{tikzpicture}
\eqndot
\end{equation}
The properties \eqref{eq: rll} and \eqref{eq: lax on vacua} guarantee that an appropriate combination of $\rr$ operators with a suitable choice of inhomogeneities acting on the tensor product of vacuum states 
$\delta_{i}^+$ and $\delta_{i}^-$ is Yangian invariant \cite{Chicherin:2013ora,Broedel:2014pia,Kanning:2014maa}; the required choice of inhomogeneities will be discussed in the following.
However, further below we will also see that Yangian invariance is broken by the insertion of the minimal form factor.
 
As discussed in \cite{Chicherin:2013ora,Broedel:2014pia,Kanning:2014maa}, the monodromy matrix \eqref{eq: def monodromy} satisfies certain exchange relations with a chain of $\rr$ operators. As a consequence of \eqref{eq: rll}, one finds
\begin{equation}
  \label{eq: rchain com}
\mono(u,\{v_i\})\,
    \rr_{i_1j_1}(z_1)\cdots \rr_{i_mj_m}(z_m)=\rr_{i_1j_1}(z_1)\cdots \rr_{i_mj_m}(z_m) \,
    \mono(u,\{v_{\sigma(i)}\}) \eqncom
\end{equation}
where $ \mono(u,\{v_{\sigma(i)}\})$ denotes the monodromy matrix in \eqref{eq: def monodromy} with the inhomogeneity at site $i$ permuted such that $v_i$ is replaced by $v_{\sigma{(i)}}$. Here, $\sigma$ is the permutation associated to the on-shell diagram and can be read off as discussed in section \ref{sec: on shell diagrams}.
The inhomogeneities $v_i$ and the spectral parameters
$z_i$ have to satisfy the relations 
\begin{equation}
    z_\ell = v_{\tau_\ell(i_\ell)}-v_{\tau_\ell(j_\ell)}\quad\quad\text{with}\quad\quad     \tau_\ell= (i_1,j_1)\permprod\cdots\permprod(i_\ell,j_\ell)\eqncom \qquad \ell=1,\ldots,m\eqncom
\end{equation}
see \cite{Chicherin:2013ora,Broedel:2014pia,Kanning:2014maa}.\footnote{Recall that $m$ denotes the number of $\rr$ operators.}
The inhomogeneities $v_i$ associated to the $i^{\text{th}}$ external leg, i.e.\ to site $i$, are related to the central charges $c_i$ via \cite{Beisert:2014qba}
\begin{equation}
    c_i=v_i-v_{\sigma(i)}\eqndot
\end{equation}

Therefore, for a planar on-shell diagram with valid deformations, we can commute the monodromy matrix through the chain of $\rr$ operators in \eqref{eq: ff function} using \eqref{eq: rchain com}:  
\begin{equation}
    \begin{split}
    \mono_n(u,\{v_{i}\})
    \ffgen&= \rr_{i_1j_1}(z_1)\cdots \rr_{i_mj_m}(z_m) \;\mono_n(u,\{v_{\sigma(i)}\})\;\ff_{2,2}^\delta(k-1,k)\eqndot
    \end{split}
    \label{eq: monodromy on ff}
\end{equation}
Since the Lax operators act diagonally on the vacua \eqref{eq: lax on vacua},
we can eliminate all Lax operators that do not act on the minimal form factor. We end up with the monodromy matrix 
\begin{equation}
\mono_2(u,\{v_{\sigma(i)}\})=\lax_{k}(u-v_{\sigma(k)})\lax_{k-1}(u-v_{\sigma(k-1)})
\end{equation}
of length two acting on the minimal form factor $\ff_{2,2}(k-1,k)$ at sites $k-1$ and $k$:
\newcommand{\ffactor}{\mathfrak{f}}
\begin{equation}\label{eq: monodromy on ff 2,2}
\begin{split}
 &\mono_n(u,\{v_{\sigma(i)}\})\;
   \ff_{2,2}^\delta(k-1,k)\\&=\ffactor(u,\{v_{\sigma(i)}\})\;
   \delta^+_1 \cdots \delta^+_{k-2} \;
        \left[
        \mono_2(u,\{v_{\sigma(i)}\})
        \ff_{2,2}(k-1,k) \;
    \right]\;
    \delta^-_{k+1}\cdots\delta^-_{n}
    \eqncom
    \end{split}
\end{equation} 
where
\begin{equation}
\label{eq: ffactor}
 \ffactor(u,\{v_{\sigma(i)}\})=\prod_{i=1}^{k-2}(u-v_{\sigma(i)}-1)\prod_{j=k+1}^{n}(u-v_{\sigma(j)}) \eqndot
\end{equation}
As we will discuss below, this significant difference to tree-level amplitudes breaks the Yangian invariance. However, as we will also see, some of the integrable structure remains.

Using the graphical language introduced earlier, we depict the formulas discussed above for the case of $\ff_{n,2}$ in figure \ref{fig: monodromy on ff n,2}.  The left picture in figure \ref{fig: monodromy on ff n,2} represents the monodromy matrix \eqref{eq: def monodromy} acting on the chain of $\rr$ operators (BCFW bridges) as introduced in \eqref{eq: action r operator} and \eqref{eq: rop} contracted with the minimal form factor \eqref{eq: form factor building block for on-shell diagrams} and the corresponding vacua \eqref{eq: vacua}. In \eqref{eq: monodromy on ff}, we commuted the monodromy matrix through the chain of $\rr$ operators (BCFW bridges) as shown in the middle picture. The action of the Lax operators on the delta functions of the on-shell vacua was discussed in \eqref{eq: lax on vacua}.
As in \eqref{eq: monodromy on ff 2,2}, we end up with a monodromy matrix of length two acting only on the minimal form factor as shown in the right picture of figure \ref{fig: monodromy on ff n,2}.

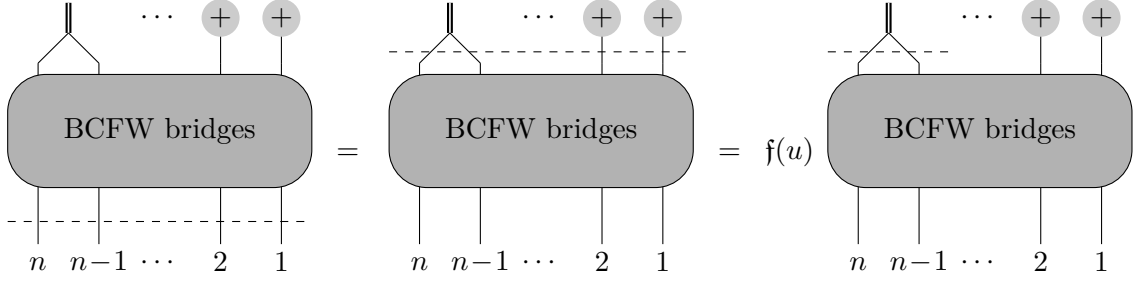
\begin{figure}[htbp]
\begin{equation*}
  \begin{aligned}
 \begin{tikzpicture}[scale=0.8]
\drawvline{1}{4}
\drawvline{2}{4}
\drawvline{4}{4}
\drawvline{5}{4}
\drawminimalff{1} 
\drawvacp{4}
\drawvacp{5}
\node at (0,-\vacuumheight-4*\bridgedistance-\labelvdist) {\strut $n$};
\node at (1,-\vacuumheight-4*\bridgedistance-\labelvdist) {$n \sminus 1$};
\node at (3,-\vacuumheight-4*\bridgedistance-\labelvdist) {$2$};
\node at (4,-\vacuumheight-4*\bridgedistance-\labelvdist) {$1$};
\node at (2,-0.25) {$\cdots$};
\node at (2,-\vacuumheight-4*\bridgedistance-\labelvdist) {$\cdots$};
\draw[dashed] (-0.5, -1*\vacuumheight-3.5*\bridgedistance) -- (4.5, -1*\vacuumheight-3.5*\bridgedistance);
\node[rectangle, rounded corners=\vacuumradius, black, fill=grayn, minimum width=4 cm, minimum height=1.5 cm, draw, inner sep=0pt] at (2,-\vacuumheight-1.5*\bridgedistance) {$\text{BCFW bridges}$};
\end{tikzpicture}
\end{aligned}
\;\;=\;\;
  \begin{aligned}
 \begin{tikzpicture}[scale=0.8]
\drawvline{1}{4}
\drawvline{2}{4}
\drawvline{4}{4}
\drawvline{5}{4}
\drawminimalff{1} 
\drawvacp{4}
\drawvacp{5}
\node at (0,-\vacuumheight-4*\bridgedistance-\labelvdist) {\strut $n$};
\node at (1,-\vacuumheight-4*\bridgedistance-\labelvdist) {$n \sminus 1$};
\node at (3,-\vacuumheight-4*\bridgedistance-\labelvdist) {$2$};
\node at (4,-\vacuumheight-4*\bridgedistance-\labelvdist) {$1$};
\node at (2,-0.25) {$\cdots$};
\node at (2,-\vacuumheight-4*\bridgedistance-\labelvdist) {$\cdots$};
\draw[dashed] (-0.5, -1*\vacuumheight+0.25*\bridgedistance) -- (4.5, -1*\vacuumheight+0.25*\bridgedistance);
\node[rectangle, rounded corners=\vacuumradius, black, fill=grayn, minimum width=4 cm, minimum height=1.5 cm, draw, inner sep=0pt] at (2,-\vacuumheight-1.5*\bridgedistance) {$\text{BCFW bridges}$};
\end{tikzpicture}
\end{aligned}
\;\;=\;\;\ffactor(u)\;
  \begin{aligned}
 \begin{tikzpicture}[scale=0.8]
\drawvline{1}{4}
\drawvline{2}{4}
\drawvline{4}{4}
\drawvline{5}{4}
\drawminimalff{1} 
\drawvacp{4}
\drawvacp{5}
\node at (0,-\vacuumheight-4*\bridgedistance-\labelvdist) {\strut $n$};
\node at (1,-\vacuumheight-4*\bridgedistance-\labelvdist) {$n \sminus 1$};
\node at (3,-\vacuumheight-4*\bridgedistance-\labelvdist) {$2$};
\node at (4,-\vacuumheight-4*\bridgedistance-\labelvdist) {$1$};
\node at (2,-0.25) {$\cdots$};
\node at (2,-\vacuumheight-4*\bridgedistance-\labelvdist) {$\cdots$};
\draw[dashed] (-0.5, -1*\vacuumheight+0.25*\bridgedistance) -- (1.5, -1*\vacuumheight+0.25*\bridgedistance);
\node[rectangle, rounded corners=\vacuumradius, black, fill=grayn, minimum width=4 cm, minimum height=1.5 cm, draw, inner sep=0pt] at (2,-\vacuumheight-1.5*\bridgedistance) {$\text{BCFW bridges}$};
\end{tikzpicture}
\end{aligned}
\end{equation*}
\caption{Action of the monodromy on $\ff_{n,2}$.}
\label{fig: monodromy on ff n,2}
\end{figure}

\subsubsection*{Minimal form factor}

One can explicitly check that the minimal from factor is not an eigenstate
of the length-two monodromy matrix \eqref{eq: def monodromy}, and thus not Yangian invariant, cf.\ \eqref{eq: def monodom}.
This can be seen since, for example,
the momentum-like generators do not contain the off-shell momentum $q$.
Acting on the minimal form factor, this produces 
\begin{equation}
    (\vll_{k-1}\vlt_{k-1}+\vll_{k}\vlt_k)\delta^4(\vll_{k-1}\vlt_{k-1}+\vll_{k}\vlt_k-q)
    \eqncom
\end{equation}
which does not vanish, as would be required for Yangian invariance in the sense of \eqref{eq: def monodom}.
However, as we will show, the minimal form factor
is annihilated by the graded sum of the Yangian generators on the diagonal of the monodromy matrix, i.e.\ the 
transfer matrix \eqref{eq: def transfer}, for equal inhomogeneities: 
\begin{equation}
    \trans_2(u-v)
    =\str\lax_{k}(u-v_{\sigma(k)})
     \lax_{k-1}(u-v_{\sigma(k-1)})
    \eqncom    \qquad
    \text{with}\quad
    v_{\sigma(k-1)}=
    v_{\sigma(k)}= v 
    \eqndot
    \label{eq: trans two hom}
\end{equation}
This can be seen as follows.
First, note that we can look at the action of the transfer matrix \eqref{eq: trans two hom} on
a single component of $\ff_{2,2}$ due to its $\mathfrak{gl}(4|4)$ invariance
\eqref{eq: transfer invariance}.
We take as an example the component
\begin{equation}
\vle^{+1}_{k-1}\vle^{+2}_{k-1}\vle^{+1}_k\vle^{+2}_k (\gamma^-)^4\delta^4(P)\eqncom
\end{equation}
which corresponds to the
minimal form factor of the operator $\tr(\phi^{++}\phi^{++})$ in \eqref{eq: def stress-tensor multiplet} with outgoing scalars label by $k-1$ and $k$.
Note that the transfer matrix does not depend on $\gamma^-$.
Second, one can check that the transfer matrix annihilates the momentum-conserving delta functions,
\begin{equation}
    \trans_2(u-v)\delta^4(\vll_1\vlt_1+\vll_2\vlt_2-q)=0
    \eqncom
\end{equation}
and thus acts only on the product of $\vle$'s, yielding
\begin{equation}
\begin{aligned}
    &\trans_2(u-v)
    \vle^{+1}_{k-1}\vle^{+2}_{k-1}\vle^{+1}_k\vle^{+2}_k 
    \\& =
    \left( 
        (u-v-1)( \hat{x}_{k-1}^\calA \hat{p}_{k-1}^{\calA} + \hat{x}_{k}^\calA \hat{p}_{k}^{\calA}  )
        +(-1)^{|\calA|} \hat{p}_{k-1}^{\calA}\hat{x}_{k-1}^{\calB}\hat{p}_k^\calB \hat{x}_k^\calA
    \right)
    \vle^{+1}_{k-1}\vle^{+2}_{k-1}\vle^{+1}_k\vle^{+2}_k 
    =0
    \eqndot
\end{aligned}
\end{equation}
This shows that the minimal form factor is an eigenstate of the transfer matrix with eigenvalue zero, i.e.\
\begin{equation}
    \trans_2(u-v)\ff_{2,2}=0
    \eqndot
\end{equation}

Moreover, due to the fact that the monodromy matrix, and therefore also the
transfer matrix, commutes through the chain of $\rr$ operators as discussed above \eqref{eq: rchain com},
the same statement applies to any planar on-shell diagram with an insertion
of the minimal form factor \eqref{eq: ff function}:
\begin{equation}
    \trans_n(u, \{v_i\})\ffgen=0
    \label{eq: transfer matrix invariance}
    \eqncom
\end{equation}
with the constraints on the inhomogeneities given in \eqref{eq: trans two hom}.
As for amplitudes, the whole argument is in particular valid for vanishing
inhomogeneities, i.e.\ for undeformed form factors:
\begin{equation}
    \trans_n \ff_{n,k}=0
    \label{eq: transfer matrix invariance undeformed}
    \eqndot
\end{equation}

Note the similarity to the Yangian invariance condition of scattering amplitudes
\eqref{eq: def monodom}: taking the super trace for them shows that they are 
also eigenstates of the transfer matrix with eigenvalue zero. 
Equation \eqref{eq: transfer matrix invariance} implies that
form factors, although not Yangian invariant, are still
annihilated by a certain combination of the Yangian generators.

\subsection{Generic operators}
In the previous section, we have shown that form factors of the stress-tensor super multiplet can be interpreted as eigenvectors of the transfer matrix \eqref{eq: def transfer} with vanishing eigenvalue.
In the following, we want to extend this to all operators. 
We will study the action of the homogeneous transfer matrix on the minimal form factors of generic single-trace operators,
and find that they too are eigenvectors provided that the single-trace operators are chosen as eigenvectors of the spectral problem.
Combining this with an $\rr$-operator construction similar to the one in the previous section, it follows that all planar on-shell diagrams that include these minimal form factors are also eigenstates of the transfer matrix. 
In general, these objects should correspond to leading singularities of loop-level form factors.
For the purposes of the following discussion it will be convenient to work in components and, in contrast to the rest of this article, we do not employ harmonic superspace variables here.

Generic operators in \nfsym\ theory can be conveniently represented via two sets of bosonic oscillators $\aoscdag^{\alpha}_i$ and $\boscdag^{\dot\alpha}_i$ and one set of fermionic oscillators $\doscdag^A_i$ acting on a suitable vacuum \cite{Gunaydin:1998sw,Beisert:2003jj}. 
The oscillators in this oscillator picture transform under $\mathfrak{psu}(2,2|4)$ in the same way as the super spinor helicity variables $\vll_i^\alpha$, $\vlt_i^{\dot\alpha}$ and $\vle_i^A$ and the algebras are formally the same if one identifies
\begin{equation}\label{eq: oscillators spinors}
 \begin{aligned}
  \aosc_i^{\dagger \alpha} &\leftrightarrow \vll_i^\alpha \eqncom & \bosc_i^{\dagger \dot\alpha}&\leftrightarrow \vlt_i^{\dot\alpha}\eqncom & \dosc_i^{\dagger A}&\leftrightarrow\vle_i^A\eqncom\\
  \aosc_{i,\alpha} &\leftrightarrow \partial_{i,\alpha}=\frac{\partial}{\partial\vll_i^\alpha} \eqncom & 
  \bosc_{i,\dot\alpha}&\leftrightarrow \partial_{i,\dot\alpha}=\frac{\partial}{\partial\vlt_i^{\dot\alpha}}\eqncom & \dosc_{i,A}&\leftrightarrow\partial_{i,A}=\frac{\partial}{\partial \vle_i^A}\eqnsem
 \end{aligned}
\end{equation}
see \cite{Beisert:2010jq} for a detailed comparison of these representations.

In \cite{Zwiebel:2011bx}, this identification was used to connect the one-loop dilatation operator to the tree-level four-point scattering amplitude based on symmetry considerations.
The field-theoretic quantities behind such an identification in the composite operators are actually form factors. Concretely, it was shown in \cite{Wilhelm:2014qua} via an explicit Feynman diagram calculation that the colour-ordered minimal tree-level super form factors of generic single-trace operators $\mathcal{O}$ can be obtained from their representation in the oscillator picture as
\begin{equation}
    \ff_{\op,L}(1,\dots,L;q) = L \,
    \delta^4\left( \sum_{i=1}^L \vll_i\vlt_i -q \right) \left[ \op \middle| {
    \begin{smallmatrix}
        \aosc_i^{\dagger \alpha} &\to& \vll_i^\alpha \\
  \bosc_i^{\dagger \dot\alpha} &\to& \vlt_i^{\dot\alpha}\\
  \dosc_i^{\dagger A} &\to& \vle_i^A
    \end{smallmatrix}
}\right] \eqncom
    \label{eq: minimal form factor oscillators}
\end{equation}
where $L$ is the number of fields in the single-trace operator which has been translated according to \eqref{eq: oscillators spinors}.

Due to the $\mathfrak{gl}(4|4)$-invariance of the transfer matrix \eqref{eq: transfer invariance}, it commutes with any function $f( \sum_{i=1}^L \jj_i^{\calA \calB})$. This in particular implies that it commutes with the momentum-conserving delta function in \eqref{eq: minimal form factor oscillators}, which becomes clear after rewriting 
\begin{equation}\label{eq:deltafc}
 \delta^4\left( \sum_{i=1}^L \vll_i\vlt_i -q \right)= \int \dd^4x\, \e^{2 \pi i (\sum_{i=1}^L \vll_i \vlt_i - q)\cdot x}
\end{equation} 
and recalling that $\jj^{\alpha\dot\alpha}_i=\vll_i^\alpha \vlt_i^{\dot\alpha}$.
Thus, the transfer matrix only acts on the operator translated into spinor helicity variables. 
Translating the spinor helicity variables in the transfer matrix $\trans$ to oscillators using \eqref{eq: oscillators spinors} yields
\begin{equation}\label{eq:transosc}
 \transosc(u)=\str \laxosc_L(u)\cdots \laxosc_1(u)\qquad\text{with}\qquad \laxosc_i(u)=\lax_i(u)\left| 
    \begin{smallmatrix}
       \partial_{i,\alpha},\;\,\vll_i^\alpha  &\to& \aosc_{i,\alpha},\;\,\aosc_i^{\dagger \alpha}\\
  \partial_{i,\dot\alpha},\;\,\vlt_i^{\dot\alpha}&\to& \bosc_{i,\dot\alpha},\;\,\bosc_i^{\dagger \dot\alpha}\\
  \partial_{i,A},\;\,\vle_i^A&\to& \dosc_{i,A},\;\,\dosc_i^{\dagger A}
    \end{smallmatrix}\right.
    \eqndot
\end{equation} 
We obtain the relation
\begin{equation}\label{eq:tactt}
    \trans_L(u) \ff_{\op,L}
    =
    \ff_{\transosc_L(u)\op,L}
    \eqncom
\end{equation}
where 
\begin{equation}
 \ff_{\transosc_L(u)\op,L}= L \,
    \delta^4\left( \sum_{i=1}^L \vll_i\vlt_i -q \right) \left[(\transosc_L(u) \op) \middle| {
    \begin{smallmatrix}
        \aosc_i^{\dagger \alpha} &\to& \vll_i^\alpha \\
  \bosc_i^{\dagger \dot\alpha} &\to& \vlt_i^{\dot\alpha}\\
  \dosc_i^{\dagger A} &\to& \vle_i^A
    \end{smallmatrix}
}\right]\eqndot
    \label{eq: minimal form factor oscillatorls}
\end{equation} 

\newcommand{\eigenval}{\mathfrak{t}}

Having expressed the action of the transfer matrix $\trans$ on the minimal form factor $\ff_{\op,L}$ in terms of the transfer matrix $\transosc_L$, cf.\ \eqref{eq:tactt}, we will now argue that $\transosc_L$ acts diagonally on $\op$, i.e.\
\begin{equation}\label{eq:tev}
 \transosc_L(u) \op=\eigenval(u) \op\eqncom
\end{equation} 
if the state corresponding to the operator $\mathcal{O}$ is an eigenstate of the spin-chain Hamiltonian $\mathbf{H}$, i.e.\ the one-loop dilatation operator of \nfsym\ theory. Here, $\eigenval(u)$ is a polynomial in the spectral parameter $u$.\footnote{In the framework of the Bethe ansatz, the eigenvalues $\eigenval(u)$ can be parametrised by the Bethe roots.}
While the particular transfer matrix $\mathbf{T}(u)$ does not contain the spin-chain Hamiltonian $\mathbf{H}$, it is commonly used to diagonalise the commuting family of operators \cite{Faddeev:1996iy}. 
Here, we show that for $v_i=0$ the transfer matrix $\transosc(u)$ commutes with $\mathbf{H}$ using a criterion by Sutherland \cite{PhysRevLett.19.103} and therefore belongs to the same family of commuting operators, see also \cite{Sklyanin:1991ss}.
This is a consequence of the Yang-Baxter equation 
\begin{equation}\label{eq: harmybe}
 \mathbf{R}_{i,i+1}(u)\laxosc_i(u+u')\laxosc_{i+1}(u')=\laxosc_{i+1}(u')\laxosc_i(u+u') \mathbf{R}_{i,i+1}(u)
\end{equation} 
studied in \cite{Ferro:2013dga} where the harmonic R matrix $\mathbf{R}$ was derived, see also \cite{Frassek:2013xza}.
The expansion of the harmonic R matrix contains the Hamiltonian density $\mathcal{H}_{i,i+1}$ at first order in the spectral parameter $ \mathbf{R}_{i,i+1}(u)=\mathcal{P}_{i,i+1}(\idm+u\mathcal{H}_{i,i+1}+\ldots)$, see e.g.\ \cite{Sklyanin:1991ss} as well as \cite{Ferro:2013dga,Kazama:2015iua} where this relation was discussed in relation to \nfsym\ theory. Taking the derivative of \eqref{eq: harmybe} with respect to $u$ and subsequently multiplying with the permutation operator $\mathcal{P}_{i,i+1}$, we obtain 
\begin{equation}\label{eq:hamcom}
 [\mathcal{H}_{i,i+1},\laxosc_i(u')\laxosc_{i+1}(u')]=\laxosc_i(u')-\laxosc_{i+1}(u')\eqndot
\end{equation} 
As a consequence, one finds that the Hamiltonian commutes with the transfer matrix
\begin{equation}
  [\mathbf{H},\transosc(u)]=0\eqncom\qquad \text{with} \qquad 
 \mathbf{H}=\sum_{i=1}^L\mathcal{H}_{i,i+1}\eqncom
\end{equation} 
where periodic boundary conditions $\mathcal{H}_{L,L+1}=\mathcal{H}_{L,1}$ are imposed.

\begin{figure}[htbp]
\begin{equation*}
        \begin{tikzpicture}[baseline=-1cm,scale=0.8]
                \node[draw,inner sep=0](ff) at (1,1) {};
                \draw[double] (1,1.4)--(ff);
                \node[] (onshell) at (1, -1) {};
                \draw (ff) to[in=65,out=305] ([shift={(0.4,0)}] onshell);
                \draw (ff) to[in=90,out=270] ([shift={(0,0)}] onshell);
                \draw (ff) to[in=115,out=235] ([shift={(-0.4,0)}] onshell);
                \draw ([shift={(-0.8,0)}] onshell) to[out=-140, in=90] (-1,-3);
                \draw ([shift={(-0.4,0)}] onshell) to[out=-115, in=90] (0,-3);
                \draw ([shift={(0.0,0)}] onshell) to[out=-90, in=90] (1,-3);
                \draw ([shift={(0.4,0)}] onshell) to[out=-65, in=90] (2,-3);
                \draw ([shift={(0.8,0)}] onshell) to[out=-40, in=90] (3,-3);
                \node[rectangle, rounded corners=15pt, draw, inner sep=22pt, scale=0.8, fill=lightgrayn] at (1, -1) {on-shell part};
                \draw[rounded corners=3pt, dashed] (-1.5,-2.7) rectangle (3.5,-2.4);
        \end{tikzpicture}
        \;\;
=
        \;\; \ffactor(u)\;
        \begin{tikzpicture}[baseline=-1cm,scale=0.8]
                \node[draw,inner sep=0](ff) at (1,1) {};
                \draw[ double] (1,1.4)--(ff);
                \node[] (onshell) at (1, -1) {};
                \draw (ff) to[in=65,out=305] ([shift={(0.4,0)}] onshell);
                \draw (ff) to[in=90,out=270] ([shift={(0,0)}] onshell);
                \draw (ff) to[in=115,out=235] ([shift={(-0.4,0)}] onshell);
                \draw ([shift={(-0.8,0)}] onshell) to[out=-140, in=90] (-1,-3);
                \draw ([shift={(-0.4,0)}] onshell) to[out=-115, in=90] (0,-3);
                \draw ([shift={(0.0,0)}] onshell) to[out=-90, in=90] (1,-3);
                \draw ([shift={(0.4,0)}] onshell) to[out=-65, in=90] (2,-3);
                \draw ([shift={(0.8,0)}] onshell) to[out=-40, in=90] (3,-3);
                \node[rectangle, rounded corners=15pt, draw, inner sep=22pt, scale=0.8, fill=lightgrayn] at (1, -1) {on-shell part};
                \draw[rounded corners=3pt, dashed] (0.2,0.15) rectangle (1.8,0.45);
        \end{tikzpicture}
        \;\;
        =
        \;\;
        \ffactor(u)\eigenval(u)\;
        \begin{tikzpicture}[baseline=-1cm,scale=0.8]
                \node[draw,inner sep=0](ff) at (1,1) {};
                \draw[ double] (1,1.4)--(ff);
                \node[] (onshell) at (1, -1) {};
                \draw (ff) to[in=65,out=305] ([shift={(0.4,0)}] onshell);
                \draw (ff) to[in=90,out=270] ([shift={(0,0)}] onshell);
                \draw (ff) to[in=115,out=235] ([shift={(-0.4,0)}] onshell);
                \draw ([shift={(-0.8,0)}] onshell) to[out=-140, in=90] (-1,-3);
                \draw ([shift={(-0.4,0)}] onshell) to[out=-115, in=90] (0,-3);
                \draw ([shift={(0.0,0)}] onshell) to[out=-90, in=90] (1,-3);
                \draw ([shift={(0.4,0)}] onshell) to[out=-65, in=90] (2,-3);
                \draw ([shift={(0.8,0)}] onshell) to[out=-40, in=90] (3,-3);
                \node[rectangle, rounded corners=15pt, draw, inner sep=22pt, scale=0.8, fill=lightgrayn] at (1, -1) {on-shell part};
        \end{tikzpicture}
\end{equation*}
\caption{Action of the transfer matrix on an on-shell diagram with an insertion of the minimal form factor of the operator $\op$.}
\label{fig: transfer}
\end{figure}
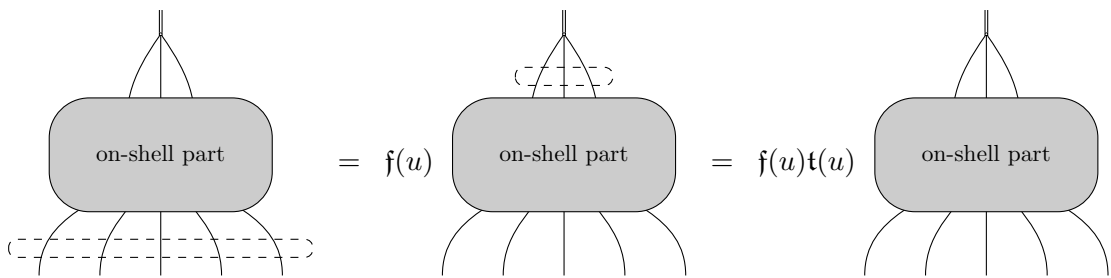

Just as for the minimal form factor of the chiral stress-tensor multiplet in section \ref{subsec: ff integrability T}, we can glue planar on-shell diagrams to the minimal form factor
\eqref{eq: minimal form factor oscillators} using $\rr$ operators, cf.\ \eqref{eq: rchain com}. By construction, this part is Yangian invariant and the monodromy matrix can be commuted through the chain of $\rr$ operators as shown in \eqref{eq: monodromy on ff}. The commutation relations for the transfer matrix built from that monodromy matrix follow immediately after taking the trace in the auxiliary space, see the first step in figure \ref{fig: transfer}.
Using \eqref{eq:tactt}, we find
\begin{equation}
    \label{eq: eigenvalue equation}
    \trans_n(u)\; \ffgen_{\op,n}
    =\ffactor(u)
    \ffgen_{\transosc_L\op,n}
    \eqncom
\end{equation}
where $\ffactor(u)$ denotes the factors arising from the action of the Lax operators on the vacuum, see \eqref{eq: lax on vacua}.
Finally, from the argument presented above, we find that these generalisations of \eqref{eq: ff function} are 
eigenstates of the transfer matrix $\trans$,
\begin{equation}
    \label{eq: eigenvalue equation2}
  \trans_n(u)\; \ffgen_{\op,n}
=
\ffactor(u)
    \ffgen_{\transosc_L\op,n}
  =\ffactor(u)
    \eigenval(u)\;\ffgen_{\op,n}
    \eqncom
\end{equation}
if the operator satisfies the eigenvalue equation \eqref{eq:tev}.
This generalises the corresponding identity \eqref{eq: transfer matrix invariance} for the stress-tensor super multiplet and can be denoted graphically as shown in figure \ref{fig: transfer}.%
\footnote{In fact, the intermediate state in figure \ref{fig: transfer} depicts the generalisation of the right hand side of \eqref{eq: monodromy on ff 2,2} to the transfer matrix, which coincides with the intermediate step in \eqref{eq: eigenvalue equation2} via \eqref{eq:tactt}.}

Note that $\ffgen_{\op,n}$ does not necessarily correspond to a
tree-level form factor of the composite operator $\op$. 
However, at least some of the on-shell diagrams correspond to leading singularities of loop-level form factors. 
It would be interesting to see whether general tree-level form factors are eigenstates of the transfer matrix,
and whether the identities \eqref{eq: eigenvalue equation} for leading singularities are hints of similar
integrability properties at loop level.

\section{Conclusion and outlook}
\label{sec: conclusion and outlook}

In this paper, we have extended many concepts that were developed in the context of the purely on-shell amplitudes to the partially off-shell form factors, 
focussing on the tree-level form factors of the chiral part of the stress-tensor multiplet as an example.

We have shown that on-shell diagrams can be used to characterise form factors by including the minimal form factor as a further building block in addition to the three-point MHV and $\MHVbar$ amplitudes. Apart from the equivalence moves for amplitudes, this requires a rotation move that reflects the cyclic invariance of the three-point form factors.
Moreover, we can extend the concept of a top-cell diagram to form factors. Whereas one top-cell diagram suffices for amplitudes, we require several ones for form factors. We have given a conjecture for the top-cell diagrams for all numbers of on-shell particles $n$ and MHV degree $k$, which is based on a relation to the amplitude case. We have explicitly checked this conjecture against the known results for all $n$ at MHV level, up to five external points in the NMHV sector and for the simplest example at NNMHV. 

Moreover, we have rewritten the previously obtained expressions into the form of a (deformed) Graßmannian integral. As we use two on-shell momenta to parametrise the off-shell momentum of the composite operator, it in general involves the Graßmannian $G(k,n+2)$, where $n$ is the number of external on-shell states. This construction geometrises the (super) momentum conservation and is based on gluing the minimal form factor to the rest of the on-shell diagram.
We have obtained the Graßmannian integral in spinor helicity as well as twistor and momentum twistor variables.
As can be seen from \eqref{eq: general grassmannian form} and \eqref{eq: general grassmannian}, there are significant differences between the Graßmannian integral formulas for scattering amplitudes and form factors. The form in the case of scattering amplitudes is expressed in terms of consecutive cyclic minors labelled by the external particles. The gluing procedure for form factors, however, results in a form that contains minors involving non-consecutive labels as well.
Moreover, we use two fictitious on-shell particles to parametrise the  off-shell momentum of the operator and hence the minors of the form factor Graßmannian are labelled not just by physical external particles but by two additional ones and it seems generally not possible to disentangle these ones from the rest. 

Introducing a central-charge deformation to form factors, we could construct them via $\rr$ operators in analogy to the amplitude case. While amplitudes are Yangian invariant and hence eigenvectors of the monodromy matrix, the behaviour of general $n$-point form factors when acting with the monodromy matrix is entirely determined by its residual action on the minimal form factor and hence the corresponding Yangian, see figure~\ref{fig: monodromy on ff n,2}. In order to obtain an eigenvalue equation for the form factor, we have taken the super trace of the homogeneous monodromy matrix, which yields the homogeneous transfer matrix. 
In particular, we find that the minimal form factor of the stress-tensor super multiplet is annihilated by this transfer matrix, which contains
a subset of the Yangian generators.
This equation is shown in \eqref{eq: transfer matrix invariance undeformed}.
Our construction of the $n$-point tree-level form factors of the chiral part of the stress-tensor multiplet via the integrability-inspired method of $\rr$ operators shows that they satisfy \eqref{eq: transfer matrix invariance undeformed} as well.
Furthermore, we have shown that the minimal form factors of all operators can be interpreted as eigenstates of the homogeneous transfer matrix and established a connection between the integrable structure of the spin chain that appeared in the spectral problem and the one that appeared in the study of tree-level amplitudes. 
Finally, we have argued that the minimal form factor of generic operators can always be dressed with a chain of $\rr$ operators without spoiling the eigenvalue equation with respect to the transfer matrix, see \eqref{eq: eigenvalue equation}.
The resulting objects can be interpreted as leading singularities of loop-level form factors.
\newline

The results discussed in this work open up very interesting directions of future research.
Our construction of the form factor top-cell diagram uses the corresponding diagram from scattering amplitudes with a box replaced with the minimal form factor, see \eqref{eq: box eater}.
This construction has worked in all our examples and can be proven for individual BCFW terms, but it is desirable to have a general proof that the conjectures \eqref{eq: box eater} and \eqref{eq: general grassmannian}, \eqref{eq: general grassmannian form} for the top-cell diagrams and the corresponding Graßmannian integrals for all N$^k$MHV form factors indeed produce all possible BCFW terms.

As mentioned earlier, we have tested our Graßmannian formula with known results based on case studies.
For the examples we have checked, we could determine the combination of the residues which gives the correct tree-level form factors, but it would be very interesting to find a general prediction of a contour which gives the right combination. 
For the scattering amplitudes, such a contour was determined by the twistor string theory formulation of scattering amplitudes \cite{Nandan:2009cc,ArkaniHamed:2009dg,Bourjaily:2010kw}, which leads to the question whether form factors also have an interpretation in terms of an underlying twistor string theory. 
The presence of non-consecutive minors as well as the fact that we have multiple top-cell diagrams
makes the classification of residues a more formidable problem. 

So far, a deeper understanding of the geometry of the Graßmannian formulation is missing. In particular, it would be interesting to see whether the form in \eqref{eq: general grassmannian} follows from a modified notion of positivity. 

Another fruitful direction to pursue would be to extend our result to loop level and to obtain a ``formfactorhedron'' as a counterpart to the amplituhedron \cite{Arkani-Hamed:2013jha}. It is known that even for planar form factors at two-loop level we need Feynman diagrams that are non-planar after removing the minimal form factor. It will be interesting to see how such apparent non-planarity for loop-level form factor plays a role in the on-shell diagrams and Graßmannian formulation.%
\footnote{For a discussion of non-planar on-shell diagrams for amplitudes, see \cite{Arkani-Hamed:2014via,Arkani-Hamed:2014bca,Chen:2014ara,Franco:2015rma,Chen:2015qna}. In particular, non-consecutive minors also appear in this context.}
Moreover, there has been very interesting progress in studying loop-level correlation functions  in \nfsym\ theory using the Lagrangian insertion techniques \cite{Eden:2010zz} and it may be interesting to understand a similar picture for form factors at loop level, or to investigate whether the minimal form factor insertions even have a direct interpretation within this framework.

While we have focused on the form factor of the chiral part of the stress-tensor super multiplet in first half of this paper, we are convinced that our results can be extended to general operators; in particular to operators that are non-protected and have a length $L\geq 3$.

We expect that the minimal form factor allows for deformations consistent with the deformation introduced through the $\rr$ operators.
The resulting eigenvalue equation for the inhomogeneous transfer matrix suggest that any given on-shell diagram with an insertion of a deformed minimal form factor is fully characterised by the inhomogeneities and Bethe roots and as such can be determined using Bethe ansatz methods along the lines of \cite{Frassek:2013xza}. Hopefully, this will yield a uniform description of the observables of planar \nfsym\ theory as an integrable model at weak coupling and beyond, see also the recent and very promising approach in \cite{Basso:2013vsa,Basso:2015zoa}.
Constructing the form factors as solutions to \eqref{eq: eigenvalue equation} at loop level should in particular yield an integrability-based approach to the eigenvalues and eigenvectors of the dilatation operator.%
\footnote{An interesting approach toward the construction of the eigenvectors was pursued in \cite{Derkachov:2013bda}. It is based on objects that are in some respects more general and in others more special than form factors of composite operators, namely special kinematic limits of minimal form factors of light-ray operators.}
While the eigenvalues can be obtained to very high loop orders via integrability, the corresponding eigenvectors are only known to one-loop order for generic operators. 

Going beyond form factors, it would be very interesting to glue the on-shell diagrams of form factors together to obtain on-shell diagrams for correlations functions, which are purely off-shell objects.
The results of \cite{Engelund:2012re} suggest that this is meaningful at least at the level of leading singularities.

Finally, it would be interesting to extend our results to other theories.
In particular, both form factors \cite{Brandhuber:2013gda,Young:2013hda,Bianchi:2013pfa,Bianchi:2013iha} as well as (deformed) on-shell diagrams and Graßmannians \cite{Huang:2013owa,Huang:2014xza,Bargheer:2014mxa} were already studied in ABJM theory \cite{Aharony:2008ug} and could be combined as was done in this paper for $\mathcal{N}=4$ SYM theory.

\section*{Acknowledgements}

It is a pleasure to thank Nils Kanning and Gregor Richter for initial collaboration and discussions.
We are grateful to Matthias Staudacher for comments on the manuscript. 
We would like to thank Burkhard Eden, Paul Heslop, Tomasz \L{}ukowski, Brenda Penante and Gang Yang for helpful discussions. 
DM thanks the Caltech Particle Theory group for their hospitality during a crucial stage of preparation of this work and DN and MW thank the C.N.\ Yang Institute for Theoretical Physics SUNY Stonybrook for their hospitality where parts of this work was done. 
DM, DN and MW acknowledge the support of the Marie Curie International Research Staff Exchange Network UNIFY of the European Union's Seventh Framework Programme [FP7-People-2010-IRSES] under Grant Agreement No 269217, which made the above visits possible.
DM is supported by GK 1504 \emph{``Masse, Spektrum, Symmetrie''}.
This work was supported in part by the SFB 647 \emph{``Raum-Zeit-Materie. Analytische und Geometrische Strukturen''} and by the Marie Curie network GATIS (\texttt{\href{http://gatis.desy.eu}{gatis.desy.eu}}) of the European Union’s Seventh Framework Programme FP7/2007-2013/ under REA Grant Agreement No 317089. MW dankt der Studienstiftung des deutschen Volkes für ein Promotionsförderstipendium.

\bibliographystyle{utphys2}
\bibliography{refs}

\end{document}